%% file: Heavy_flavor_Tevatron.tex
\documentclass[final]{ws-ijmpa}
\pdfoutput=1
\usepackage[super,compress]{cite}
\usepackage{graphicx}
\usepackage{url}
\usepackage{hyperref}
\usepackage{xspace}

\begin{document}

\input{definitions.tex}

\markboth{Jonathan Lewis, Rick Van Kooten}
{Heavy Flavor Physics at the Tevatron}

%
\catchline{}{}{}{}{}
%

\title{REVIEW OF PHYSICS RESULTS FROM THE TEVATRON: \\
       HEAVY FLAVOR PHYSICS}

\author{JONATHAN LEWIS}

\address{Fermi National Accelerator Laboratory\\
MS-318, P.O. Box 500\\
Batavia, IL 60510, United States\\
jdl@fnal.gov}

\author{RICK VAN KOOTEN}

\address{Department of Physics, Indiana University, Swain Hall West 117\\
Bloomington, IN 47405-7105,
United States\\
rvankoot@indiana.edu}

\maketitle

\begin{history}
\end{history}

\begin{abstract}
We present a review of heavy flavor physics results from the CDF and D\O\ Collaborations
operating at the Fermilab Tevatron Collider.  A summary of results from Run 1 is included, 
but we concentrate on legacy results of charm and $b$ physics from Run 2, including results up to Summer 2014.


\end{abstract}

\ccode{PACS numbers: 12.38.Qk, 12.39.Hg, 12.39.Jh, 12.39.Ki ,12.39.Mk,
13.20.Fc, 13.20.Gd, 13.20.He,
13.25.Ft, 13.25.Gv, 	13.25.Hw,
13.30.Ce, 13.30.Eg,
13.60.Le, 13.60.Rj,
13.87.Ce, 13.87.Fh,
14.20.Lq, 14.20.Mr,
14.40.Lb, 14.40.Nd, 14.40.Pq, 14.40.Rt,
14.65.Dw, 14.65.Fy,
}

\newpage
\tableofcontents

\newpage
\input{introduction.tex}

\input{experimental.tex}

\input{run1.tex}

\input{production.tex}

\input{spectroscopy.tex}

\input{lifetimes.tex}
\input{decays.tex}

\input{mixing.tex}

\input{CPViolation.tex}

\input{rare_decays.tex}

\section{Summary}

When planning began for the Tevatron, the bottom quark had only
recently been discovered.  The measurement of $b$ quark cross sections
and full reconstruction of a handful $B$ meson decays in the early
data were already an achievement.  While conventional wisdom held that
$b$ physics could not be done in the busy hadron collision
environment, early in Run 1, CDF and \d0\ were breaking new ground and
making world-leading measurements, often surprising their LEP competitors
at the time with their capabilities. 
The improvements in technology and
increases in luminosity that came with Run~2 coupled with the
ingenuity of analyzers led to an explosion of heavy flavor results
including the discovery of $B_s$ mixing and new baryons, world-leading
measurements of \CP\ violation and $b$ hadron properties, and
precision results on heavy-quark and quarkonium production and
spectroscopy.
          
\section*{Acknowledgments}

We thank Marjorie Corcoran, Mark Williams, and Daria Zieminska for reviewing and providing valuable
comments.
We also thank the Fermilab staff and technical staffs of
the participating institutions for their vital contributions.
We acknowledge support from the DOE and NSF
(USA), ARC (Australia), CNPq, FAPERJ, FAPESP
and FUNDUNESP (Brazil), NSERC (Canada), NSC,
CAS and CNSF (China), Colciencias (Colombia), MSMT
and GACR (Czech Republic), the Academy of Finland,
CEA and CNRS/IN2P3 (France), BMBF and DFG (Germany),
DAE and DST (India), SFI (Ireland), INFN
(Italy), MEXT (Japan), the KoreanWorld Class University
Program and NRF (Korea), CONACyT (Mexico),
FOM (Netherlands), MON, NRC KI and RFBR (Russia),
the Slovak R\&D Agency, the Ministerio de Ciencia
e InnovaciÂ´on, and Programa Consolider--Ingenio 2010
(Spain), The Swedish Research Council (Sweden), SNSF
(Switzerland), STFC and the Royal Society (United
Kingdom), the A.P. Sloan Foundation (USA), and the
EU community Marie Curie Fellowship contract 302103. 

\bibliographystyle{ws-ijmpa}
\bibliography{Heavy_flavor_Tevatron}
\end{document}

%% file: definitions.tex

\def\ra{\rightarrow}

\newcommand{\particle}[1]{\ensuremath{#1}\xspace}
\newcommand{\ee}{\particle{e^+e^-}}
\newcommand{\ppbar}{\particle{p\bar{p}}}
\newcommand{\Jpsi}{\particle{J/\psi}}
\newcommand{\Ups}{\particle{\Upsilon(4S)}}
\newcommand{\Upsone}{\particle{\Upsilon(1S)}}
\newcommand{\Upstwo}{\particle{\Upsilon(2S)}}
\newcommand{\Upsthree}{\particle{\Upsilon(3S)}}
\newcommand{\Upsfive}{\particle{\Upsilon(5S)}}
\newcommand{\bp}{\particle{b}}
\newcommand{\B}{\particle{B}}
\newcommand{\Bd}{\particle{B^0}}
\newcommand{\Bs}{\particle{B^0_s}}
\newcommand{\Bu}{\particle{B^+}}
\newcommand{\Bc}{\particle{B^+_c}}
\newcommand{\Bdbar}{\particle{\bar{B}^0}}
\newcommand{\Bsbar}{\particle{\bar{B}^0_s}}
\newcommand{\BL}{\particle{B_{\mathrm{L}}}}
\newcommand{\BH}{\particle{B_{\mathrm{H}}}}
\newcommand{\BsL}{\particle{B_{s\mathrm{L}}}}
\newcommand{\BsH}{\particle{B_{s\mathrm{H}}}}
\newcommand{\Lb}{\particle{\Lambda_b^0}}
\newcommand{\Xib}{\particle{\Xi_b}}
\newcommand{\Xibd}{\particle{\Xi_b^-}}
\newcommand{\Xibu}{\particle{\Xi_b^0}}
\newcommand{\Omegab}{\particle{\Omega_b^-}}
\newcommand{\Lc}{\particle{\Lambda_c^+}}
\newcommand{\XJpsi}{\particle{X(3872)}}

\newcommand{\fBs}{\ensuremath{f_{\particle{s}}}\xspace}
\newcommand{\fBd}{\ensuremath{f_{\particle{d}}}\xspace}
\newcommand{\fBu}{\ensuremath{f_{\particle{u}}}\xspace}
\newcommand{\fbb}{\ensuremath{f_{\rm baryon}}\xspace}
\newcommand{\fLb}{\ensuremath{f_{\Lambda_{b}}}\xspace}
\newcommand{\fXib}{\ensuremath{f_{\Xi_{b}}}\xspace}
\newcommand{\fOb}{\ensuremath{f_{\Omega_{b}}}\xspace}

\newcommand{\dmd}{\ensuremath{\Delta m_{\particle{d}}}\xspace}
\newcommand{\dms}{\ensuremath{\Delta m_{\particle{s}}}\xspace}
\newcommand{\dmq}{\ensuremath{\Delta m_{\particle{q}}}\xspace}
\newcommand{\xd}{\ensuremath{x_{\particle{d}}}\xspace}
\newcommand{\xs}{\ensuremath{x_{\particle{s}}}\xspace}
\newcommand{\yd}{\ensuremath{y_{\particle{d}}}\xspace}
\newcommand{\ys}{\ensuremath{y_{\particle{s}}}\xspace}
\newcommand{\chibar}{\ensuremath{\overline{\chi}}\xspace}
\newcommand{\chid}{\ensuremath{\chi_{\particle{d}}}\xspace}
\newcommand{\chis}{\ensuremath{\chi_{\particle{s}}}\xspace}
\newcommand{\GL}{\ensuremath{\Gamma_{\mathrm{L}}}\xspace}
\newcommand{\GH}{\ensuremath{\Gamma_{\mathrm{H}}}\xspace}
\newcommand{\GsL}{\ensuremath{\Gamma_{s\mathrm{L}}}\xspace}
\newcommand{\GsH}{\ensuremath{\Gamma_{s\mathrm{H}}}\xspace}
\newcommand{\Gd}{\ensuremath{\Gamma_{\particle{d}}}\xspace}
\newcommand{\DGd}{\ensuremath{\Delta\Gd}\xspace}
\newcommand{\DGGd}{\ensuremath{\DGd/\Gd}\xspace}
\newcommand{\Gs}{\ensuremath{\Gamma_{\particle{s}}}\xspace}
\newcommand{\DGs}{\ensuremath{\Delta\Gs}\xspace}
\newcommand{\DGGs}{\ensuremath{\Delta\Gs/\Gs}\xspace}
\newcommand{\ASLd}{\ensuremath{{\cal A}_{\rm SL}^\particle{d}}\xspace}
\newcommand{\ASLs}{\ensuremath{{\cal A}_{\rm SL}^\particle{s}}\xspace}
\newcommand{\ASLb}{\ensuremath{{\cal A}_{\rm SL}^\particle{b}}\xspace}

\newcommand{\DG}{\ensuremath{\Delta\Gamma}\xspace}
\newcommand{\phiccbars}{\ensuremath{\phi_s^{c\bar{c}s}}\xspace}

\newcommand{\BR}[1]{\particle{{\cal B}(#1)}}
\newcommand{\CL}[1]{#1\%~\mbox{CL}}
\newcommand{\Qjet}{\ensuremath{Q_{\rm jet}}\xspace}

\def\d0{D\O}

\newcommand{\CP}{{\em{CP}}\xspace}

\newcommand{\GeVc}{{\rm GeV}}  
\newcommand{\GeVcc}{{\rm GeV}} 
\newcommand{\MeVcc}{{\rm MeV}} 
\newcommand{\ipb}{\ensuremath{{\rm pb}^{-1}}\xspace}
\newcommand{\ifb}{\ensuremath{{\rm fb}^{-1}}\xspace}
\newcommand{\microbarn}{\ensuremath{\mu{\rm b}}\xspace}

%

%% file: introduction.tex
\section{Introduction and Context}

The flavor sector is that part of the standard model (SM) that arises from
the interplay of quark weak gauge couplings and quark-Higgs
couplings. The misalignment of these in the mass eigenstate basis
gives rise to the Cabbibo-Kobayashi-Maskawa (CKM) matrix that encodes
the physics of the weak flavor-changing decays of quarks. There are
three generations of quarks and a very wide range of quark couplings
and masses; however, we observe hadrons, not quarks. The top quark is
so massive that it decays on timescales shorter than the typical
hadronization time into other quarks. Bottom and charm quarks are therefore
the most massive quarks that can comprise observable particles, and
these are termed the ``heavy flavor" hadrons. Their large mass often
allows for relatively precise theoretical predictions via symmetries,
perturbative QCD and the operator product expansion, heavy quark
expansion, lattice QCD calculations, and QCD sum rules. Along the way,
a great deal can be learned in the area of strong interactions that
permits the probing of the weak physics of heavy flavors and the CKM
matrix.

The production of heavy flavor hadrons tests QCD theory, and
spectroscopy explores the interactions and dynamics of quarks inside
of hadrons. Lifetimes and branching fractions straddle the boundary of
weak decays and hadronic physics effects. The $3 \times 3$ CKM matrix
allows for a \CP-violating phase, and with enough independent
measurements, it is possible to test SM predictions for \CP\
violation. We know that the SM level of \CP\ violation is not
sufficient to explain the current matter-antimatter asymmetry of the
universe, so we are continually searching for additional sources of \CP
violation due to physics beyond the standard model. The handy
interferometric system of neutral heavy meson mixing and
oscillations is another avenue to test \CP violation as well as
provide powerful constraints on CKM matrix elements.  Finally, quantum
effects in flavor loops can occur in any of the above areas and in
cleanly predicted rare decays of heavy hadrons that in turn provide
exquisitely sensitive probes for new physics. Following a description
of the characteristics of heavy flavor physics at the Tevatron, all of
the heavy flavor results from the CDF and \d0\ Collaborations in each
of these areas are reviewed.

%% file: experimental.tex
\section{Experimental Characteristics of Flavor Physics at the
  Tevatron}

The most appealing feature of hadron machines as tools to study $b$
physics is their very high cross section for $b\bar{b}$ production.
The Tevatron was a copious source of $b$ hadrons with a production
cross section four orders of magnitude greater than that at $e^+e^-$
$B$ factories.  $B$ factories operating at the $e^+e^- \rightarrow
\Upsilon(4S)$ only produce $B^0$ and $B^{\pm}$ mesons, and after LEP
ceased operations, the Tevatron was a unique source for all the other
heavier $b$ hadrons such as $B^0_s$, $B_c^+$, $b$ baryons, and all their
excited states for more than a decade.\footnote{Unless otherwise
stated, references to particles and processes include the charge
conjugates as well.}

Despite a very large production rate, $b$-quark events were a
minuscule fraction of the overall $p\bar{p}$ event rate at the
Tevatron.  While the cross section for hadronic interactions at the
Tevatron is about 50\,mb, for the production of $b$ and $c$ quarks in
the central region where the experiments have sensitivity, it is only
about 10\,\microbarn.  Two features of heavy quarks make it possible
to observe meaningful levels of signal among the enormous backgrounds.
The first is the long lifetime of $\simeq$1.5\,ps for $b$ hadrons, so
that these boosted hadrons are likely to decay at secondary vertices a
significant distance (of the order of a millimiter) from the beamline and
interaction point of the $p\bar{p}$ beams.  The reconstruction of
these secondary vertices or the observation that a charged particle
track is inconsistent with pointing back to the beamline is a powerful
signature for identifying heavy flavor decays.  Secondly, $b$ hadrons
have semileptonic branching ratios that are about 10\% while it is
rare for leptons to be produced in the prompt decays of light hadrons.
Leptons are produced from interactions or decays at larger distances
such as electrons from photon conversions, or muons from decays in
flight of kaons and pions, but these backgrounds are small compared to
inclusive hadron rates and can be studied and controlled.  The
situation for charm particles is more complicated.  The lifetime and
semileptonic branching ratio of $D^+$ mesons is similar to those for
$b$ hadrons; however, due to the large number of decay modes available
to $D^0$ and $D_s^+$ mesons and to charm baryons, their lifetimes are
considerably shorter and semileptonic branching ratios are smaller.
Nonetheless, these signatures can be used for charm physics, albeit
with more difficulty than in $b$-quark studies.

A cornerstone of $b$ physics in hadron collider experiments is the
signature provided by $B \rightarrow J/\psi X$ or $\psi^{\prime} X$
with a branching fraction of approximately 1\% followed by the decay
of the $\psi$ meson into $\mu^+\mu^-$ or $e^+e^-$.  While the product
branching fraction $B \ra \Jpsi \ra \mu^+\mu^-$ is only
$\simeq$$6\times10^{-4}$, these decays provide a distinctive signature
and are a rich source of information about the $b\ra c\bar{c}s$
transition as well as providing triggering and tagging for the study
of global properties of $b$ hadrons.

Due to huge event rates, effective triggers and quality detectors are
essential for extracting physics results.  Heavy flavor analyses
typically require detector strengths in three aspects of the
experiments: triggering, reconstruction, and flavor tagging.  Heavy
quarks are produced in hadronic colliders preferentially at small
polar angles $\theta$ (with respect to the beam axis) and at large
absolute values of pseudorapidity $\eta \equiv -\ln[\tan(\theta/2)]$.
Both experiments employed muons from $b \ra \mu$ and $b \ra c \ra \mu$
for triggering. With a muon acceptance\cite{Abazov2005372} in rapidity
$|\eta| < 2$ that is twice as large as CDF's, the \d0\ detector has a
distinct advantage in inclusive muon and dimuon triggering and
studies.  With less material before the first set of muon chambers,
the CDF detector allows the study of dimuons with lower momenta.
CDF's deadtimeless data acqusition system allowed for high-rate
triggers based only on tracking information in the first level of the
three-level system.  Events so selected could be analyzed by the
online silicon vertex tracker (SVT)\cite{Bardi:2001uv}, the first
trigger processor developed to measure track impact parameters.  The
impact parameter resolution of the SVT was similar to that for offline
reconstruction.  The SVT gave CDF access to a host of decay modes that
include only charged hadrons in the final state.
Electrons from semileptonic decays were included in the triggers of
both experiments, but played a significantly lesser role in the
physics program.

The upgraded CDF and \d0\  detectors\cite{Affolder:2003ep,Abazov:2005pn}
also bring different strengths to reconstruction.  With a much larger
radius tracker, the CDF detector has superior momentum resolution
leading to significantly better resolution on the reconstructed
invariant masses of particles.  This helps not only in the measurement
of the masses themselves, but also with background rejection.  The
\d0\ detector gains from the larger rapidity acceptance of the
tracker\cite{Abazov:2005pn} that extends out to $|\eta| < 3$.  The
impact parameter resolution of the CDF silicon
detector\cite{Aaltonen:2013uma} and the \d0\ silicon microvertex
tracker (SMT)\cite{Ahmed20118} (with a ``Layer
0"\cite{Angstadt2010298} added in 2006 partway through Run~2) are
similar, approximately 30\,$\mu$m for tracks with typical momenta from
$b$ hadron decay with an asymptotic resolution of 10--15\,$\mu$m for
high momentum single tracks.  CDF also includes hadron identification
with both specific ionization ($dE/dx$) measurement in the drift
chamber\cite{Affolder:2003ep} and a time-of-flight (TOF)
system\cite{Acosta:2004kc} comprising scintillator bars between the
drift chamber and the solenoid.  TOF measurements give unambiguous
kaon identification for $p_T<0.7$\,\GeVc, and at higher momenta the
combination of $dE/dx$ and TOF can be used on a statistical basis to
separate particle types in reconstructed decays.

Flavor tagging in the context of heavy flavor physics is the
determination of whether a particle with the potential of mixing or in
a decay that is a \CP\ eigenstate (e.g., $D^0\ra\pi^+\pi^-$) was
created as a particle or anti-particle.  In $b$ physics, there are two
types of flavor tags: away-side and same-side.  In an away-side tag,
the flavor of the other produced $b$ hadron is used to determine the
flavor of the one under study.  These can be in the form of either a
lepton presumably from a semileptonic decay or from the weighted sum
of the charges of particles with displaced impact parameters.  The
rapidity of a $b$ and $\bar{b}$ are only weakly correlated, so the
efficiency of making a tag is substantially better with \d0's larger
$\eta$ range.  However, these tags suffer from background and
resolution effects that make the probability of an incorrect tag large
as well.  Same-side tags seek to identify the last particle created in
the fragmentation process before the $b$ hadron emerges.  Thus a
$\pi^+$ would be associated with a $B^0$, a $K^+$ with a $B_s^0$, and
the negative tagging hadrons with a $\bar{B}^0$ or $\bar{B}^0_s$
meson.  Same-side tags have high efficiency because they are
associated with the $B$ already within the detector acceptance.  Given
the large population of background pions, the TOF system in CDF was
instrumental in employing the same-side kaon flavor tag in studies of
$B_s$ mixing.
For charm physics, the key tag is the exploitation of the sign of the
charge of the soft pion from $D^{*+}\ra D^0\pi^+$ decays to determine
the charm flavor.

For placing constraints on \CP\ violation, the Tevatron also has an
advantage of having a \CP-invariant initial state of the $p\bar{p}$
beams and hence almost perfect \CP-symmetric production in contrast to
the LHC where production in $pp$ collisions is not \CP\ symmetric.  In
addition, the \d0\ detector has solenoidal and toroidal magnetic
fields and consistently switched polarities of the magnets at regular
two-week intervals. As a result, differences in reconstruction
efficiency between positively and negatively charged particles cancel
to first order allowing high-precision measurements of charge
asymmetries in the study of \CP\ violation.

%% file: run1.tex
\section{Historical Review: Run 1 Results} 
\label{run1}

Early heavy flavor measurements at hadron colliders focused on
production, which was a natural question given the new energy regime.
Furthermore, it was widely believed that studies of bottom and charm
particle decays would not be possible as a result of the many
background particles present in hadron collisions.  Access to heavy
flavor events was limited as they could be found only in semileptonic
decays that led to single high-$p_T$ lepton triggers or in decays
with quarkonium that produced dimuons which in turn provided the
trigger signature.

The first high-energy collider was the $Sp\bar{p}S$ at CERN which
provided $p\bar{p}$ collisions at $\sqrt{s}=546$\,GeV where the UA1
collaboration pioneered the field using the spectra of transverse
momentum $p_T$ of leptons to deduce the spectrum of $b$-quark
production.\cite{Albajar:1986iu,Albajar:1988th,Albajar:1990zu} CDF
made similar measurements in Run 0 (1987--89) at $\sqrt{s}=1800$\,GeV
using inclusive electrons\cite{Abe:1993sj} and muons.\cite{Abe:1993hr}
After correcting for backgrounds, Monte Carlo simulations were used to
derive an effective quark momentum threshold.  Cross sections were not
truly differential, but were quoted as the momentum $p_{T,b}$ such
that 90\% of leptons is a sample with some $p_{T,\ell}$ threshold
resulted from quarks with momenta exceeding $p_{T,b}$.  Because
prompt production of \Jpsi\ and $\psi^\prime$ mesons was thought to be
minimal, these particles were also used to derive $b$ hadron cross
sections.\cite{Abe:1992ww} The apparent discrepancies between the CDF
and UA1 results and between the lepton and charmonium results were
strong motivation for the production studies that would come in Run 1.
In charm production, early measurements from both
UA1\cite{Arnison:1984rj} and CDF\cite{Abe:1989tc} were limited to the
fraction of jets containing $D^{*+}$ mesons.

CDF pioneered the full reconstruction of $B$ mesons with a
measurement\cite{Abe:1992fc} of the $\Bu$ meson cross section using
the decay $\Bu\ra\Jpsi\,K^+$.  The measurement\cite{Abe:1992cc} of
the average $B\bar{B}$ mixing fraction in dilepton events served as
a determination of fraction of \Bs\ mesons in $b$-hadron
production based on the $\Bd\bar{B}^0$ mixing fraction observed in
$e^+e^-$ colliders at the $\Upsilon(4S)$ and the expectation of
complete mixing for \Bs\ mesons.

Run 1 (1992--95) led to a revolution in thinking about heavy-quark
physics at hadron colliders.  While studies of heavy-quark production
continued, CDF's installation of the SVX,\cite{Haber:1990ip} the first
silicon microvertex detector at a hadron collider, provided the
opportunity not only to measure $b$ hadron lifetimes\cite{Abe:1993ks,
Abe:1994ve, Abe:1995ka, Abe:1996es, Abe:1996df, Abe:1996ir,
Abe:1997bd, Abe:1998wt, Abe:1998cj,Acosta:2002nd} but also a
substantial reduction in backgrounds that was critical to measurements
of $b$ hadron masses.\cite{Abe:1995jp,Abe:1996tr} The riddle of large
\Jpsi\ and $\psi^\prime$ was at once solved and renewed with the
measurement\cite{Abe:1997jz} of differential production cross sections
where the SVX could be used to separate prompt charmonium from that
arising from $b$-hadron decays and showed the prompt fraction to be
much larger than had been anticipated.  Further work toward
understanding quarkonium production included measurements of the
$\Upsilon$ production
spectrum;\cite{Abe:1995an,Affolder:1999wm,Acosta:2001gv} polarization
in \Jpsi, $\psi^\prime$, and $\Upsilon$
decays;\cite{Affolder:2000nn,Acosta:2001gv} and the fractions of
\Jpsi\ and $\Upsilon$ that result from $\chi$
production.\cite{Abe:1997yz,Affolder:1999wm,Affolder:2001ij}
Measurements of $b$ hadron cross sections were
refined,\cite{Abe:1994qk, Abe:1995dv,Acosta:2001rz} and
correlations\cite{Acosta:2002qk} between leptons and displaced tracks
at both $\sqrt{s}=1800$ and 630\,GeV showed that theory properly
described the energy scaling and that that the apparent discrepancies
between CDF and UA1 had to come from another source.  The production
fractions of $b$ hadron species were also
studied.\cite{Affolder:1999iq, Abe:1999ta}

The low backgrounds afforded by the SVX also brought 
limits\cite{Abe:1996et, Abe:1998bc, Abe:1998ah, Acosta:2001gy, Acosta:2002fh} on rare decays and 
measurements\cite{Abe:1995aw, Abe:1996kc, Abe:1996yya, Affolder:2000ec, Affolder:2001qi,
Acosta:2002pw, Abe:1998yu, Abe:1995yd} of $B$ meson decay properties 
such as the amplitudes in $\Bu\ra\Jpsi
K^{(*)}$.  A highlight of Run 1 is the
discovery\cite{Abe:1998wi,Abe:1998fb} of the $B_c$ meson with the
measurement of its lifetime showing that the decay of the charm quark
dominated the decay width.  Based on the development of tagging
techniques used in $B\bar{B}$ mixing measurements,\cite{Abe:1998sq,
Abe:1998qj, Abe:1999ds, Affolder:1999cn, Abe:1999pv} CDF made and the
first non-trivial limit on the CKM angle $\beta$ from the
measurement\cite{Abe:1998qja,Affolder:1999gg} of $\sin 2\beta$ in
$\Bd\ra\Jpsi K_S^0$ decays which set the stage for the mixing and
\CP-violation measurements that would come in Run 2.

The \d0\ detector did not participate in the first runs of the
Tevatron (Run 0), but did join CDF in Run 1 operations beginning in
1992.  The Run~1 \d0\ detector was optimized to meet the goals of
excellent identification of electrons and muons, good measurement of
parton jets at large $p_T$ through a highly segmented calorimeter with
very good energy resolution, and a well-controlled measure of missing
transverse energy. The central design features\cite{D0Run1} thus
incorporated a compact, non-magnetic inner tracking volume with
reasonable spatial resolution and particular emphasis on suppression
of backgrounds to electrons. Via a toroidal magnetic field, the
transverse momenta of muons could be determined, so invariant masses
of only dimuon pairs could be found. The Run 1 \d0\ detector included
a drift chamber vertex detector, so was able to reconstruct the impact
parameters of tracks, but could not reconstruct invariant masses of
long-lived heavy-flavor particles creating secondary vertices.  $b$
hadrons were identified via their semileptonic decays to muons, with
the $p_T$ of the muon relative to the jet axis used to separate charm
semileptonic decays. As a result, the \d0\ heavy-flavor program in
Run~1 was limited to $b$-quark and $b$-jet inclusive and differential
cross section
measurements\cite{Abachi:1994kj,Abbott:1999wu,Abbott:1999se,Abbott:2000iv},
$J/\psi$ cross section measurements\cite{Abachi:1996jq,Abbott:1998sb},
and a search for rare $b$ decay\cite{Abbott:1998hc}.

To prepare for a vigorous heavy-flavor program for Run 2, the \d0\
detector upgrade\cite{Abazov:2005pn} maintained its excellent muon
coverage and calorimetry, but now included a central magnetized
tracker comprised of a superconducting solenoid surrounding a
scintillating fiber tracker over a wide range of detector $\eta$ plus
a silicon microvertex tracker (SMT)\cite{Ahmed20118,Angstadt2010298}
which provided the benefits described above.

%% file: production.tex
\section{Production}

\subsection{Inclusive \texorpdfstring{$b$ and $b\bar{b}$}{b and bb}}
\label{sec:inclb}
     
In hadron collisions, most production happens as $b\bar{b}$ pairs,
either via $s$-channel production or gluon splitting, with
a smaller fraction of $b$ quarks produced by flavor
excitation.\cite{Agashe:2014kda} The total $b$ production cross section is an interesting
test of our understanding of QCD processes.  As described in
Section~\ref{run1}, past measurements of inclusive $b$ quark
production in the central rapidity region at Run~1 indicated a general
agreement in shape with the calculated transverse momentum ($p_T$)
spectrum, but were systematically higher than the NLO QCD predictions
at the time by up to a factor of 2.5. With improved measurements, more
accurate input parameters, and more advanced calculations, the
discrepancy between theory and data is now much reduced.

Previous studies of $b$-quark production exploited the kinematic
relationship between $b$ quarks and daughter (semileptonic decay)
muons and electrons to extract integrated $b$ quark production rates. To avoid
fragmentation and unfolding uncertainties, CDF chose to measure
differential spectra for final-state bottom hadrons, while \d0\  references
$b$ jets rather than $b$ quarks in Run~2, where $b$ jets are defined
as hadronic jets carrying $b$ flavor. As opposed to quarks, jets or hadrons are
directly observable and therefore reduce model dependence when
comparing experimental data with theory and are in direct
correspondence with a NLO QCD calculations. For instance, large
logarithms that appear at all orders in the open quark calculation
(due to hard collinear gluons) are avoided when all fragmentation
modes are integrated.

\begin{figure}[tb]
\centerline{\includegraphics[width=0.90\linewidth]{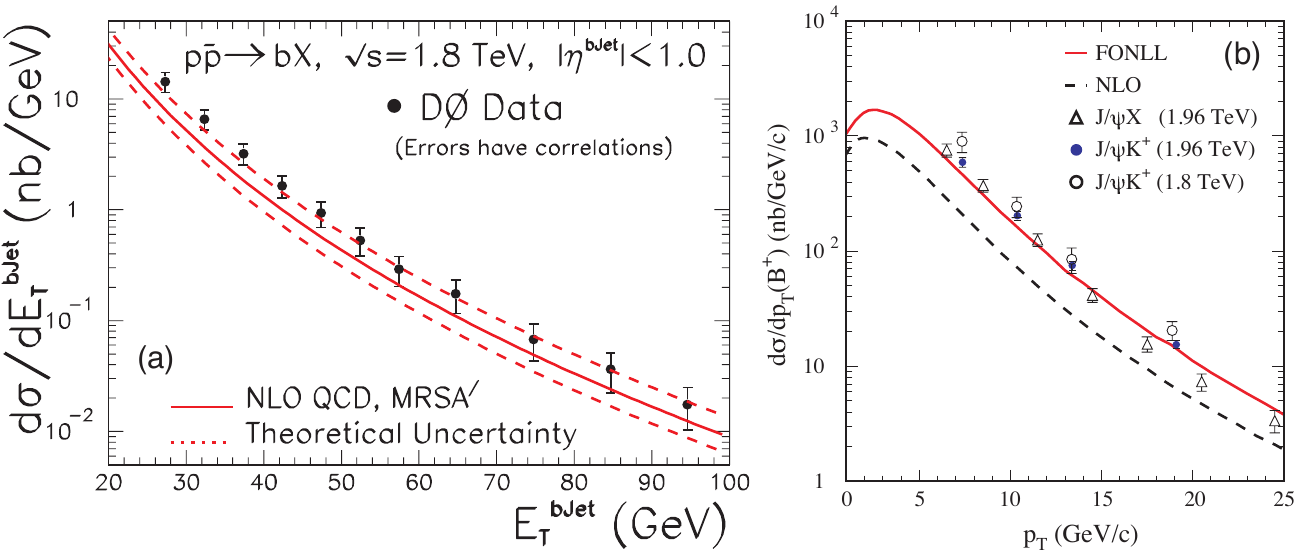}}
\hfill
\caption[]{(a) \d0\ measurement of the $b$-jet cross section; (b) CDF
  measurement of the \Bu cross section as well as the cross section of
  $b$ hadrons inferred from displaced \Jpsi mesons.  For the later
  measurement and the theoretical predictions, the fragmentation
  fraction $f_u=0.389$ is applied.}
\label{fig:d0_bjets}
\end{figure}

The \d0\ Collaboration measured\cite{Abbott:2000iv} the inclusive
$b$-jet cross section by tagging the jets using $b \ra \mu$
semileptonic decay and the relatively high $p_T$ of muons with respect
to the jet axis to distinguish jets from charm and light flavor
jets. As shown in Fig.~\ref{fig:d0_bjets}, within experimental and
theoretical uncertainties, \d0\ results are found to be higher than,
but compatible with, next-to-leading-order QCD predictions. This
result supersedes older Run~1
results\cite{Abbott:1999wu,Abachi:1994kj} for $b$-quark
production (rather than jets).


CDF measured\cite{Abulencia:2006ps} the differential \Bu\ meson
production spectrum using the decay mode $\Bu\rightarrow J/\psi\,K^+$
for $|y|<1$.  The total cross section was found to be $2.78\pm
0.24\,\mu $b for $p_T>6\,\GeVc $.  Comparison to theory is more robust
than in the case of the derived quark cross sections because the hard
production and fragmentation can be calculated in a consistent
framework.  CDF also measured\cite{Acosta:2004yw} the differential
cross section of \Jpsi mesons produced in $b$ hadron decays.  The
fraction of \Jpsi mesons from $b$ hadron decays is determined by
considering the displacement of the \Jpsi decay points.  Prompt mesons
have a decay point consistent with the beamline, while those from the
$b$ decays will be displaced as a result of the long $b$ lifetime.
While event-by-event identification is not possible, the distributions
are sufficiently distinct that prompt fractions can be measured in
each momentum bin.  Monte Carlo simulations are used to correct for
the fraction of the parent $b$ hadron momentum carried by the \Jpsi
mesons to yield a differential cross section for bottom hadrons $H_b$.
The results of both the \Bu and \Jpsi measurements are shown in
Fig.~\ref{fig:d0_bjets}(b) along with results from Run~1.

CDF also measured the differential $H_b$ cross
section\cite{Aaltonen:2009xn} using correlations of muons with $D^0$
or $D^{*+}$ mesons.  The $D^{(*)}\mu$ spectra are corrected in an
analogous way to the displaced \Jpsi spectra.  
%
%
The \Jpsi, $\mu D^{(*)}$, and \Bu results are all in good agreement with
each other and with the fixed-order leading log calculation of
Refs.~\citen{Cacciari:1998it} and \citen{Cacciari:2003uh}.

Measurements of both $b$ quarks or $b$ jets in the same event and
their correlations have also been made at the Tevatron.  Using samples
of single muon and dimuon events from $b \ra \mu$, the \d0\
Collaboration made an additional measurement of the $b$-quark cross
section in a given kinematic range. As for previous $b$-quark
measurements, the results agree in shape with the next-to-leading
order QCD calculation of heavy flavor production but are greater than
the central values of these predictions. The angular correlations
between $b$ and $\bar{b}$ quarks, measured from the azimuthal opening
angle between their decay muons, agree in shape with the
next-to-leading order QCD prediction.

CDF used muon pairs to measure\cite{Aaltonen:2007zza} correlation in
the production of $b$ and $\bar{b}$ quarks.  The analysis considers
the impact parameter distribution of muons with tight particle
identification cuts to separate the component arising from $b\bar{b}$
from backgrounds including $c\bar{c}$ production and prompt-hadron
fakes.  For muons with $p_T \geq 3\,\GeVc$ that are produced by quarks
with $p_T \geq 2\,\GeVc$ and $|y|<1.3$, the measured cross section is
$\sigma_{b\rightarrow\mu,\bar{b}\rightarrow\bar{\mu}}=1549\pm133$\,pb.
This result is in general agreement with theoretical predictions and
previous measurements.

The more precise cross section measurements of Run~2 demonstrated
that the discrepancies between the Run~1 measurements and NLO
predictions were significant and motivate the theoretical advances that
can accurately model the production spectra.

\subsection{Charm}

\begin{figure}[tb]
\centerline{\includegraphics[width=0.90\linewidth]{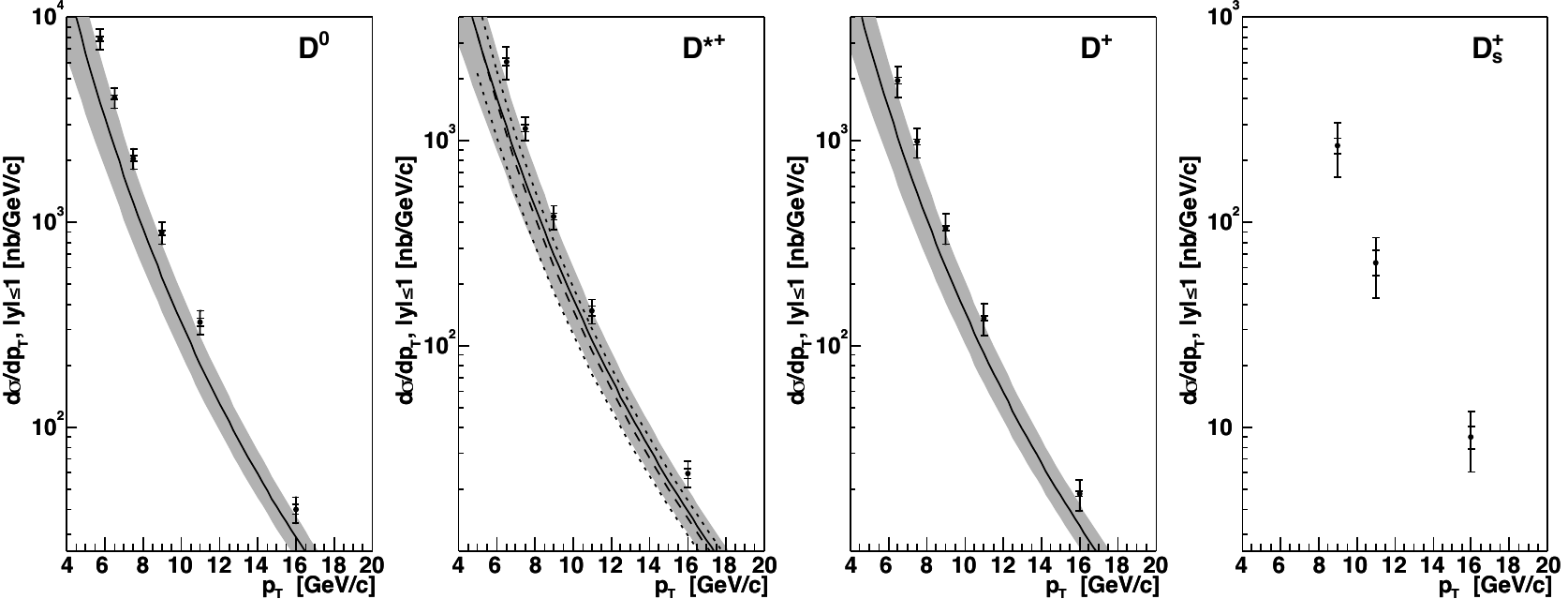}}
\hfill
\caption[]{CDF measurements of the differential cross sections of 
$D^0, D^{*+}, D^+,$ and $D_s^+$ mesons}
\label{fig:cdf_charm}
\end{figure}

The differential charm meson cross
section\cite{Acosta:2003ax} was one of CDF's first
measurements in Run~2.  The measurement is made using only 6\,\ipb of
data collected with a displaced-track trigger.  In the small dataset, the acceptance could be tightly
controlled and therefore well modeled.  There were no significant
changes to the active channels in the detector over the course of the
data used in the measurement.  Even with the small dataset, the
dominant uncertainties are systematic.  As in the case of $J/\psi$
cross section, secondary decays are accounted for on the basis of the
impact parameter distribution.  
%
%
The differential cross sections are shown in Fig.~\ref{fig:cdf_charm}
compared to the theory of Ref.~\citen{Cacciari:2003zu}.

   \subsection{Quarkonia}
      
The study of heavy quarkonium production provides important
information for perturbative and nonperturbative QCD, particularly
factorization methods since heavy quark masses are larger than
$\Lambda_{\mathrm{QCD}}$, the typical scale where nonperturbative
effects become significant.  The nonperturbative evolution of the
$Q\bar{Q}$ heavy-quark pair into a quarkonium has been discussed
extensively in terms of models such as the color-singlet model (CSM),
the color-evaporation model (CEM), the nonrelativistic QCD (NRQCD)
factorization approach, and the fragmentation-function approach (see
review of Ref.~\citen{Brambilla:2010cs}).


In high-energy $p\bar{p}$ collisions, \Jpsi\ mesons can be produced in
three ways: direct production, from the prompt decays of heavier
charmonium states such as $\chi_{cJ}$ via $\chi_{cJ} \ra \Jpsi\,
\gamma$, or from the decays of $b$ hadrons, {\it i.e.,} $B \ra \Jpsi$.

\begin{figure}[tb]
\centerline{\includegraphics[width=0.9\linewidth]{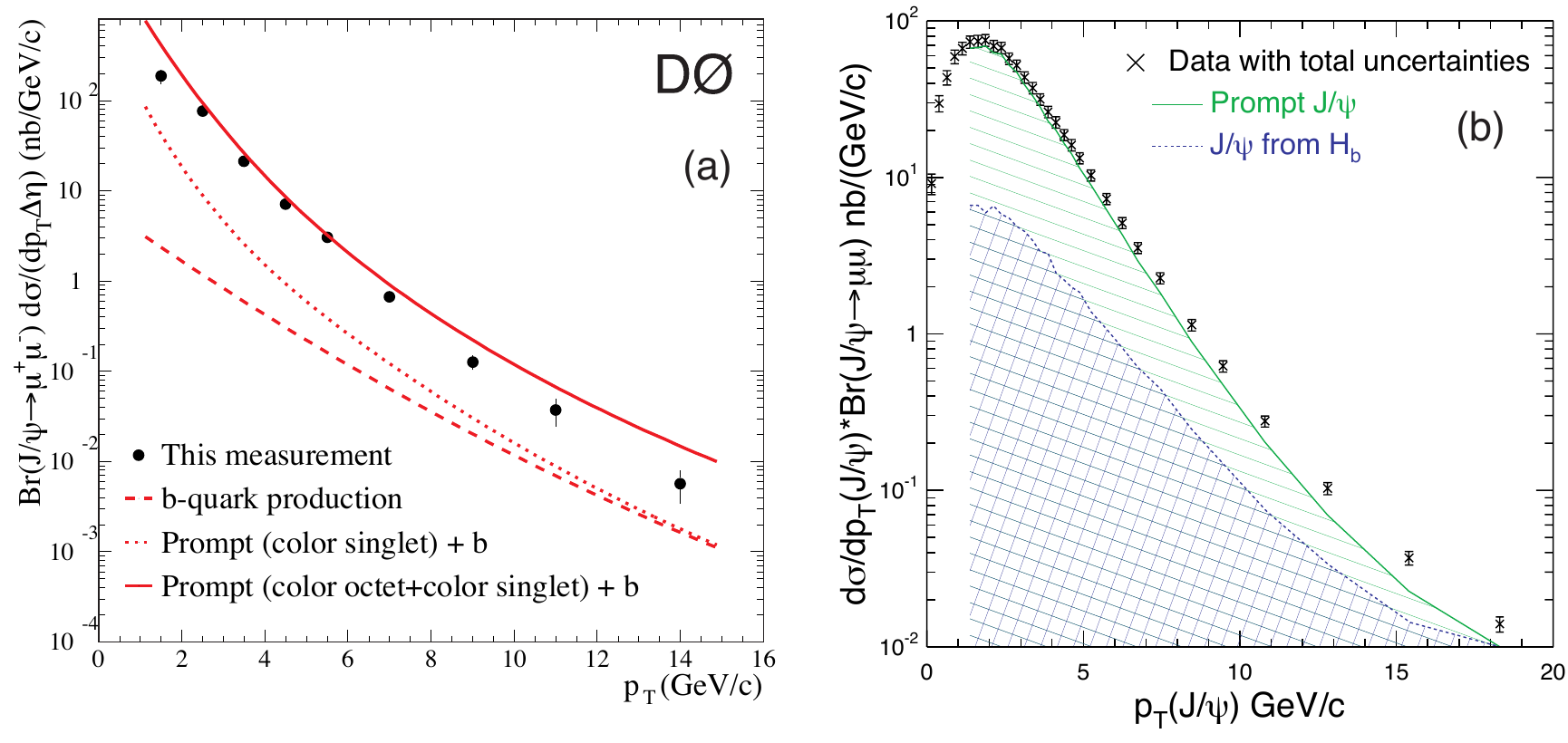}}
\hfill
\caption[]{(a) $p_T$ dependence of the \Jpsi\ differential cross
section at \d0\ and its theoretical predictions with only
statistical uncertainties. The correlated uncertainty across all
points is approximately 20\%, including varying the \Jpsi\
polarization between 0 and 100\%. (b) CDF measurement of the inclusive \Jpsi\ cross section as a function of \Jpsi\ $p_T$ integrated over the rapidity range
$|y| < 0.6$.}
\label{fig:d0_Jpsixsec}
\end{figure}

A measurement\cite{Abazov:2014qba} by the \d0\ Collaboration of
double \Jpsi\ production also included a measurement of the single \Jpsi\
cross section. The decay length from the primary \ppbar interaction
vertex to the \Jpsi\ production vertex determined via $\Jpsi \ra \mu^+
\mu^-$ was used to distinguish prompt from non-prompt \Jpsi mesons to
measure a cross section consistent with value calculated in the
$k_T$-factorization approach\cite{Baranov:2012re} that includes direct
production and that via $\chi_{cJ}$.  Two earlier measurements by the
\d0\ Collaboration using muon impact parameters to separate prompt
from non-prompt production in the central region\cite{Abachi:1996jq}
and at far forward angles\cite{Abbott:1998sb} of $2.5 \leq
|\eta^{\Jpsi}| \leq 3.7$ showed rough agreement with the CEM in both
$p_T$ and $\eta$ of the \Jpsi\ and ruled out the CSM (see
Fig.~\ref{fig:d0_Jpsixsec}(a).  The former analysis measured a fraction
of \Jpsi\ mesons from $\chi_{cJ}$ decays to be significantly less than
that predicted by direct charmonium production and gluon
fragmentation.

 CDF's measurement\cite{Acosta:2004yw} of the inclusive \Jpsi spectrum
 has been described above in Sect.~\ref{sec:inclb} with results shown
 in Fig.~\ref{fig:d0_Jpsixsec}(b). While the
 inclusive spectrum is measured down to zero transverse momentum, the
 prompt component can be separated only for $p_T>1.25\,\GeVc$.  Because
 there are no $c\bar{c}$ states between the $\psi^\prime$ and the
 open-charm threshold, the $\psi^\prime$ cross section provides a clean test
 of production models compared to the \Jpsi that is polluted by
 $\chi_c$ feed down.  CDF measured the $\psi^\prime$
 spectrum\cite{Aaltonen:2009dm} using dimuon events.  The method for
 separating prompt and secondary components is the same as in the
 \Jpsi measurement. The results are consistent with gluon-tower
 models\cite{Khoze:2004eu} which have large uncertainties.  However, like the
 \Jpsi cross section, $\psi^\prime$ production is poorly fit by NNLO QCD
 descriptions.\cite{Lansberg:2008gk}


Bottomonium states are produced either promptly or indirectly as a
result of the decay of a higher mass state, {\it e.g.,} in a radiative decay
such as $\chi_b \ra \Upsone \gamma$, but with the advantage over
charmonium of not being produced via the decays of other heavy flavor
states leading to a simpler analysis as all states are prompt.

By reconstructing the \Upsone\ through its decay $\Upsone \ra \mu^+
\mu^-$, the \d0\ Collaboration has determined\cite{Abazov:2005yc}
production cross section of the \Upsone\ as a function of its
transverse momentum, in three rapidity ranges.  These are reasonably
consistent within theoretical and experimental uncertainties with
theoretical models of the time\cite{Berger:2004cc,Berger:2004ct},
which are similar to the CEM, as are the ratios of the cross sections
in different rapidity ranges. These results provide greater precision
than the CDF Run~1 result.\cite{Acosta:2001gv} CDF's other important
result in the study of bottomonium from Run~1 is the
fraction\cite{Affolder:1999wm} of \Upsone mesons that are produced
from $\chi_b$ decays which is found to be $(49.1\pm 8.2\, ({\mathrm{stat}}) \pm
9.0 \, ({\mathrm{syst}}))\%$ for $p_T>8\,\GeVc$.

   \subsection{\texorpdfstring{$\Upsilon$}{Upsilon} Polarization}
      
Theoretical models that were constructed to accommodate the
surprisingly large production cross section of \Jpsi\ and $\Upsilon$
mesons beyond the initial CSM also make specific predictions about
their production polarization but were generally in poor agreement
with initial experimental measurements. The angular distribution of
muons from $\Upsilon \ra \mu^+\mu^-$ decays are described by the
distribution:
\begin{equation}
\frac{d\Gamma}{d\Omega}  \propto 1 + \lambda_{\theta}\cos^2\theta + 
\lambda_{\varphi}\sin^2\theta \cos 2\varphi + \lambda_{\theta \varphi} 
\sin 2\theta \cos\phi
\label{Upspoleqn}
\end{equation}
in the $\Upsilon$ rest frame where the angles refer to the positive
lepton with respect to the direction of the $\Upsilon$.  A convenient
measure of the polarization is the variable $\alpha \equiv
\lambda_{\theta} = (\sigma_T - 2\sigma_L)/(\sigma_T + 2\sigma_L)$
where $\sigma_T$ and $\sigma_L$ are the transversely and
longitudinally polarized components of the production cross section,
respectively.

\begin{figure}[tb]
\centerline{\includegraphics[width=1.0\linewidth]{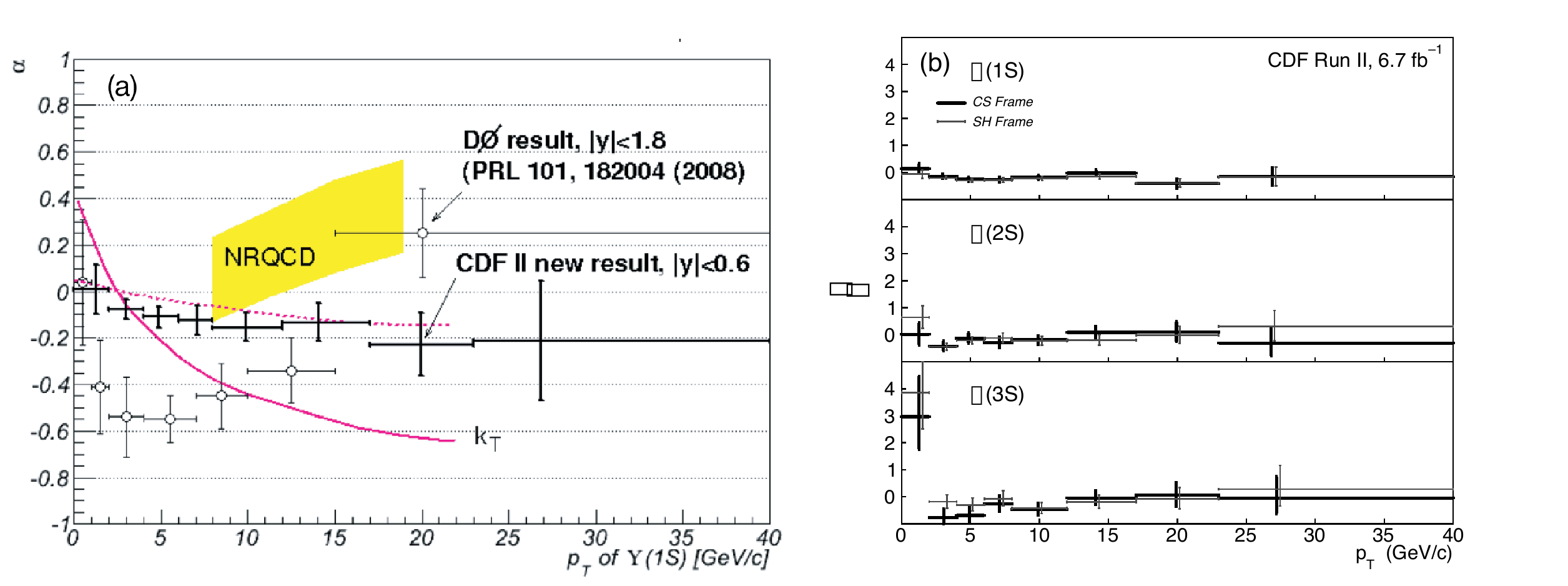}}
\hfill
\caption[]{(a) CDF and \d0\ results on $\Upsilon$ polarization compared to
theoretical predictions for \Upsone. The yellow
bands indicates the range of $\alpha ( \equiv \lambda_{\theta}$) predicted
by NRQCD\cite{PhysRevD.63.071501}. The two magenta curves show two
extreme cases of the $k_T$ factorization
model\cite{Baranov:2007ay}. The upper (dashed) curve assumes complete
de-polarization of the \Upsone\ when produced in the decay of $\chi_b$
states, while the lower (solid) curve assumes that the polarization is
preserved. (b) CDF results on the frame-invariant quantity $\tilde\lambda$ in 
the Collins-Soper frame and $s$-channel helicity frame.}
\label{fig:Upspol}
\end{figure}

The \d0\ Collaboration measured\cite{Abazov:2008aa} the
$\lambda_\theta$ polarization variable for the \Upsone\ meson using
decays to $\mu^+\mu^-$ as a function of $p_T(\Upsilon)$ as shown in
Fig.~\ref{fig:Upspol}(a), indicating strong longitudinal polarization
for lower values of $p_T$.  Discrepancies between results for
$\lambda_\theta$ obtained by different experiments suggest that
quarkonia might be strongly polarized when produced, but that
different experimental acceptances can impact the final
measurement.  Early analyses measured only $\lambda_{\theta}$ as a
function of $p_T(\Upsilon)$ in one reference frame; however,
 polarization could be manifested by significantly non-zero
values of $\lambda_{\varphi}$ or $\lambda_{\theta \varphi}$ even when
$\lambda_{\theta}$ near zero\cite{PhysRevD.81.111502}.

While the observed lack of transverse polarization at high momentum in
the helicity basis is inconsistent with NRQCD-inspired models, it is
not a definitive demonstration of a lack of polarization in $\Upsilon$
production.  To demonstrate that the production is truly unpolarized,
CDF performed a full three-dimensional decomposition to
measure\cite{CDF:2011ag} the three components of polarization
$\lambda_\theta$, $\lambda_\varphi$, and $\lambda_{\theta\varphi}$ in
both the helicity and Collins-Soper frames for $p_T<40\,\GeVc$ for
\Upsone, \Upstwo, and \Upsthree decays to muon pairs.  From the
measured components, one can form the frame-invariant quantity
$\tilde{\lambda}=(\lambda_\theta+3\lambda_\varphi)/(1-\lambda_\varphi)$.
The results for $\tilde\lambda$ are shown in Fig.~\ref{fig:Upspol}(b).
The agreement between measurement in the two frames demonstrates a
lack of systematic bias. The values near zero indicate that indeed the
production of \Upsone, \Upstwo, and \Upsthree mesons is
unpolarized.

   \subsection{\texorpdfstring{$b$}{b} Fragmentation fractions}

$b$ quarks produced in \ppbar\ collisions are accompanied by
$q\bar{q}$ pairs created in the color field in the process of
fragmentation where anti-quarks combine with the $b$ quark to form a
$B$ meson $|b\bar{q}\rangle$ or with di-quarks to form a $b$ baryon
$|b q_1 q_2 \rangle$.  In contrast with the $B$ factories operating at
the \Ups\ peak where only $B^0$ and $B^{\pm}$ are produced, in
high-energy collisions, all species of weakly decaying $b$ hadrons may
be produced, either directly or in strong and electromagnetic decays
of excited $b$ hadrons. The probabilities that the fragmentation of a
$b$ quark will result in a $B^+$ $|\bar{b}u \rangle$, $B^0$ $|\bar{b}d
\rangle$, $B^0_s$ $|\bar{b}s \rangle$, or $\Lambda_b$ $|b u d \rangle$
are denoted as \fBu, \fBd, \fBs, and \fLb, respectively.

The \d0\ Collaboration has measured\cite{Abazov:2011wt} the product 
$\fLb \cdot \BR{\Lb \ra \Jpsi\, \Lambda} = (6.01 \pm 0.88) \times 10^{-5}$.

CDF has measured\cite{Aaltonen:2008zd}
relative production fractions using the yields of $\ell D^0, \ell D^+,
\ell D^{*+}, \ell D_s^+$, and $\ell \Lambda_b^0$.
After
correcting for branching ratios updated\cite{PDG2012} since the
publication of the paper, the relative fractions are
\begin{align*}
{f_u\over{f_d}}=& 1.054\pm0.018 \, ({\mathrm{stat}})^{+0.025}_{-0.045} \, ({\mathrm{syst}})\pm0.058 \, (Br), \\
{f_s\over{f_u + f_d}}=& 0.128\pm0.005^{+0.009}_{-0.008}\pm0.011(Br),
  {\rm and} \\
{f_{\Lambda_b}\over{f_u + f_d}}=&0.281\pm0.012 \, ({\mathrm{stat}})^{+0.058}_{-0.056}\, ({\mathrm{syst}})^{+0.128}_{-0.086} \, (Br).
\end{align*}
The first result is consistent with expectations from isospin, and the
second agrees with results from LHCb\cite{Aaij:2011jp,Aaij:2013qqa}.
The baryon fraction is also in general agreement with LHCb; however,
because it depends on momentum, a more detailed evaluation is required.
     
The Heavy Flavor Averaging group has used these as inputs to a global
fit giving average and independent fragmentation fractions 
both for \ppbar\ collisions at the Tevatron and for high energies
({\it i.e.,} LEP, Tevatron, LHC).\cite{Amhis:2012bh}


%% file: spectroscopy.tex
\section{Spectroscopy}

   \subsection{\texorpdfstring{$B$}{B} mesons}

\begin{figure}[tb]
\centerline{
\includegraphics[width=0.60\linewidth]{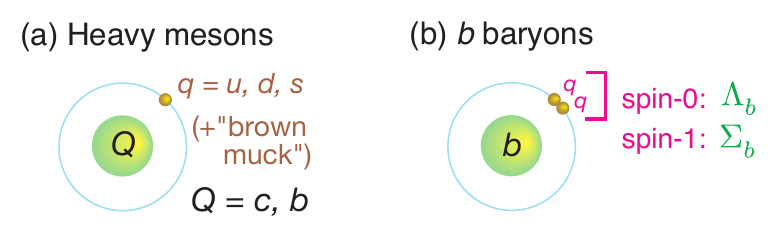}
}
\caption{(a) Atomic analogy of HQET for heavy mesons; (b) for $bud, buu, bdd$ baryons.}
\label{fig:HQET_cartoon}
\end{figure}

Heavy flavor spectroscopy provides the opportunity to test the theory of QCD bound 
states in the simplest system. 
Thus, heavy-quark
hadrons can be considered the hydrogen atom of QCD, and $b$ hadrons
offer the heaviest quarks in bound systems.  In the framework of Heavy
Quark Effective Theory (HQET)\cite{HQET}, a $b$ hadron can be roughly described
by the heavier $b$ quark being analogous to the nucleus of an atom
with lighter $u$, $d$, or $s$ quarks orbiting the nucleus similar to
the electrons of an atom, but surrounded by a complicated, strongly
interacting cloud of light quarks, antiquarks, and gluons sometimes
referred to as ``brown muck''\cite{Isgur:1991wq} as shown in
Fig.~\ref{fig:HQET_cartoon}(a). Studies of these states provide very sensitive
tests of potential models, HQET, and many regimes of QCD in general,
including lattice gauge calculations and QCD strings.

%
%

The Tevatron has the capability of producing heavier states not
accessible at the $B$ factories running at the $\Upsilon(4S)$:
\begin{itemize}
\item bottom-strange mesons: $B^0_s$ ($\bar{b}s$, the ground state with the
spins of the quarks anti-aligned) and $B^*_s$ ($\bar{b} s$, with the
spins of the quarks aligned);
\item bottom-charm mesons $B_c$ ($\bar{b} c$, the ground state);
\item excited mesons $B^{**}$ / $B_s^{**}$ ($\bar{b} u$, $\bar{b} d$,
and $\bar{b}s$ with the quarks having relative orbital angular
momentum); and
\item the $b$ baryons $\Lambda_b^0$ ($bud$), $\Sigma_b^{(*)\pm}$
  ($buu$ and $bdd$), $\Xi_b^-$ and $\Xi_b^0$ ($bsd$ and $bsu$), and
  $\Omega_b^-$ ($bss$).
\end{itemize}

\begin{figure}[tb]
\centerline{\includegraphics[width=0.8\linewidth]{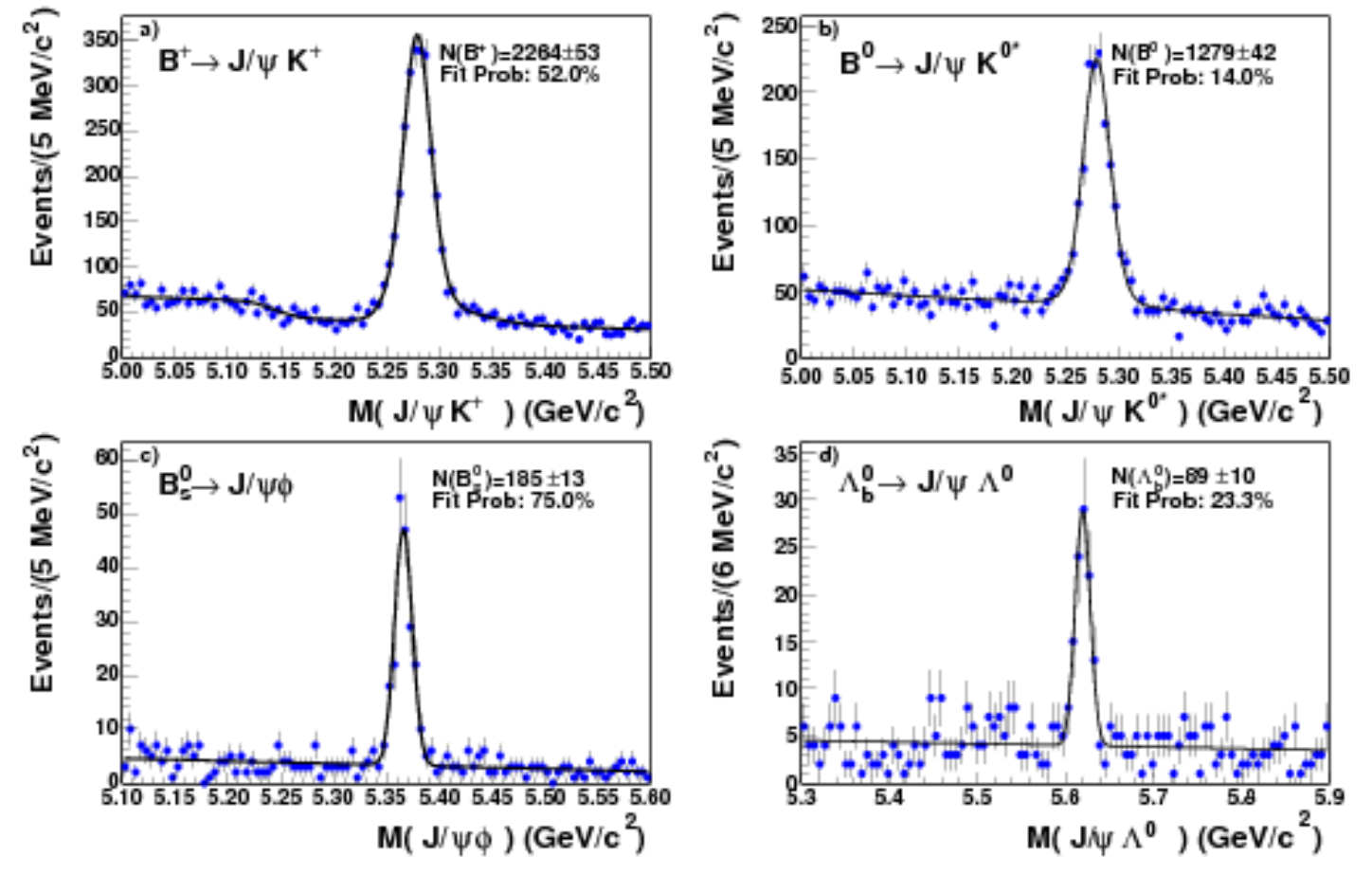}}
\caption{The invariant mass distributions for $B^+\rightarrow\Jpsi\,K^+,
 B^0\rightarrow\Jpsi\,K^{*0}, \Bs\rightarrow\Jpsi\,\phi$ and 
$\Lb\rightarrow\Jpsi\,\Lambda^0$ candidates. The results of log-likelihood
fits are superimposed. The fit probabilities obtained from a $\chi^2$ test are
shown.}
\label{fig:CDF_mass}
\end{figure}

Using $b$ hadron decay events with a \Jpsi in the final state, CDF made
what was at the time of publication the world's best
measurement\cite{Acosta:2005mq} of the masses of the \Bu, \Bd, and \Bs
mesons, as well as of the \Lb baryon.  Fig.~\ref{fig:CDF_mass} shows the
mass distributions for the various decay modes used in the analysis.
The key to the measurement is a precise calibration of the momentum
measurement for charged particles which is achieved using samples of
$\Jpsi\rightarrow\mu^+\mu^-$ and
$\psi^\prime\rightarrow\Jpsi\,\pi^+\pi^-$ decays.  In the study of
spectroscopy, often it is the difference between particle masses that
is most significant.  That also has the advantage experimentally as
many systematic uncertainties cancel.  In the case of these
measurements, the mass differences involving \Bs and \Lb were more
precise than the existing world averages.  The measured masses and
differences are:
\begin{align*}
m(B^+) =& 5279.10 \pm 0.41 \, ({\mathrm{stat}}) \pm 0.36 \, ({\mathrm{syst}}) \, \MeVcc, \\
m(B^0) =& 5279.63 \pm 0.53  \pm 0.33\, \MeVcc, \\
m(B^0_s) =& 5366.01 \pm 0.73  \pm 0.33\, \MeVcc, \\
m(B^+) - m(B^0) =& -0.53 \pm 0.67 \pm 0.14 \, \MeVcc, \\
m(B^0_s) - m(B^0) =& 86.38 \pm 0.90 \pm 0.06 \, \MeVcc.\\
\end{align*}



\Bc\ mesons are predicted by the quark model to be members of the $J^P
= 0^−$ pseudo-scalar ground-state multiplet and to have
zero isospin as the lowest-lying bound state of a $\bar{b}$ anti-quark
and a $c$ quark. This meson is of special interest because of its
unique status as a short-lifetime bound state of heavy but
different-flavor quarks. Measurements of its mass, production,
lifetime (see Sect.~\ref{Bmesonlife}), and decay (see
Sect.~\ref{subsec:modesbr}) therefore allow for tests of theories under
new approximation regimes or extended validity ranges beyond quarkonia
which is formed from bound states of same-flavor quarks.

%
%
\begin{figure}[tb]
\centerline{
\includegraphics[width=0.95\linewidth]{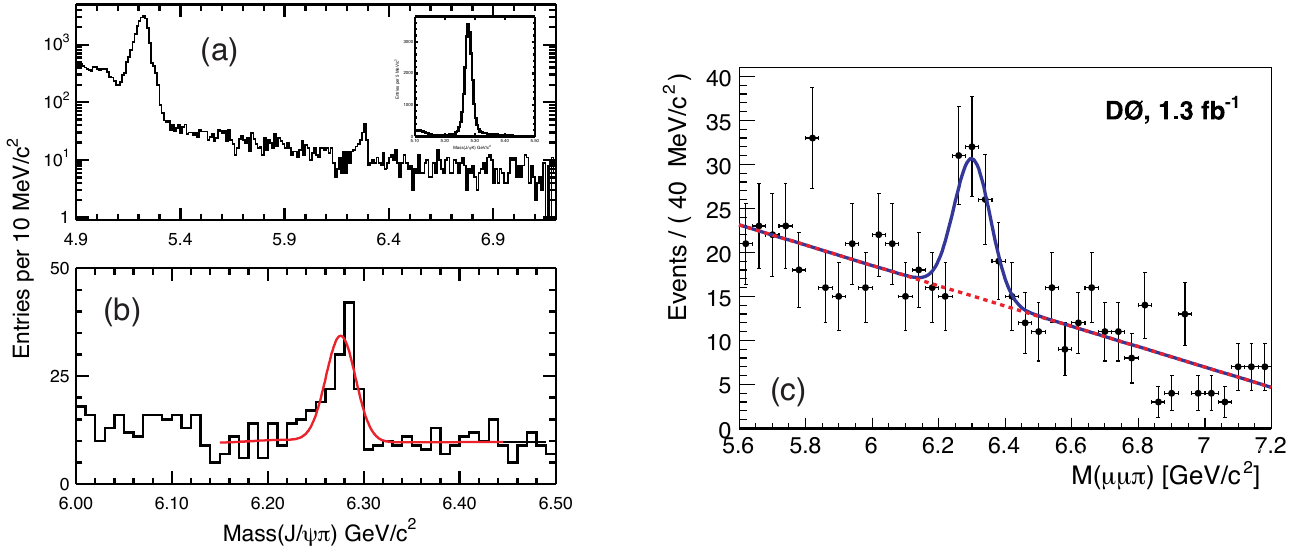}
}
\caption{(a) (inset) In CDF, the $\Jpsi\,K^+$ mass distribution has an excellent
  signal-to-background ratio as a result of the optimized
  cuts. The $\Jpsi\,\pi^+$ mass distribution shows a clear excess
  for the \Bc as well as the \Bu reflection;  (b) the \Bc mass signal peak as
  determined in a binned likelihood fit; (c) \d0\ \Bc signal peak in the same channel.}
\label{fig:bc}
\end{figure}

The CDF Collaboration made the first
observation\cite{Abulencia:2005usa,Aaltonen:2007gv} of the \Bc\ meson
in the fully reconstructed mode $\Bc\ra\Jpsi\,\pi^+$.  Cuts on
lifetime, momenta, resolutions, and other quantities were optimized in
the topologically similar $\Bu\ra\Jpsi\,K^+$ sample. 
Fig.~\ref{fig:bc}(a) shows the effect of the tight cuts on the
background to the \Bu signal (inset) as well as the relative size of the \Bu
and \Bc signals.  The \Bc mass is determined in a binned
log-likelihood fit and found to be $6275.6\pm 2.9 \, ({\mathrm{stat}}) \pm 2.5
({\mathrm{syst}})\, \MeVcc$.  The signal is observed with $8\sigma$ significance.   

Using lifetime cuts and requiring a pion at large transverse momentum
with respect to an identified $\Jpsi \ra \mu^+\mu^-$, the \d0\
Collaboration established a signal\cite{Abazov:2008kv} shown in
Fig.~\ref{fig:bc}(b) for $\Bc \ra
\Jpsi\,\pi^+$. with greater than $5\sigma$ significance 
and measured its mass to be $6300 \pm 14 \thinspace {\mathrm{(stat)}} \pm 5
\thinspace {\mathrm{(syst)}} \thinspace \MeVcc$.

\begin{figure}[tb]
\centerline{\includegraphics[width=0.80\linewidth]{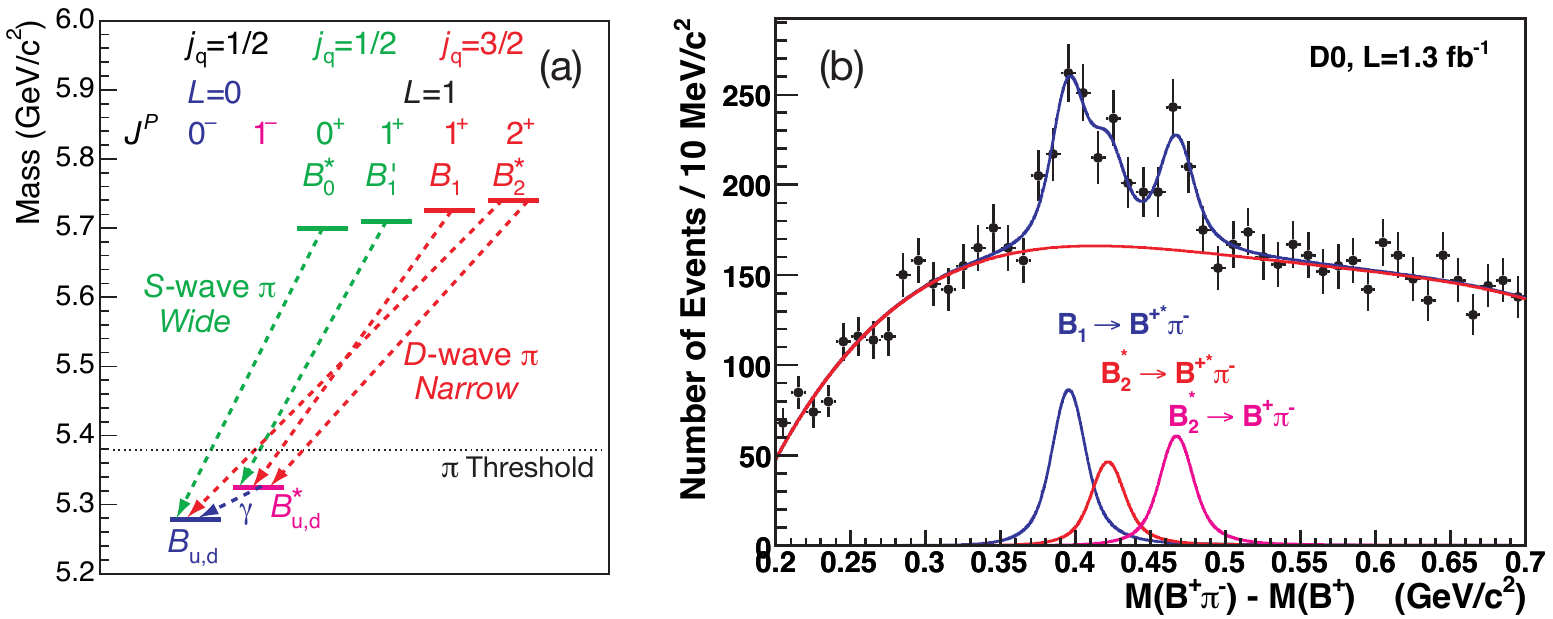}}
\hfill
\caption[]{(a) Spectroscopy of orbitally excited $B$ meson states; (b) \d0\ signal peaks for excited $B$ meson states.}
\label{fig:B**1}
\end{figure}

$B^{**}$ and $B^{**}_s$ mesons (also denoted $B_{J}$ and $B_{sJ}$,
respectively) are composed of a heavy $b$ quark and a lighter $u$,
$d$, or $s$ quark in a $L=1$ state of orbital momentum, with four possible
states in each case as shown in Fig.~\ref{fig:B**1}(b), where only the
two states that decay via $D$-wave transitions are narrow enough to be
resolved above backgrounds. The mass splittings shown are analogous to
the fine and hyperfine splittings in hydrogen.

By examining mass differences in the decay $B_J \rightarrow B^{(*)}
\pi$, the \d0\ Collaboration observed\cite{Abazov:2007vq} the $B_1^0$
and $B_2^{*0}$ states for the first time as separated states (see
Fig.~\ref{fig:B**1}(c) and measured their masses, mass splittings and
production rates. The CDF Collaboration in a similar analysis
measured\cite{Aaltonen:2008aa} the masses with higher precision and
also measured the width of the $B^{0*}_2$ for the first time. There
are some discrepancies between the two measurements, the largest being
a $2.7\sigma$ difference in the mass splittings between the two
$B^{0**}$ states.

Similarly, searching for $B^*_{s2} \ra \Bu K^-$ and $B_{s1} \ra B^{*+}
K^-$ decays, the CDF Collaboration reported\cite{Aaltonen:2007ah} the
first observation of the narrow $j_q \equiv s_q + L = 3/2$ states of
the orbitally excited \Bs\ mesons with mass and mass splitting values
consistent with theoretical predictions.  The \d0\ Collaboration also
observed\cite{Abazov:2007af} the $B^*_{s2}$ excited state but, given
the data set at the time, was not able to make any conclusions about
the presence of the $B_{s1}$.

\begin{table}[tb]
\tbl{Measured masses and widths of $B^{**}_{(s)}$ mesons from CDF.}{
    \begin{tabular}{lccc}
      \hline \hline
       & $Q$ (MeV) & & $\Gamma$ (MeV)\\
      \hline
      $B_{1}^{0}$   & $262.7 \pm 0.9 \  ^{+1.1}_{-1.2}$ 	& & $23 \pm 3 \pm 4$ \\
      $B_{2}^{*0}$  & $317.9 \pm 1.2  \ ^{+0.8}_{-0.9}$ 	& & $ 22 \ ^{+\ 3}_{-\ 2} \ ^{+\ 4}_{-\ 5} $ \\
      $B_{1}^{+}$   & $262\ \ \,  \pm 3\ \ \,   \ ^{+1}_{-3}\ \ \,  $ 	& & $49 \ ^{+12}_{-10}  \ ^{+\ \,2}_{-13}$ \\
      $B_{2}^{*+}$  & $317.7 \pm 1.2  \ ^{+0.3}_{-0.9}$ 	& & $11 \ ^{+\ 4}_{-\ 3} \ ^{+\ 3}_{-\ 4} $ \\
      $B_{s1}^{0}$  & $10.35 \pm 0.12 \pm 0.15$ 	& & $0.5 \pm 0.3 \pm 0.3$ \\
      $B_{s2}^{*0}$ & $66.73 \pm 0.13 \pm 0.14$ 	& & $1.4 \pm 0.4 \pm 0.2$ \\
      \hline \hline
     \end{tabular}
\label{tab:CDF_bstarstar}
}
\end{table}

\begin{figure}[tb]
\centerline{\includegraphics[width=1.0\linewidth]{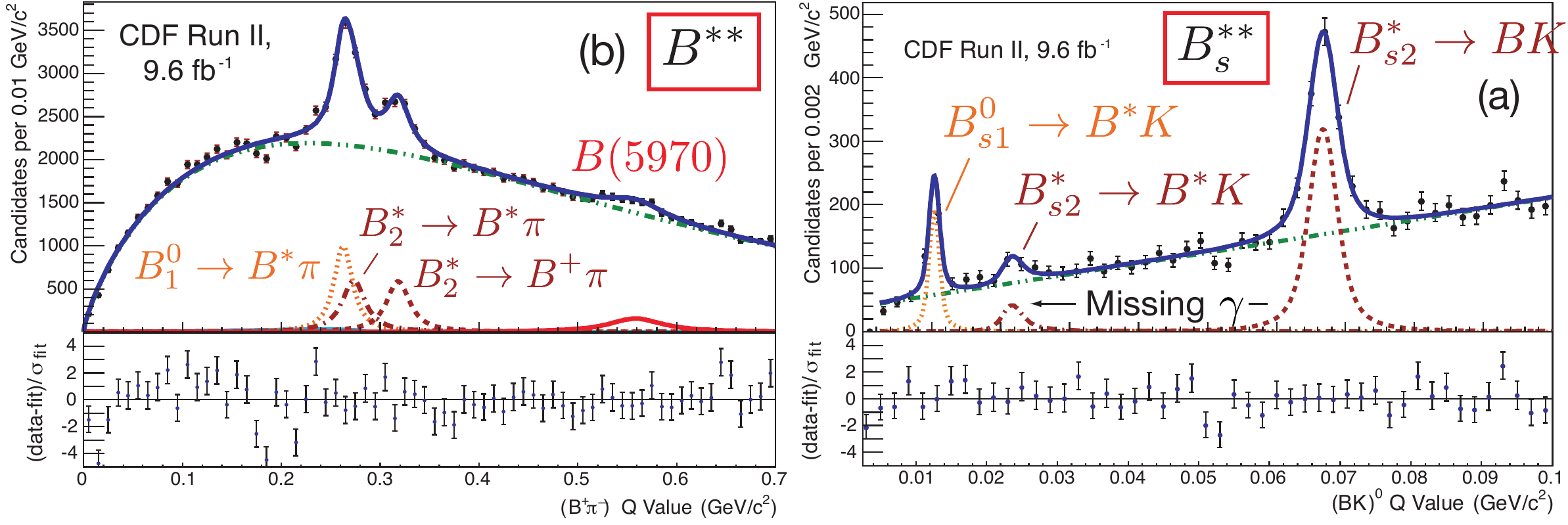}}
\hfill
\caption[]{CDF (a) $B^{**}$ (and radial excitation); and (b) $B_s^{**}$ states ($Q=M(B^{**})-M(B)-M(\pi^+)$).}
\label{fig:B**2}
\end{figure}

Using the full Run~2 data set, the CDF Collaboration updated their
measurements of the narrow orbitally excited
states~\cite{Aaltonen:2013atp} as shown in Fig.~\ref{fig:B**2}.  The
measurements include the masses and widths of the $B_1$ and $B_2^*$
for all three meson flavors: $B^{+**}$, $B^{0**}$, and $B_s^{0**}$.
%
%
Fig.~\ref{fig:B**2} shows the $B^-\pi^+$ and
$B^-K^+$ mass distributions including the results of fits.
The results are summarized in
Table~\ref{tab:CDF_bstarstar}.  This is the first observation of the
$\B^{+**}$ resonances.  In addition, CDF observes an additional excess in
both the charged and neutral channel at a $B\pi$ mass of approximately
5970\, \MeVcc with a width of about 70\, \MeVcc.  The significance exceeds
$4.4\sigma$, and the mass is consistent with expectations for a radial
excitation.

   \subsection{Charm mesons}
\label{sec:charmspect}   
\begin{figure}[tb]
%
%
\centerline{\includegraphics[width=0.90\linewidth]{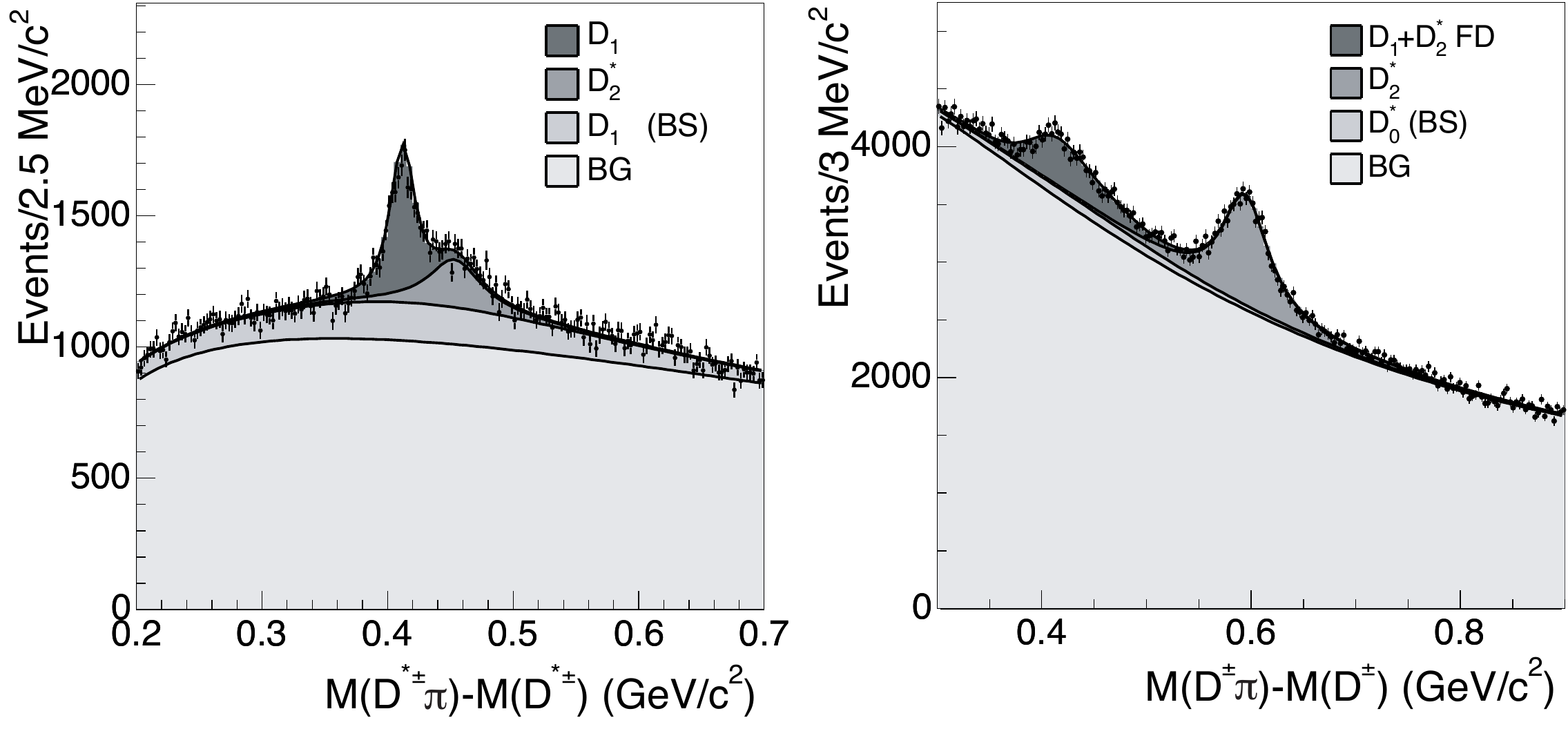}}
\caption{Mass distributions for (a) $D^{*+}\pi^-$ and (b) $D^{+}\pi^-$
  indicating the signals for $D_1^{\prime0}$ and $D_2^{*0}$.  In (b),
  the left peak arises from $D^{**0}\ra D^{*+}\pi^-$ decays where a
  $\pi^0$ from the $D^{*}$ decay is not observed.}
\label{fig:cdf_Dstarstar}
\end{figure}

Charm mesons display an similar set of excited states as described
above for bottom mesons.  While in the $B$ system, decays to the $B^*$
can be seen by a mass shift from the missing photon, in the charm
system, 
isospin violation in $D^*$ decays requires that the full final state be reconstructed.  
Absent efficient $\pi^0$ detection,
measurements are limited to $D^{(*)+}\pi^-$ decays.  CDF has
measured\cite{Abulencia:2005ry} masses and widths of the two neutral
$j_q=3/2$ states.  The mass distributions for $D^+\pi^-$ and
$D^{*+}\pi^-$ are shown in Fig.~\ref{fig:cdf_Dstarstar} where the
various components of the fits are indicated showing clear signals
for both the $D_1^{0}$ and $D_2^{*0}$, 
%
%

Using semileptonic decays of $B$ mesons to orbitally excited charm
states, the \d0\ Collaboration has measured the masses of the
$D^0_1$ and $D^{*0}_2$ via the $B^0 \ra D^{**} \mu \nu
X$ decay\cite{Abazov:2005ga} and of the $D_{s1}^{\pm}(2536)$ via the $B_s^0 \ra
D_s^{**} \mu \nu X$ decay.\cite{Abazov:2007wg}  The latter is particularly
interesting given the surprisingly light masses of the $j_q=1/2$
states plus the observation of new $D_{sJ}$ systems that may be quark
molecular states.\cite{Aubert:2006mh,Brodzicka:2007aa}

%
%

    \subsection{\texorpdfstring{$c$}{c} and \texorpdfstring{$b$}{b}-flavored baryons}
\label{sec:baryon_mass}

\begin{figure}[tb]
\centerline{\includegraphics[width=0.8\linewidth]{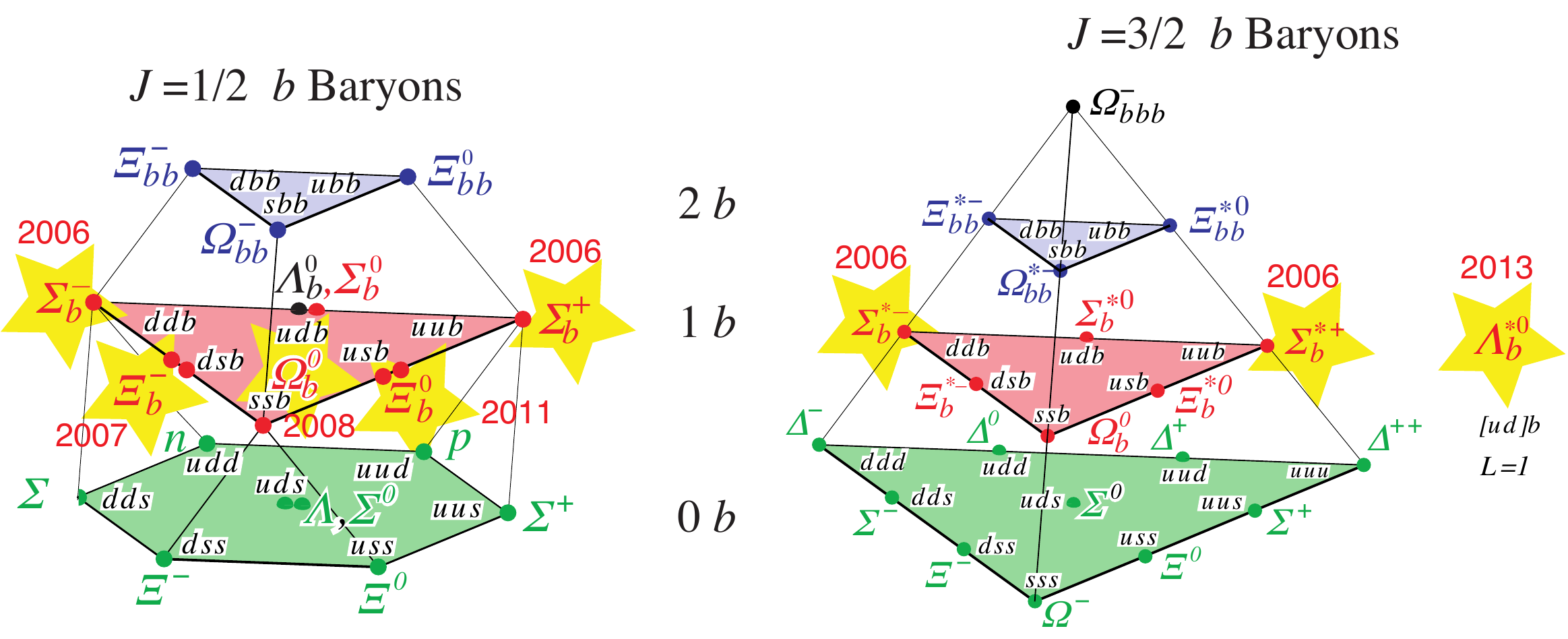}}
\hfill
\caption[]{
Baryon mulitplets with new $b$-flavored baryons discovered at the Tevatron indicated by the star symbols.}
\label{fig:multiplets}
\end{figure}

In a continued analogy with atomic systems, baryons containing a $b$
quark can be approximated as a heavy quark filling in for the nucleus
orbited by a light diquark.  Examples of the $L=0$ system would then
be the iso-singlet $J=1/2$ $\Lambda_b$ with the $qq$ spins
anti-aligned and the iso-triplet $J=3/2$ $\Sigma_b$ with the light
quark spins aligned as shown in Fig.~\ref{fig:HQET_cartoon}(b). Before Run~2
of the Tevatron, only the ground-state $\Lambda_b$ had been
identified, but the Tevatron has gone on to discover a host of new $b$
baryons as shown in Fig.~\ref{fig:multiplets}. With more data, the
properties of these states were then measured with more precision.

%
%

\begin{table}[bt]
\tbl{CDF final results on masses of $\Xi_c$ and $b$ baryons.}{
\begin{tabular}{cccc}
\hline \hline
Baryon &  Mass (MeV/$c^2$) \\
\hline
$\Xi_c^{0}$  & $2470.85\pm0.24\pm0.55$  \\
$\Xi_c^{+}$  & $2468.00\pm0.18\pm0.51$  \\
$\Lambda_b$  & $5620.15\pm0.31\pm0.47$  \\
$\Xi_b^-$    & $5793.4\pm1.8 \ \, \pm0.7$  \\
$\Xi_b^0$    & $5788.7\pm4.3 \ \, \pm1.4$  \\
$\Omega_b^-$ & $6047.5\pm3.8 \ \, \pm0.6$  \\
\hline
$M(\Xi_c^0)-M(\Xi_c^+)$ & $\ \ \ \ 2.85\pm0.30\pm0.04$ \\
$M(\Xi_b^-)-M(\Xi_b^0)$ & $\ \ \ 4.7 \, \pm4.7 \ \, \pm0.7$\\
\hline
\hline
\end{tabular}
\label{tab:cdf_baryon_final}
}
\end{table}

\begin{figure}[tb]
\centerline{\includegraphics[width=0.8\linewidth]{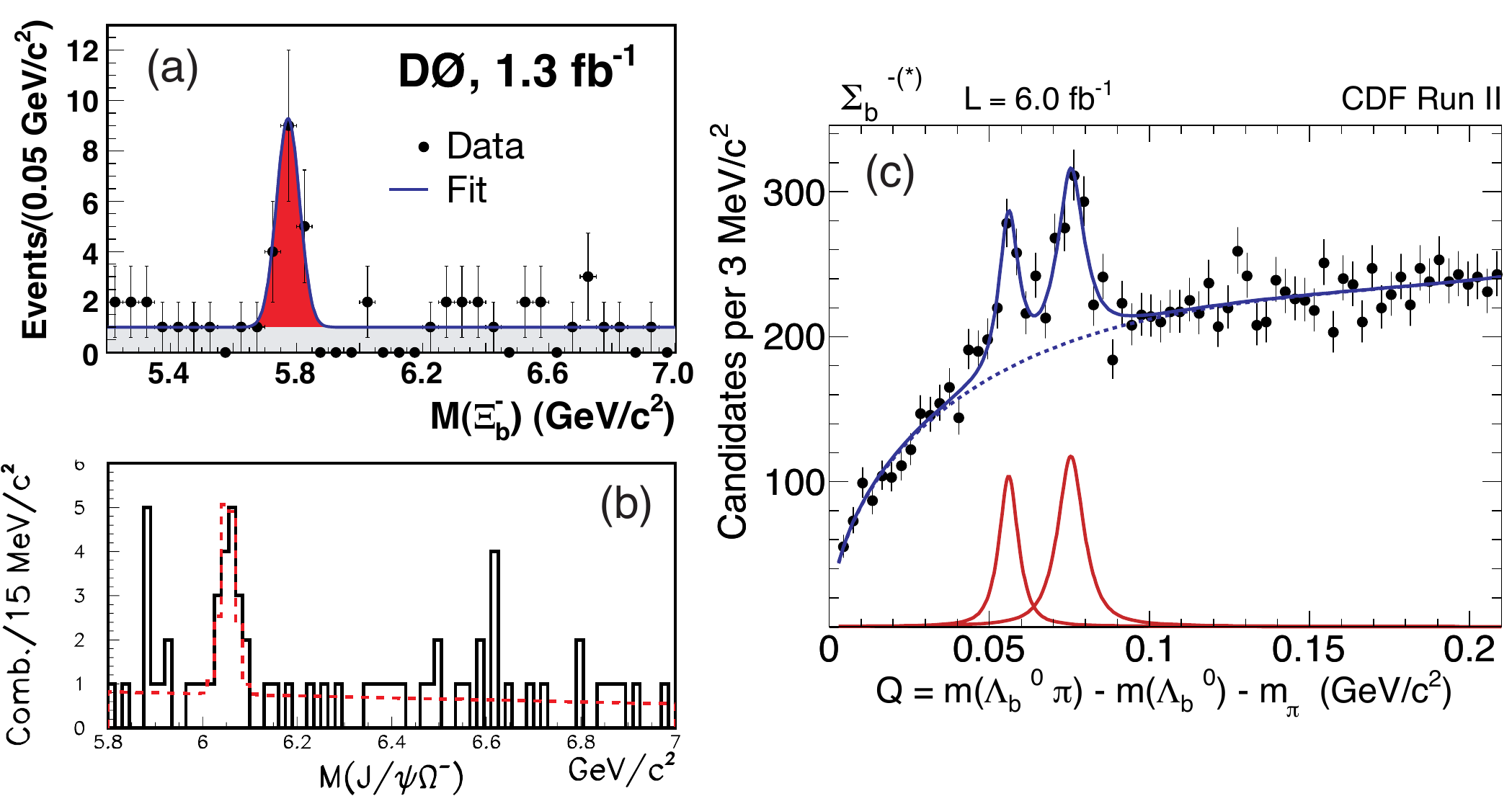}}
\hfill
\caption[]{ (a) \d0\ signal peak for $\Xi_b^{\pm} \ra \Jpsi\,\Xi$; (b)
CDF signal peak for $\Omega_b \ra \Jpsi\,\Omega$; and (c) CDF updated
signal peaks for $\Sigma_b^-$ and $\Sigma_b^{-*}$.}
\label{fig:bbaryon_peaks}
\end{figure}

Using the displaced track trigger samples, CDF made the first
observation\cite{Aaltonen:2007ar} with a significance of $5.2\sigma$ 
of the $\Sigma_b^{(*)}$ baryons in
the decay mode $\Sigma_b^{(*)\pm}\ra\Lambda_b^0\pi^\pm$ where
$\Lambda_b^0\ra\Lambda_c^+\pi^-$ and $\Lambda_c^+\ra p K^+\pi^-$.  
With approximately four times more data, CDF refined their
measurements\cite{CDF:2011ac} and included measurements of the natural
widths of each of the four charged $\Sigma_b$ baryons (see
Fig.~\ref{fig:bbaryon_peaks}(c)).

The \d0\ Collaboration, using 1.3 fb$^{-1}$ of data,
reported\cite{Abazov:2007am} the first direct observation of the
strange $b$ baryon \Xibd\ ($bsd$) via the fully reconstructed decay
$\Xibd \ra \Jpsi\,\Xi^-$ where $\Xi^-\ra\Lambda^0\pi^-$ (see
Fig.~\ref{fig:bbaryon_peaks}(a)) and measured its mass and production
rate with respect to $\Lb \ra \Jpsi\,\Lambda$.  The CDF Collaboration
also observed\cite{Aaltonen:2007ap} this state.  By reconstruction the
trajectory of the $\Xi^-$ in the silicon vertex detector, CDF was able
to to substantially decrease backgrounds and refine the $\Xi^-$
momentum measurements.  They measured a mass consistent with the \d0\
result, but with a significant improvement in precision.  With
additional data, CDF was able to extend the charged hyperon
reconstruction technique to discover\cite{Aaltonen:2011wd} the
$\Xi_b^0$ in the decay mode $\Xi_b^0 \ra \Xi_c^+\pi^-,
\Xi_c^+\ra\Xi^-\pi^+\pi^+$ and measure its mass.  This is a particle
that could not be observed in a decay mode including \Jpsi mesons that
provide clean channels for other bottom particle discoveries at the
Tevatron.  In addition, CDF made the first observation of the decay
mode $\Xi_b^- \ra \Xi_c^0\pi^-, \Xi_c^0\ra\Xi^-\pi^+$.  As a
by-product of the analyses that include fully reconstructed $\Xi_c$
baryons, CDF has made the most precise
measurements\cite{Aaltonen:2014wfa} of the $\Xi_c^0$ and $\Xi_c^+$
masses.

In the decay mode of $\Omegab \ra \Jpsi\,\Omega^-$, both the
CDF (see Fig.~\ref{fig:bbaryon_peaks}(b)) and \d0\ Collaborations reported
observations\cite{Aaltonen:2009ny,Abazov:2008qm} of
the doubly strange $b$ baryon \Omegab\ ($bss$). The mass measurements
of the \Omegab\ baryon were significantly different, $\Delta M = 111
\pm 12 \thinspace {\mathrm{(stat)}} \pm 14 \thinspace
{\mathrm{(syst)}} \thinspace {\mathrm{MeV}}$.  Furthermore, CDF observed a
substantially lower production rate relative to \Xibd\ consistent with the
$\sim15\%$ expected for producing an additional $s\bar{s}$ pair while \d0
observed a relative rate close to unity. This
CDF mass measurement, the updated mass
measurement\cite{Aaltonen:2014wfa}, and a measurement from the LHCb
Collaboration\cite{Aaij:2013qja} are all consistent within small
uncertainties. The \d0\ Collaboration continues to investigate this
signal with their full data set.  CDF's final results on $\Xi_b$ and
$\Omega_b$ masses and lifetimes using the full Run~2 dataset can be
found in Ref.~\citen{Aaltonen:2014wfa}.  The mass measurements are
listed in Table~\ref{tab:cdf_baryon_final}.

CDF has also studied an excited state of a $b$ baryon.  The
observation \cite{Aaltonen:2013tta} of
$\Lambda^{*0}_b\ra\Lb\pi^+\pi^-, \Lb\ra\Jpsi\Lambda^0$ with a
significance of $3.5\sigma$ confirms the initial
observation\cite{Aaij:2012da} made by LHCb.  The mass splitting
between $\Lambda^{*0}_b$ and \Lb is $299.82\pm0.35\pm0.30\,\MeVcc$.
Once again taking advantage of the displaced-track hadronic decay
sample, CDF has measured\cite{Aaltonen:2011sf} the masses and widths
of a variety of excited charm baryons in decay modes of the form
$\Sigma_c^*\ra\Lambda_c\pi$ and $\Lambda_c^*\ra\Lambda_c\pi\pi$, which
are typically the world's most precise and provide important
constraints on QCD models of baryon structure.
%
%

     \subsection{Exotic States}
        
Starting with the discovery\cite{Choi:2003ue} of a $\Jpsi\,\pi^+ \pi^-$
resonance at around 3872 MeV by the Belle Collaboration in 2003,
experimenters have uncovered a host of exotic charmonium-like
particles with a variety of quantum numbers.  The \XJpsi, was first
observed in exclusive decays $B^{\pm} \ra X K^{\pm}, \thinspace X \ra
\Jpsi\,\pi^+ \pi^-$ from $B$ mesons produced in \ee\ collisions. The
value of its mass very close to the $D^0 \bar{D}^{*0}$ mass threshold
along with failure of models of conventional higher-mass $c\bar{c}$
charmonium states to match its mass has generated a great deal of
interest in this state, with suggestions of it being a weakly bound
$D^0$-$\bar{D}^{*0}$ ``molecular" state\cite{Bignamini:2009sk}, a
possible four-quark state\cite{Maiani:2004vq}, or a $c\bar{c}g$
hybrid.
 
\begin{figure}[tb]
\centerline{\includegraphics[width=0.9\linewidth]{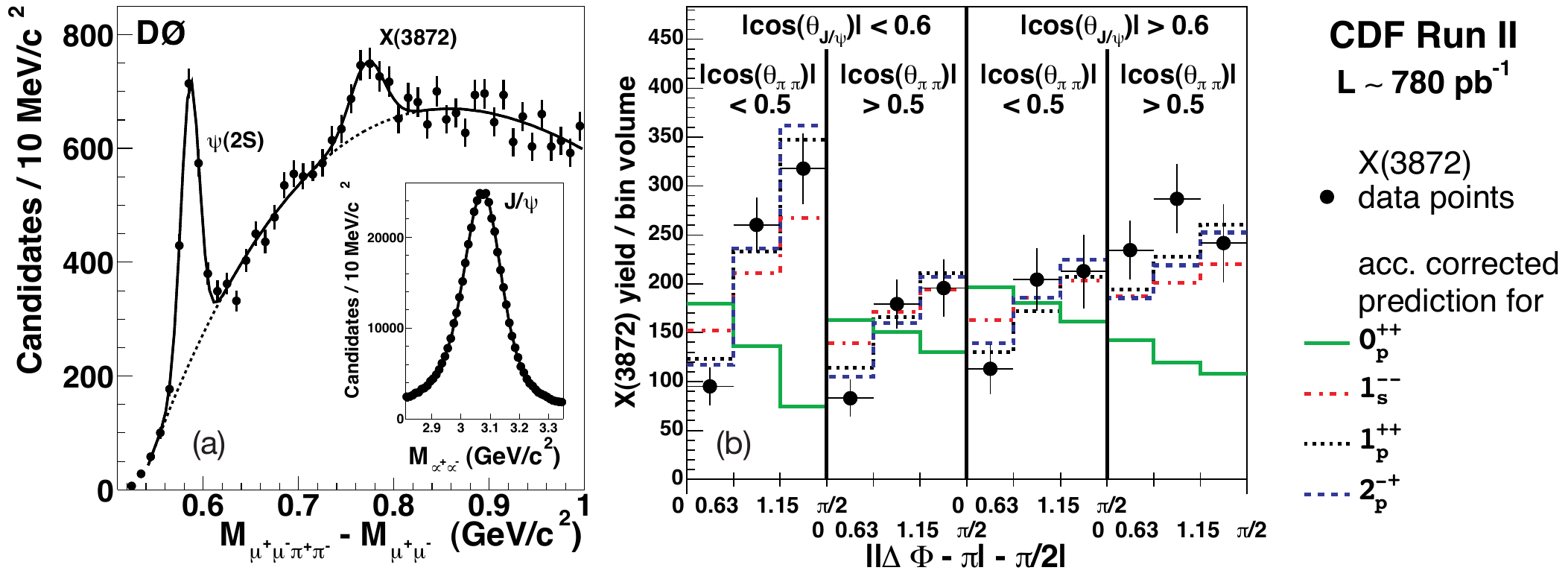}}
\hfill
\caption[]{(a) \d0\ signal peak for \XJpsi; (b) CDF angular analysis
  of the dipion system in \XJpsi\ decays.}
\label{fig:X3872}
\end{figure}

\begin{figure}[tb]
\centerline{\includegraphics[width=0.5\linewidth]{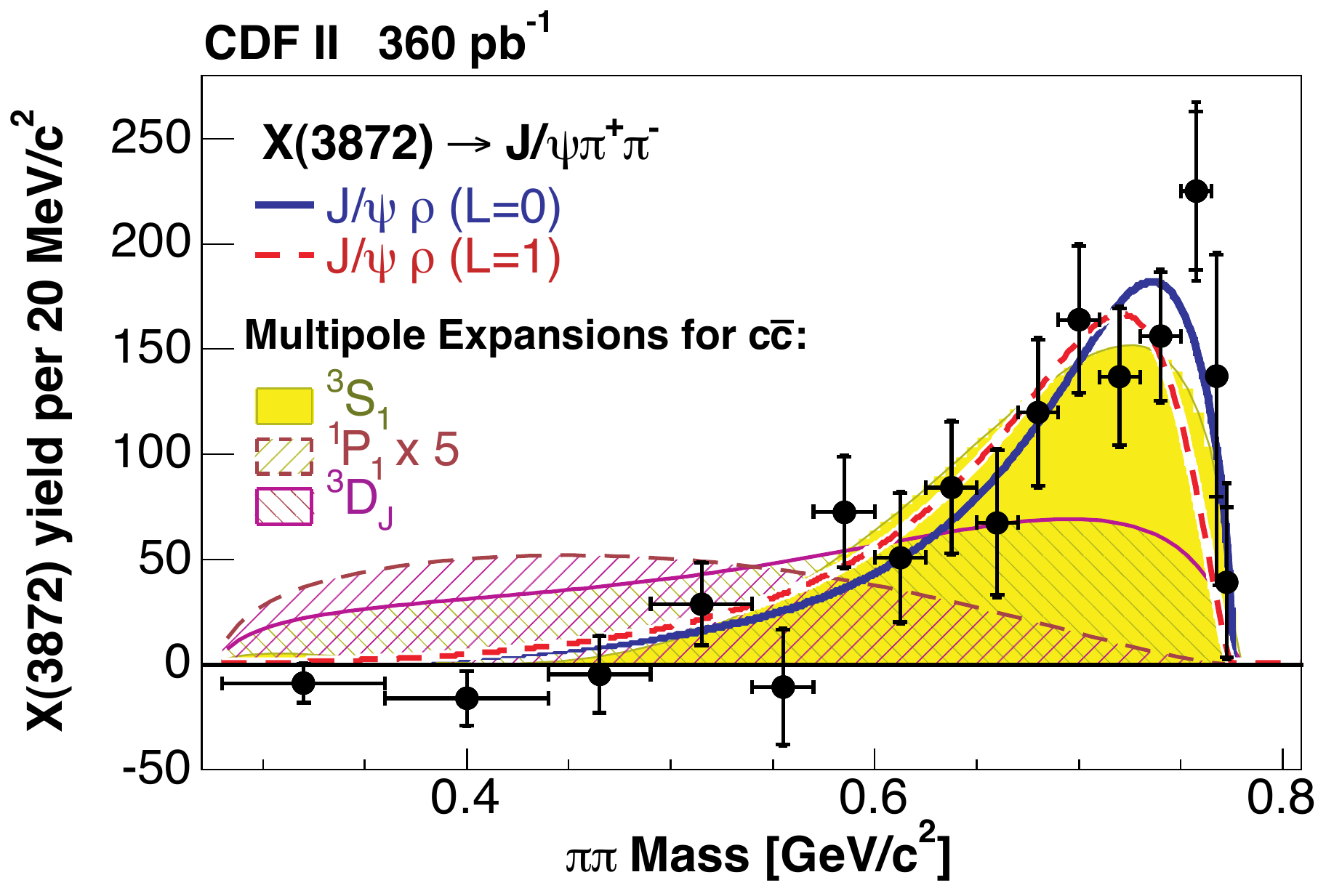}}
\hfill
\caption{CDF fit to the dipion mass distribution in $X(3872)$ decays.}
\label{fig:X_dipion}
\end{figure}

Shortly following this discovery, both the CDF and \d0\ Collaborations
also observed\cite{Acosta:2003zx,Abazov:2004kp} inclusive production
of this narrow state in \ppbar\ collisions through the decay $\XJpsi
\ra \Jpsi\,\pi^+ \pi^-$ as shown in Fig.~\ref{fig:X3872}(a). \d0\
compared characteristics of the decay with that of the $\psi(2S)$ and
observed no differences,\cite{Abazov:2004kp} while CDF, with larger data sets, 
made more
detailed measurements of its properties.  One hypothesis for the
$X(3872)$ is that it is a four-quark state.  Therefore, there would be
two different states $c\bar{c}u\bar{u}$ and $c\bar{c}d\bar{d}$ with
slightly different masses.\cite{Maiani:2004vq} CDF placed an upper
limit\cite{Aaltonen:2009vj} of $3.6\,\MeVcc$ for the mass splitting of
the two states assuming equal abundance, effectively excluding the
four-quark hypothesis.  Given that the $X(3872)$ is a single narrow
state, CDF then made the most precise measurement of its mass:
$3871.61\pm0.16\pm0.19\,\MeVcc$.  A CDF study\cite{Abulencia:2006ma}
of the dipion mass distribution placed constraints on its $J^{PC}$
assignment. Fig.~\ref{fig:X_dipion} compares the observed dipion
mass spectrum to the lowest $C$-odd possibilities.
%
%
The data give a good fit to
$\Jpsi\rho^0$ models, but it is not possible to differentiate between
$J=0$ and $J=1$.  
To determine the spin-parity of the state, CDF used
a larger data sample and performed an angular analysis as shown in
Fig.~\ref{fig:X3872}(b) where the fit clearly favors the $\Jpsi\rho^0$
hypothesis, but finds $J^{PC}=1^{++}$ and $2^{-+}$ to be equally
likely, each with a $\chi^2$ probability of 28\%.

In 2009, the CDF Collaboration reported\cite{Aaltonen:2009tz} evidence
of a narrow structure near the $\Jpsi\,\phi$ threshold in exclusive
$\Bu \ra \Jpsi\,\phi K^+$ decays and measured its mass and width.
This state does not fit conventional expectations for charmonium which
above the threshold for open charm decays should decay predominantly
to a pair of charm particles rather than to $\Jpsi\,\phi$.  This
anomalous state, known as $Y(4140)$ has been given theoretical
interpretations similar to those for \XJpsi.\cite{YI:2013vba} The \d0\
Collaboration has confirmed\cite{Abazov:2013xda} (calling it the
$X(4140)$) this enhancement at greater than $3\sigma$
significance.  Observation\cite{Chatrchyan:2013dma} by the CMS
Collaboration supports the Tevatron measurements.  Belle has searched
for $Y(4140)$ state and observes\cite{Shen:2010iu} no signal, although
does not rule it out, while LHCb has searched for the state and has
also observed\cite{Aaij:2012pz} no evidence, in disagreement
with the CDF result at the 2.4$\sigma$ level.

\begin{figure}[tb]
\centerline{\includegraphics[width=0.9\linewidth]{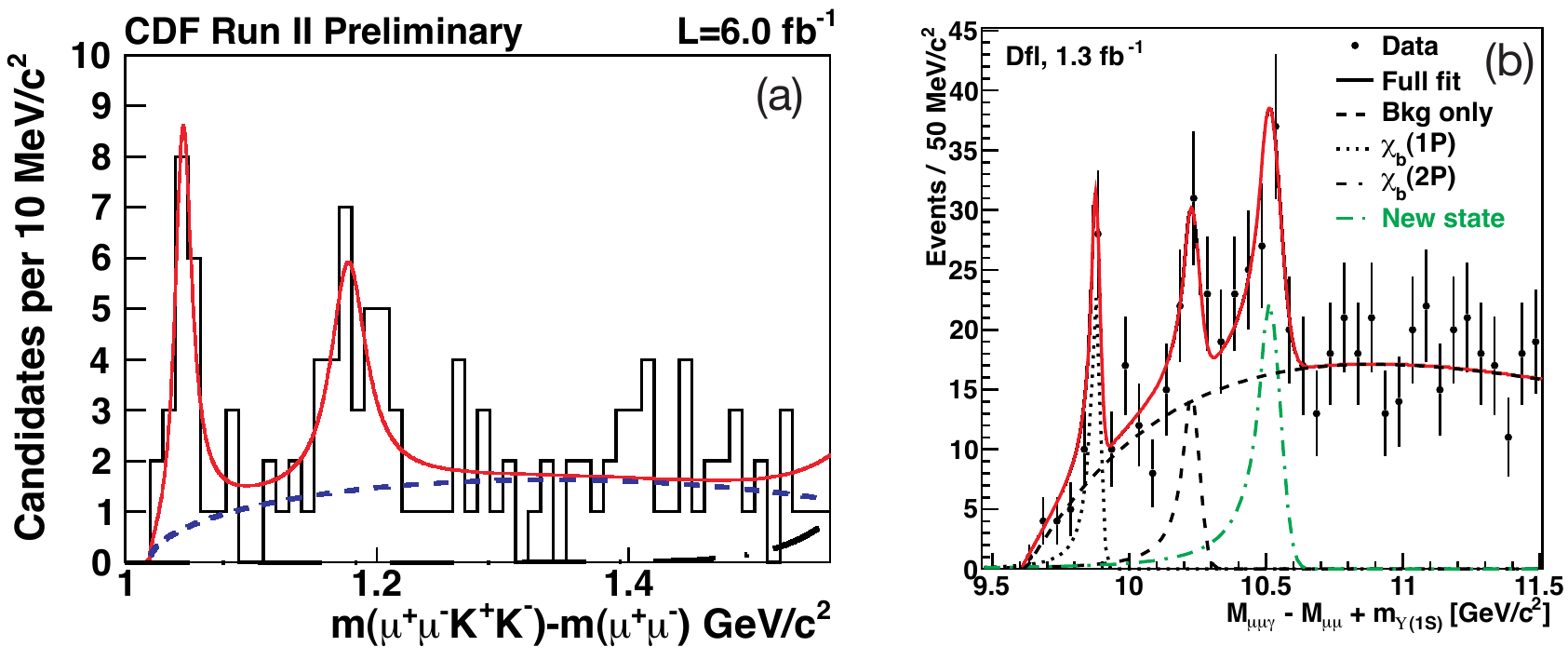}}
\hfill
\caption[]{(a) CDF signal peak for $Y(4140)$; and (b) \d0\ peaks for decays into $\Upsone \gamma$.}
\label{fig:YUpspeaks}
\end{figure}

Discoveries of more $b$-quark counterparts to the possible exotic
states above could shed light on their underlying structure. The \d0\
Collaboration has searched for particles decaying into $\Upsone +
\gamma$ and observes\cite{Abazov:2012gh} the $\chi_b(1P)$,
$\chi_b(2P)$, plus a third higher-mass structure (see
Fig.~\ref{fig:YUpspeaks}(b)) consistent with a state
observed\cite{Aad:2011ih} by the ATLAS Collaboration. Further analysis
is needed to determine whether this structure is due to the
$\chi_b(3P)$ system or some exotic $b$-quark state.

Models\cite{Nappi:1981ft,Moxhay:1985dy} of supersymmetry that include
light scalar quarks would include bound states that decay to dimuons
similarly to quarkonium.  CDF searched\cite{Aaltonen:2009rq} for
narrow states below the $\Upsilon$.  
 By comparing to the
$\Upsilon(1S) \ra\mu^+\mu^-$ yield, CDF is able to set limits in a sample of
 $630\,{\rm pb}^{-1}$ on
narrow resonances
%
%
in the
range $6.3<M_{\mu\mu}<9.0$\,\GeVcc, excluding models with $1^{--}$
bound states in that mass region.

%% file: lifetimes.tex
%
%

\section{Decays and Lifetimes}

Precision lifetime measurements of $b$ hadrons are needed in the
extraction of the weak parameters that are important for understanding
the role of the CKM matrix throughout heavy flavor physics, including
from \CP violation.  In the simplest spectator model for weakly
decaying $b$ hadrons, there is only the flavor-changing direct $b \ra
W q, \thinspace (q=c,u)$ decay, without any involvement or
interactions with the other anti-quark in a $B$ meson or quarks in a
$b$ baryon, and the lifetimes of all $b$ hadrons would be
identical. Non-spectator effects, such as $W$ exchange and
the interference between contributing amplitudes, modify this simple
picture and give rise to a lifetime hierarchy for $b$ hadrons similar
to the one observed for charm hadrons, although variations are
expected to be smaller since lifetime differences are expected to
scale as $1/m_Q^2$.  The expected hierarchy is:
\begin{equation}
\tau(\Bu) \geq \tau(\Bd) \simeq \tau(\Bs) > \tau(\Lb) \gg \tau(\Bc).
\label{lifehierarchy}
\end{equation}
Lifetime differences thus test our understanding of quark dynamics in
$b$ hadrons with comparisons to predictions of heavy quark expansions
 (HQE)\cite{Bigi:1992su,Lenz:2014jha}. 

   \subsection{\texorpdfstring{$B$}{B} meson lifetimes}
   \label{Bmesonlife}
      
Despite the huge samples of $\Bd$ and $\Bu$ mesons at the \Ups\ $B$
factories, the Tevatron experiments' measurements of
$\tau(\Bd)$ and $\tau(\Bu)$, including their ratio, were the best in
the world prior to the first results from LHCb,
 with a precision of better than 1\%. These measurements
are also important benchmarks for measuring the lifetimes of the
heavier $b$ hadrons at the Tevatron.

Both the CDF and \d0\ Collaborations have
measured\cite{Abazov:2008jz,Abazov:2005sa,Abazov:2012iy,Abazov:2007sf,Abazov:2004bn,Aaltonen:2010pj}
$\tau(\Bd)$ and $\tau(\Bu)$ through the exclusive decays $B \ra \Jpsi\,
K$, including when ratios are made to other lifetimes.  The \d0\ Collaboration also measured the ratio
$\tau(\Bu)/\tau(\Bd)$ by examining the time evolution of the number of
$\Bu \ra \bar{D}^0 \mu^+ \nu X$ decays compared to $\Bd \ra D^{*-}
\mu^+ \mu X$ decays.  CDF has measured\cite{Aaltonen:2010ta} the
\Bu lifetime in the decay mode $B^+ \rightarrow \bar{D}^0 \pi^+$ with
data from the displaced hadron trigger using a novel technique that
accounts for the decay-time bias of the trigger without simulation.

Neutral $B$ mesons contain short- and long-lived components, since
their light (L) and heavy (H) eigenstates, \BL\ and \BH, differ not
only in their masses, but also in their total decay widths, with a
decay width difference defined as $\DG = \GL - \GH$. Neglecting \CP
violation in $B$-$\bar{B}$ mixing, which is expected to be very small,
the mass eigenstates are also \CP eigenstates with the light \BL\
state being \CP-even and the heavy \BH\ state being \CP-odd. While the
decay width difference \DGd\ is tiny and can be neglected in the \Bd\
system, the \Bs\ system exhibits a significant value of \DGs\ with
$\DGGs \simeq 10\%$. For the \Bs meson, the mean lifetime, defined as
$1/\Gs$ where $\Gs = (\GsL + \GsH)/2$, and \DGs\ are the fundamental
parameters, and the measured lifetime will depend on the final state.

By isolating the decay $\Bs \ra \Jpsi\, f_0(980)$, the CDF Collaboration
made the first measurement\cite{Aaltonen:2011nk} of the \Bs\ lifetime
with a \CP-eigenstate as a final state, thus directly
determining $1/\GH = 1/(\Gs - \DGs/2)$ (neglecting \CP\ violation).

Flavor-specific decays, such as semileptonic $\Bs \ra D^-_s \ell^+
\nu$ or $\Bs \ra D^- \pi^+$ have equal fractions of \BL\ and \BH\ at
time $t=0$. If the resulting superposition of two exponential
distributions is fitted with a single exponential function, one
measures the flavor-specific lifetime from which \Gs\ and \DGs\ can
be extracted (see Ref.~\citen{Hartkorn:1999ga}). Using an exclusive
final state $\Bs \ra D_s^- \pi^+$, the CDF Collaboration
measured\cite{Aaltonen:2011qsa} the flavor-specific lifetime. The \d0\
Collaboration measured\cite{Abazov:2006cb} what was at that time the
world's best flavor-specific
lifetime in the semileptonic mode $\Bs \ra D_s^- \mu^+ \nu X$, finding
the boost by correcting for the missing neutrino energy using
correction distributions determined by MC.

The final state in the decay $\Bs \ra \Jpsi\, \phi$ contains a
 mixture of \CP-even and \CP-odd components. Early
analyses\cite{Abazov:2004ce,Abazov:2008jz} focused on measuring the
effective single lifetime of this decay channel, and the CDF
Collaboration performed the first analysis\cite{Acosta:2004gt} that
separated the two \CP\ components through a full angular study to
determine directly $1/\Gs$ and \Gs.  Subsequent analyses by both CDF
and \d0\ focusing on extracting \DGs\ specifically, and then including
the \CP-violating phase in \Bs\ mixing and are discussed in more
detail in Sections~\ref{DmsDGs} and \ref{CPV_mixing}, respectively.

For the \Bc\ meson, both quarks decay weakly, so the lifetime is
expected to be much shorter. Early measurements of the \Bc\ meson
lifetime, from CDF\cite{Abulencia:2006zu} and
\d0\cite{Abazov:2008rba}, use the semileptonic decay mode $\Bc \ra
\Jpsi\, \ell X$ and are based on a simultaneous fit to the mass and
lifetime using as a decay point the vertex formed with the leptons
from the decay of the \Jpsi and a third lepton.  Because the \Bc decay
is not fully reconstructed, these analyses require significant
background subtractions and correction factors to estimate the boost
due to the missing neutrino.  The recent
determination\cite{Aaltonen:2012yb} of the \Bc\ lifetime, from the CDF
Collaboration is based on fully reconstructed $\Bc \ra \Jpsi\, \pi^+$
decays.  Notwithstanding the limitations of the semileptonic
measurements, for a given luminosity, they still have significantly
better resolution resulting from the much larger sample sizes.  The
Tevatron average of these measurements is $\tau(\Bc) = 0.458 \pm 0.030
\thinspace {\mathrm{ps}}$, significantly shorter, as expected, than
the approximately $1.5 \thinspace {\mathrm{ps}}$ lifetime of the other
$b$ hadrons.
           
   \subsection{\texorpdfstring{$b$}{b} baryon lifetimes}
      
\begin{figure}[tb]
\centerline{\includegraphics[width=0.95\linewidth]{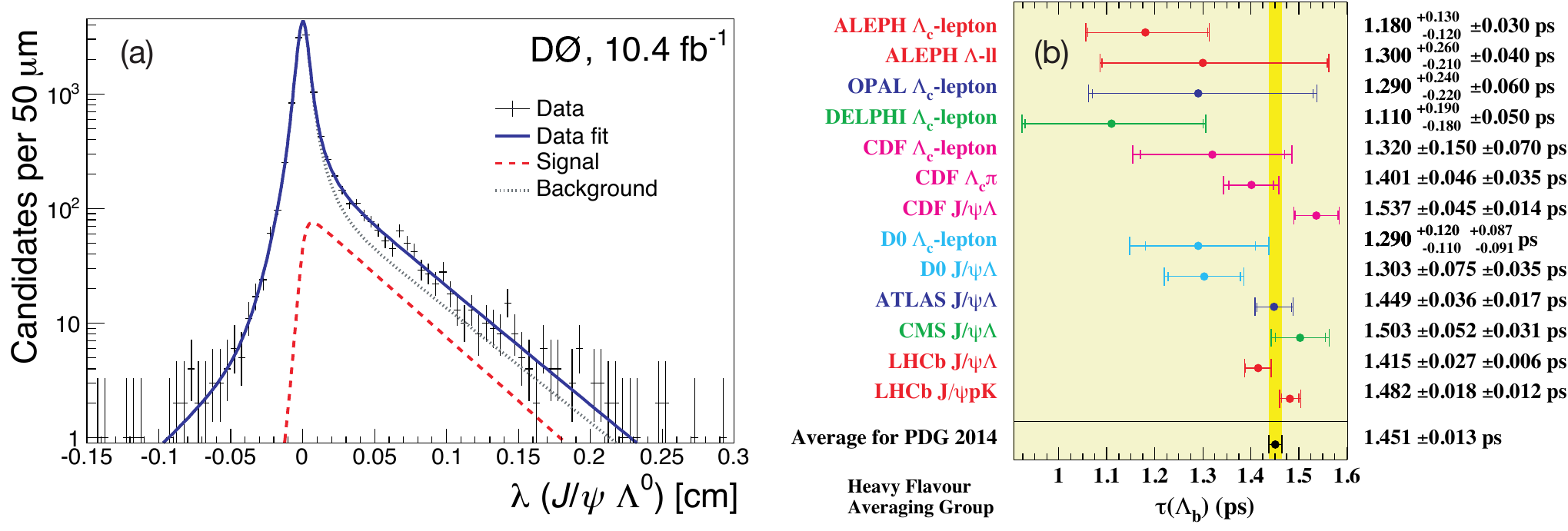}}
\hfill
\caption[]{(a) Decay length distribution for $\Lb \ra \Jpsi\, \Lambda$
decays in \d0. (b) Measurements of
$\tau(\Lb)$.}
\label{fig:lambdab_life}
\end{figure}

The ratio $\tau(\Lb)/\tau(\Bd)$ has gained a great deal of attention
since predictions of this ratio ranged from above 0.90 to only being a
few percent below unity\cite{Lenz:2014jha,Stone:2014pra}, whereas in the early
2000's, the average value was far below this prediction, with an
experimental world average\cite{PDBook2002} in 2002 being $0.800 \pm
0.081$ (and $0.786 \pm 0.034$ for a $b$-baryon mixture). The \d0\
Collaboration has measured\cite{Abazov:2007al} this lifetime in the
semileptonic mode $\Lb \ra \Lambda_c^+ \mu^- \nu X$, and CDF has
measured\cite{Aaltonen:2009zn} the lifetime in the exclusive mode $\Lb
\ra \Lambda_c^+ \pi^-$, while both collaborations have
measured\cite{Abazov:2012iy,Abazov:2007sf,Abazov:2004bn,Abulencia:2006dr,
Aaltonen:2010pj, Aaltonen:2014wfa} it via $\Lb \ra \Jpsi\, \Lambda^0$
with an example of a fit to the decay length distribution shown in
Fig.~\ref{fig:lambdab_life}(a). The evolution of measurements of
$\tau(\Lb)$ can be seen in Fig.~\ref{fig:lambdab_life}(b), where the
world average lifetime has now been pulled higher, with Tevatron
measurements being consistent with each other, the current world
average, and current theoretical predictions.

\begin{figure}[tb]
%
%
\centerline{
\includegraphics[width=0.95\linewidth]{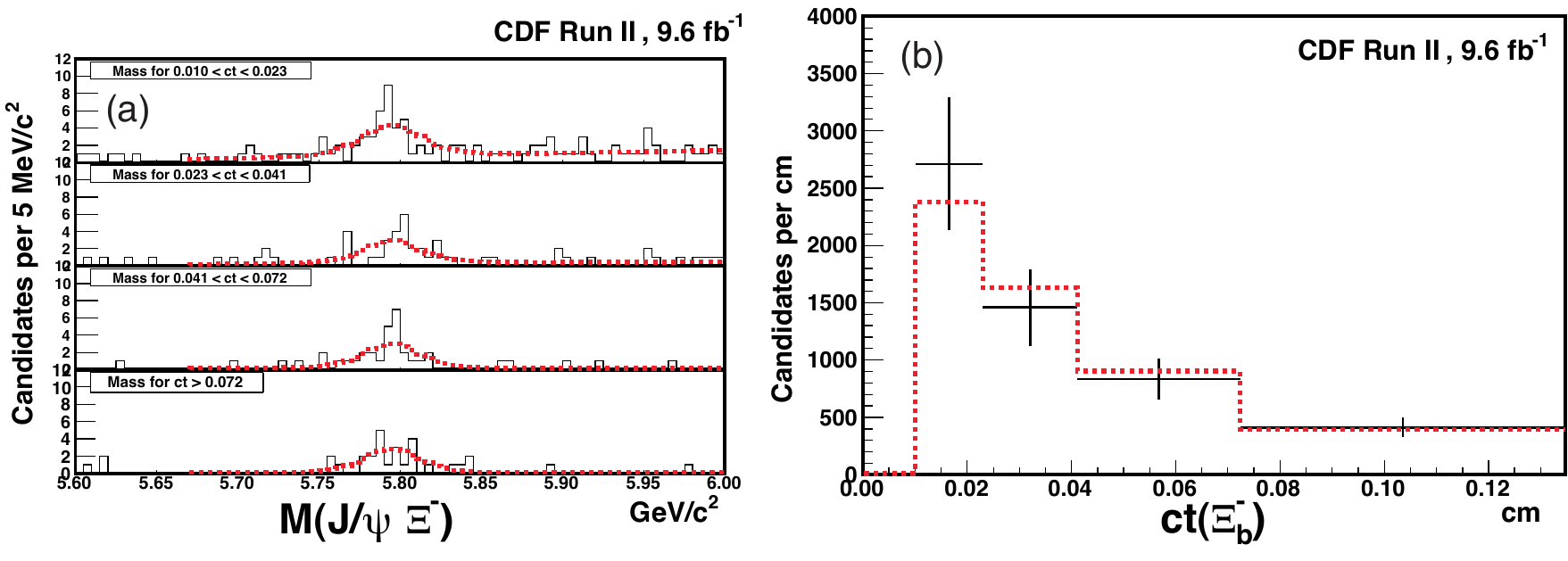}
}
\caption[]{(a) Mass distributions of $\Jpsi\,\Xi^-$ in four bins of
  proper decay time from CDF. The red dotted line shows the result of
  an unbinned likelihood fit for the $\Xi_b^-$ yield.  (b) Lifetime
  distribution using the yields from the fits to the mass
  distributions.  The red dashed line shows the fit for the $\Xi_b^-$
  lifetime.}
\label{fig:Xi_b_life}
\end{figure}

CDF  has measured\cite{Aaltonen:2009ny,Aaltonen:2014wfa} the lifetime
of the $\Xi_b^-$ and $\Omega_b^-$ baryons in the decay modes 
$
\Xi_b^-    \ra  \Jpsi\,\Xi^-,   \,\,  \Xi^-    \ra  \Lambda^0 \pi^-$ and
$\Omega_b^- \ra  \Jpsi\,\Omega^-, \,\, \Omega^- \ra  \Lambda^0 K^- $
where the charged hyperons are tracked in the silicon detector as
described in Section~\ref{sec:baryon_mass}.  The lifetime is
determined by measuring the yield in $\Jpsi\,\Xi^-$ or
$\Jpsi\,\Omega^-$ mass distributions for several regions of proper
time and then fitting the derived proper time distribution for the
lifetime.  The results for $\Xi_b^-$ are shown in Fig.~\ref{fig:Xi_b_life}.

%% file: decays.tex
%
%
%
  \subsection{Decay modes and branching ratios}
  \label{subsec:modesbr}

In many cases, specific branching fractions
are measured along the road to other results, such as \CP\ violation
and precise measurements of neutral $B$ meson mixing parameters, or
 in the process of exploring the properties of one of final-state
particles, using the final state of $b$-hadron decays to provide a clean
sample.

The \d0\ Collaboration has investigated\cite{Abazov:2005ga} the
properties of semileptonic decays of \Bd\ mesons to orbitally excited
states of the $D$ meson that have small decay widths, {\it i.e.,} those that
decay via $D$-wave transitions. Using mild assumptions on the
subsequent branching fractions of the excited mesons $D_1$ and
$D_2^*$, world's-best measurements of the branching fractions
${\cal{B}}(B \ra \bar{D}_1^0 \ell^+ \nu X)$ and ${\cal{B}}(B \ra
\bar{D}_2^{*0} \ell^+ \nu X)$ were made along with the first measurement of
their ratio $R$, one of the least model-dependent
predictions\cite{Eichten:1989zv,Dai:1998ca} of HQET for these
states. Similarly, \d0\ made a first measurement\cite{Abazov:2007wg}
of the branching fraction of ${\cal{B}}(\Bs \ra D^-_{s1}(2536) \mu
\nu_{\mu} X)$, while providing measurements of the properties of the
$D^-_{s1}(2536)$, as described in Sect.~\ref{sec:charmspect}.

In preparation for \CP-violation measurements, both the CDF and \d0\
Collaborations measured\cite{Acosta:2004gt,Abazov:2008jz} the linear
polarization amplitudes in the decay $\Bd \ra \Jpsi\, K^{*0}$, and
measurements of the strong phases indicated evidence for the presence
of final-state interactions for this decay. Comparisons with same
measurements in the analogous decay $\Bs \ra \Jpsi\, \phi$ were
consistent with SU(3) symmetry, {\it i.e.,} that the strong phases are
consistent with being the same in both systems.

The study of $B$ meson decays into several charmonium states can also
be used to constrain the long-distance parameters associated with
color-octet production which are important for the understanding of
both mixing induced and direct \CP\ violation. It also allows for possible
additional channels, and provides a test of quark-hadron duality in
the comparison with \Bd\ and \Bu\ decays. Both the CDF and \d0\
Collaborations have measured\cite{Abulencia:2006jp,Abazov:2008jk} the
ratio of branching fractions of $B^0_s \ra \psi^\prime \phi$ to $B^0_s
\ra \Jpsi\, \phi$ to be consistent with the ratio of branching fractions
of $B^{\pm} \ra \psi^\prime K^{\pm}$ to $B^{\pm} \ra \Jpsi\, K^{\pm}$.

The decay products in $\Bs \ra \Jpsi\, f_0(980)$ are in a \CP-odd
eigenstate, which can provide a more direct measurement of the
\CP-violating phase $\phi_s$ as compared to the decay $\Bs \ra \Jpsi\,
\phi$ where the decay products are in an indefinite \CP\ state (see
Sect.~\ref{CPViolating}). This can provide a useful additional
channel, although the branching fraction of the former is expected to
be smaller. Also, the decay chain $\Bs \ra \Jpsi\, f_0(980), f_0(980)\ra
K^+K^-$ forms a background to $B^0_s
\ra \Jpsi\, \phi$ that must be constrained in studies of \CP\
violation.  The CDF and \d0\ Collaborations measured\cite{Aaltonen:2011nk,Abazov:2011hv} the
ratio of the $\Bs \ra \Jpsi\, f_0(980)$ and $\Bs \ra \Jpsi\,
\phi$ branching fractions to be consistent with
expectations.

$B^0_s \rightarrow J/\psi\,K_S^0$ is a CP eigenstate, and measurement
of its lifetime would directly probe the lifetime
$\tau_{B_{s{\mathrm H}}}$. Additionally, large samples can be used to
extract the angle $\gamma$ of the unitary triangle.
CDF made the first observation\cite{Aaltonen:2011sy} of the
Cabibbo-suppressed decay modes $B^0_s \rightarrow J/\psi\,K_S^0$ and
$B^0_s \rightarrow J/\psi\,K^{*0}$, observing both with greater than
$7\sigma$ significance.

The \d0\ Collaboration studied\cite{Abazov:2012dz} the decays $B^0_s
\ra J/\psi\, K^+K^-$, and from the invariant mass and spin of the
$K^+K^-$ system, found evidence for the two-body decay $B^0_s \ra
J/\psi\, f_2^{\prime}(1525)$ and measured the relative branching
fraction of the decays with respect to the rate into $J/\psi\, \phi$.

In Run 2, CDF made the first studies of hadronic decays of \Bs\
mesons.  Taking advantage of the samples and techniques developed for
the measurement of \Bs\ mixing (Sect. \ref{sec:BsMix}), CDF
measured\cite{Abulencia:2006aa}  the ratios of branching
fractions $B(\Bs \ra D_s^- \pi^+ \pi^+ \pi^-) / B(B^0 \ra D^- \pi^+
\pi^+ \pi^-)$ and $B(\Bs \ra D_s^- \pi^+) / B(B^0\ra D^-\pi^+)$

In studies of decay modes that can be used for future studies of \CP\
violation, the \d0\ Collaboration reported the first
evidence\cite{Abazov:2007rb,Abazov:2008ig} for the decay $\Bs \ra
D_s^{(*)} D_s^{(*)}$ with CDF quickly following with the first
observation\cite{Abulencia:2007zz} of the decay $B_s^0 \to D_s^+
D_s^-$ with a $7.5\sigma$ significance and measured its branching
ratio.  Subsequently, CDF made the world's best
measurements\cite{Aaltonen:2012mg} of the $B_s^0 \to D_s^{(*)+}
D_s^{(*)-}$ branching ratios.  These modes are also important as they
can be used to measure lifetimes of the \CP-even eigenstate of \Bs\
that complements the measurements of the \CP-odd state $\Bs \ra
\Jpsi\, f_0(980)$.  In another mode with potential for future CPV
studies, CDF made the first observation\cite{Aaltonen:2008ab} of
$\bar{B}^0_s \to D_s^\pm K^\mp$ and measured the ratio of branching
fraction relative to that for $\bar{B}^0_s\to D_s^+ \pi^-$.

CDF also made great strides in the studies of hadronic decays of $b$
baryons, making the first observation\cite{Abulencia:2006df} of a
fully hadronic \Lb\ decay and measuring the ratio of cross section
times branching ratio for the mode $\Lambda_b^0 \to \Lambda_c^+ \pi^-$
relative to $\bar{B}^0 \to D^+ \pi^-$.  More data brought the ability
to do detailed analysis of baryon decays including a 
measurement \cite{CDF:2011aa} of the branching fraction
${\mathcal{B}}(\Lambda^0_b\rightarrow \Lambda^+_c\pi^-\pi^+\pi^-)$
including of the resonant substructure of this multibody decay mode.

%% file: mixing.tex
%
%
                            
\section{Mixing and Oscillations of Heavy Neutral Mesons}
\label{DmsDGs}

\begin{figure}[tb]
\centerline{\includegraphics[width=0.85\linewidth]{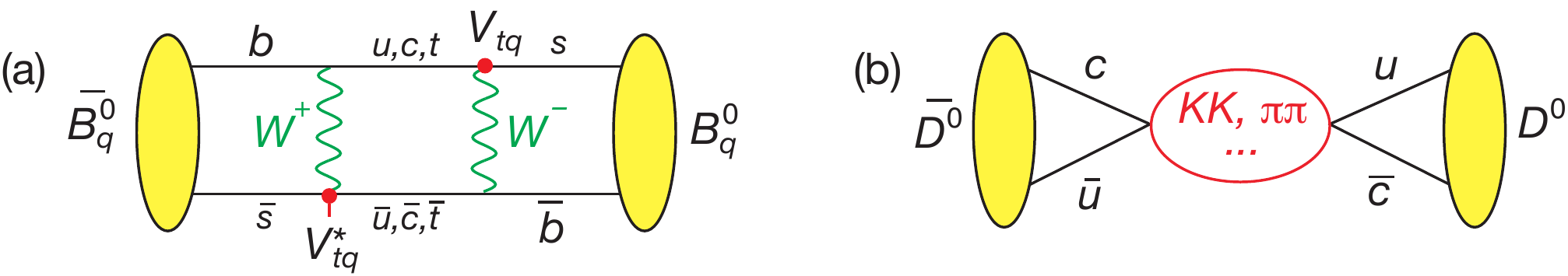}}
\hfill
\caption[]{(a) Example of second-order box diagram responsible for
neutral meson mixing, in this case for neutral $B$ mesons with
$q=d,s$. (b) Long-range contribution for charm mixing.}
\label{fig:mixbox}
\end{figure}

Neutral meson systems exhibit the phenomenon of particle-antiparticle
oscillations that can proceed by a second-order weak box diagram as
shown in Fig.~\ref{fig:mixbox}(a). The intense interest in this
process is due to the access provided to CKM matrix elements (e.g.,
$V_{td}$ and $V_{ts}$), as well as probing for new heavy particles
that could also be participate in the box diagram. The time evolution
for a meson $M$
of such a $M^0$-$\bar{M}^0$ system is then governed by $2 \times 2$
mass and decay matrices. In each of these systems, the light (L) and
heavy (H) mass eigenstates,
\begin{equation}
|M_{\mathrm{L,H}} \rangle = p|M^0 \rangle \pm q |\bar{M}^0 \rangle,
\label{eq:mixing}
\end{equation}
have a mass difference $\Delta m = m_{\mathrm{H}} - m_{\mathrm{L}}$
and a decay width difference $\Delta \Gamma = \Gamma_{\mathrm{L}} -
\Gamma_{\mathrm{H}}$ with values determined by the off-diagonal terms
of the mass and decay matrix. We further define $x = \Delta m/\Gamma$
and $y = \Delta\Gamma/2\Gamma$, where $\Gamma = (\Gamma_{\mathrm{L}} +
\Gamma_{\mathrm{H}})/2$. In the absence of \CP\ violation in the
mixing, $|q/p| = 1$. Neglecting \CP\ violation, the probability
density ${\cal{P}}_+$ (${\cal{P}}_-$) for a $\bar{M}^0$ produced at
proper time $t=0$ to decay as a $\bar{M}^0$ ($M^0$) at time $t$ is
then
\begin{equation}
{\cal{P}}_{\pm}(t) = \frac{\Gamma}{2}e^{-\Gamma t}[1 \pm \cos(\dmq t)].
\label{eq:oscillate}
\end{equation}

     \subsection{Charm mixing}

In the standard model, $D^0$-$\bar{D}^0$ mixing is a weak interaction
process that occurs primarily through long-range virtual intermediate
states that consist of common decay channels for particle and
antiparticle, such as $\pi^+\pi^-$ (see Fig.~\ref{fig:mixbox}(b).)
Mixing could also result from exotic particles that appear as virtual
states in a short-range box diagram.  The decay $D^0 \ra K^+ \pi^- $
can arise from mixing of a $D^0$ state to a $\bar{D}^0$ state,
followed by a Cabibbo-favored (CF) decay, or from a doubly
Cabibbo-suppressed (DCS) decay of a $D^0$.  Several experiments have
measured\cite{Staric:2007dt,Lees:2012qh,Aaltonen:2007ac} the ratio $R$
of $D^0\ra K^+ \pi^-$ to $D^0\ra K^- \pi^+$ decay rates and have seen
$3\sigma$ evidence for mixing.  $R$ can be approximated\cite{PDG2012}
as a quadratic function of $t/\tau$, where $t$ is the proper decay
time and $\tau$ is the mean $D^0$ lifetime:
\begin{equation}
R(t/\tau) = R_D + \sqrt{R_D}y^\prime (t/\tau) +
\frac{x^{\prime2}+y^{\prime2}}{4}(t/\tau)^2,
\label{eq:dmix}
\end{equation}
where $R_D$ is the squared modulus of the ratio of DCS to CF
amplitudes, $x^\prime= x\cos\delta+y\sin\delta$ and
$y^\prime=-x\sin\delta+y\cos\delta$, $x$ and $y$ are defined as above,
and $\delta$ is the strong-interaction phase difference between the
DCS and CF amplitudes.  In the absence of mixing, $x=y=0$ and
$R(t/\tau) = R_D$.

\begin{figure}[tb]
\centerline{
\includegraphics[width=0.45\linewidth]{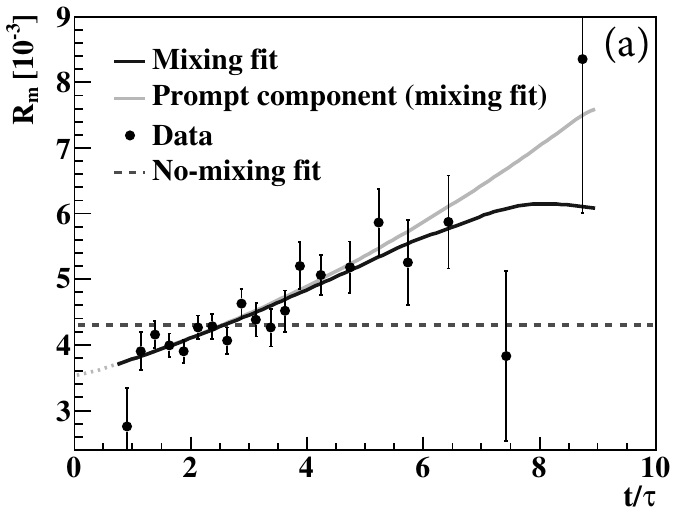}
\hfill
\includegraphics[width=0.45\linewidth]{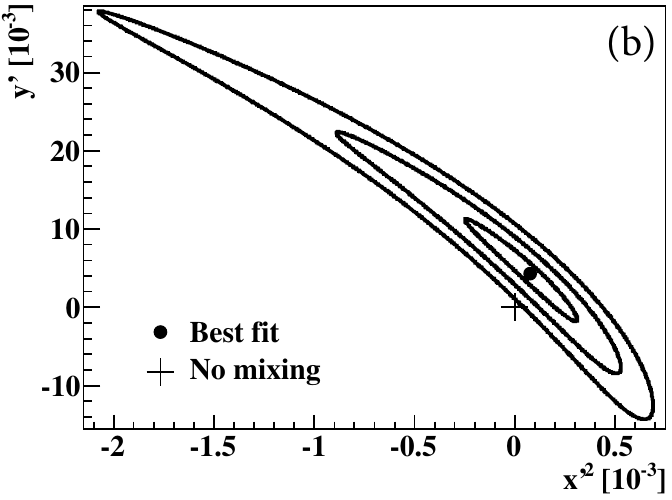}
}
\caption{(a) Measured ratio $R_m$ of wrong-sign to right-sign $D^0$
decays as a function of normalized proper decay time. The results of a
least-squares fit are shown for the mixing fit and for the prompt
component. A fit assuming no mixing is clearly incompatible with the
data. (b) Contours in $x^{\prime2}$-$y^\prime$ parameter space
bounding regions with Bayesian posterior probability corresponding to
1, 3, and 5 Gaussian standard deviations.  }
\label{fig:dmix}
\end{figure}

 Following first $>5\sigma$ observation\cite{Aaij:2012da} of $D^0$
 mixing by LHCb, CDF confirmed\cite{Aaltonen:2013pja} the effect,
 observing mixing with $6\sigma$ significance.  The deviation of
 $R(t/\tau)$ from a constant is clearly visible in Fig.\ref{fig:dmix}(a)
 which shows the data as well as the results of a fit to
 Eq.~\ref{eq:dmix}, while Fig.~\ref{fig:dmix}(b) shows contours for the
 values of $x^{\prime2}$ and $y^\prime$ returned by the fit.

      \subsection{\texorpdfstring{$B^0$}{B0} mixing and oscillations}
         
The identification of \Bd\ oscillations and extraction of \dmd\ use
Eq.~\ref{eq:oscillate} and the measurement of probability density
functions that describe the measured time development of
\Bd\ mesons that decay with the same or opposite flavor as their
flavor at production. The flavor, {\it i.e.} $B^0$ or $\bar{B}^0$, at
the time of decay is determined by the charge of the decay
products. Since the dominant production mechanisms at the Tevatron
produce $b\bar{b}$ pairs, the flavor at the time of production can be
determined by the charge of the lepton from semileptonic decays or a
momentum-weighted charge of the decay products of the second $b$
hadron produced in the collision. 

Since \Bd\ oscillations were definitively measured at the $B$
factories, measurements at the Tevatron, {\it e.g.} by the \d0\
Collaboration\cite{Abazov:2006qp} were mostly pursued as controls
comparing to the world average value (currently $\dmd = 0.507 \pm
0.004 \thinspace {\mathrm{ps}}^{-1}$) in preparation for searching for
\Bs\ oscillations and to calibrate the tagging power of flavor-tag
techniques.  A measurement of $\DGGd = 0.0079 \pm 0.0115$ has been
made at the Tevatron by \d0\cite{Abazov:2013uma}, as described in
Sect.~\ref{CPV_mixing}.

\subsection{\texorpdfstring{$B^0_s$}{Bs0} mixing and oscillations}
\label{sec:BsMix}


The determination of the the $B^0_s$-$\bar{B}^0_s$ oscillation
frequency \dms\ has been a major goal of experimental particle physics
since the first observation of \Bd\ mixing in
1987\cite{Albajar:1986it,Albrecht:1987dr}. Since $|V_{ts}|$ is larger
than $|V_{td}|$, the oscillation frequency \dms\ was expected to be
much greater than that for $B^0$-$\bar{B}^0$ oscillations, 
requiring the large data samples and excellent proper time resolution
available at the Tevatron and its detectors. In addition to the
opposite-side tagging described above, same-side tagging was also used to
take advantage of fragmentation on the reconstructed side of the
event. For an example, if an associated $K^+$ containing a strange antiquark
is found, the strange quark will have likely hadronized with a $\bar{b}$,
tagging the flavor at production

\begin{figure}[tb]
\centerline{\includegraphics[width=1.1\linewidth]{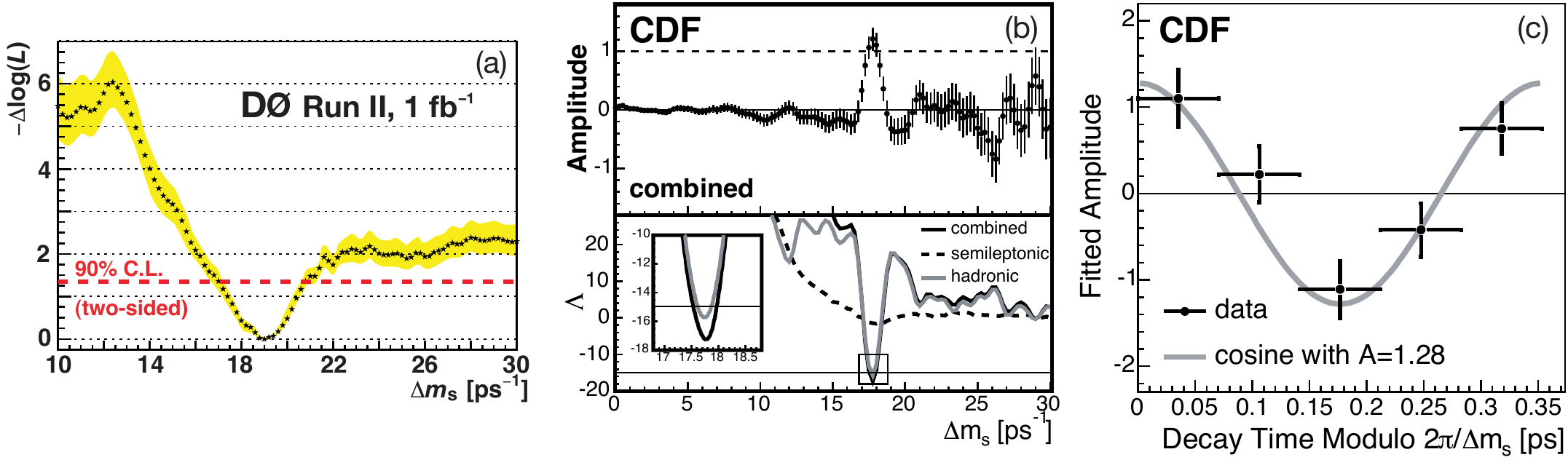}}
\hfill
\caption[]{(a) Result of a \d0\ likelihood
  analysis\cite{Abazov:2006dm} showing $\Delta \log {\cal{L}}$ as a
  function of \dms; (b) measured amplitude values
  and uncertainties (upper) and the logarithm of the ratio of
  likelihoods for amplitude equal to one and amplitude equal to zero
  for the CDF result\cite{Abulencia:2006mq}; and (c) for the same
  analysis, the oscillation signal measured in five bins of proper
  decay time modulo the measured oscillation period. }
\label{fig:bsmix}
\end{figure}

Until the Tevatron Run~2 started, there were only lower limits on the
value of \dms. The \d0\ Collaboration placed\cite{Abazov:2006dm} the
first two-sided limit (at 90\% C.L.) on \dms\ (see
Fig.~\ref{fig:bsmix}). Shortly after, the CDF Collaboration made the
first $>$$5\sigma$ observation of $B^0_s$-$\bar{B}^0_s$ oscillations
and the first measurement\cite{Abulencia:2006mq} of \dms. A follow-up
publication\cite{Abulencia:2006ze} from CDF with subsequent
improvements then reported a measurement of
\begin{equation}
\dms = 17.77 \pm 0.10 \thinspace {\mathrm{(stat)}} \pm 0.07 \thinspace {\mathrm{(sys)}} \thinspace {\mathrm{ps}}^{-1},
\end{equation}
as shown in Fig.~\ref{fig:bsmix}(b,c).

\begin{figure}[tb]
\centerline{\includegraphics[width=1.0\linewidth]{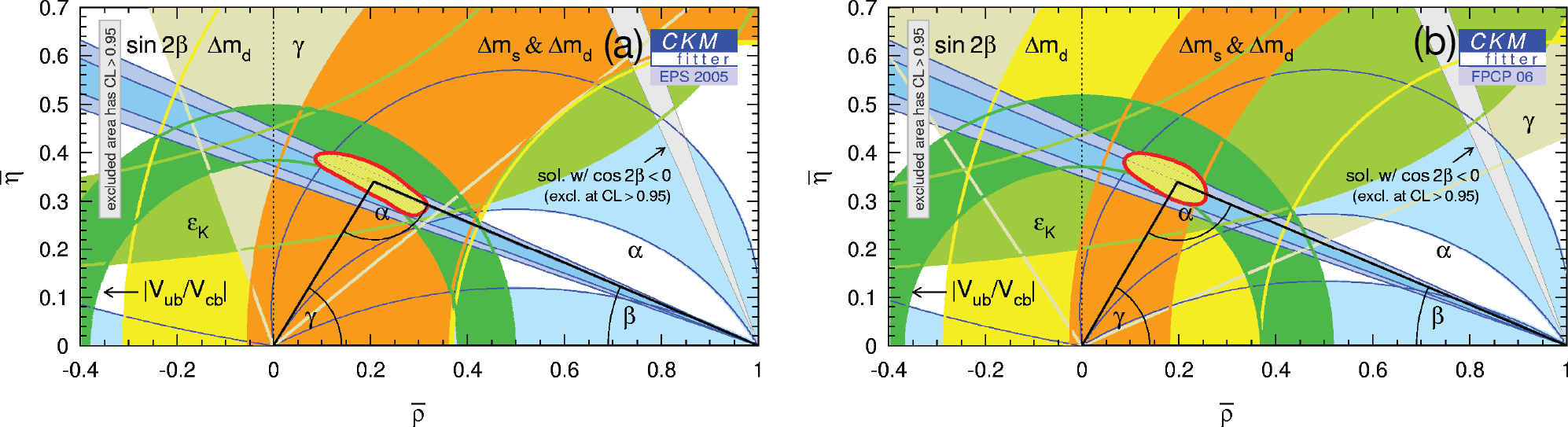}}
\hfill
\caption[]{Constraints on the apex ($\bar{\rho},\bar{\eta})$ of the $B$-meson
CKM unitarity triangle (a) before and (b) after
the Tevatron result on \dms. From Ref.~\citen{Charles:2004jd} (see for definition
of $\bar{\rho}$ and $\bar{\eta}$).}
\label{fig:ckmfitter}
\end{figure}

An important goal of heavy flavor physics is constraining the CKM
matrix using experimental results on observables together with
theoretical inputs and unitarity conditions. The constraint from our
knowledge on the ratio $\dms / \dmd$ and hence $|V_{td}/V_{ts}|$ is
more effective in limiting the position of the apex of the CKM
$B$-meson unitarity triangle than the one obtained from \dmd\ measurements
alone, due to the reduced hadronic uncertainty as shown comparing
constraints on the CKM matrix in
Fig.~\ref{fig:ckmfitter}(a) before and (b) after the measurement of
\dms\ from the Tevatron\cite{Charles:2004jd}. The measured value of
\dms\ is also consistent with the Standard Model prediction at the
time of $\dms = 19.0 \pm 1.5 \thinspace {\mathrm{ps}}^{-1}$
obtained\cite{Bona:2006ah} from CKM fits where no experimental
information on \dms\ is used. 

The existence of final states to which both the $B^0_s$ and
$\bar{B}^0_s$ can decay, such as $\Bs \ra D_s^+ D_s^-$ and $\Bs \ra
\Jpsi\, \phi$ which involve $b \ra c\bar{c}s$ transitions, results in
a relatively large value of the width difference between mass
eigenstates $\DGGs \simeq 10\%$.  In contrast the $b \ra c\bar{c}d$
transition is Cabibbo-suppressed so that \DGGd is tiny in the standard
model. The large value of \DGs allows for an additional richness of
physics in the $B^0_s$ system.

Because other $CP$ eigenstate decay modes are either helicity or CKM
suppressed, then under certain theoretical assumptions\cite{Aleksan:1993qp} the
semi-inclusive decays $\Bs \ra D_s^{(*)} D_s^{(*)}$ saturate the
\CP-even eigenstates in the \Bs.  Thus the partial width for this mode
accounts for the difference between the widths of \BsL\ and \BsH\ and
$\Delta\Gamma^{\mathrm{CP}}_s$ can be measured using information from
branching ratios without lifetime fits.\cite{Dunietz:2000cr}  The branching
fraction\cite{Abazov:2007rb,Abazov:2008ig,Abulencia:2007zz,Aaltonen:2012mg}
measurements by CDF and \d0\ described in Sec.~\ref{subsec:modesbr}
can be used to find $\Delta\Gamma^{\mathrm{CP}}_s = \DGs/\cos\phi_s$,
where $\phi_s$ is the \CP-violating mixing phase (see
Sect.\ref{CPViolating}) that can constrain models of new physics.

\begin{figure}[tb]
\centerline{\includegraphics[width=1.0\linewidth]{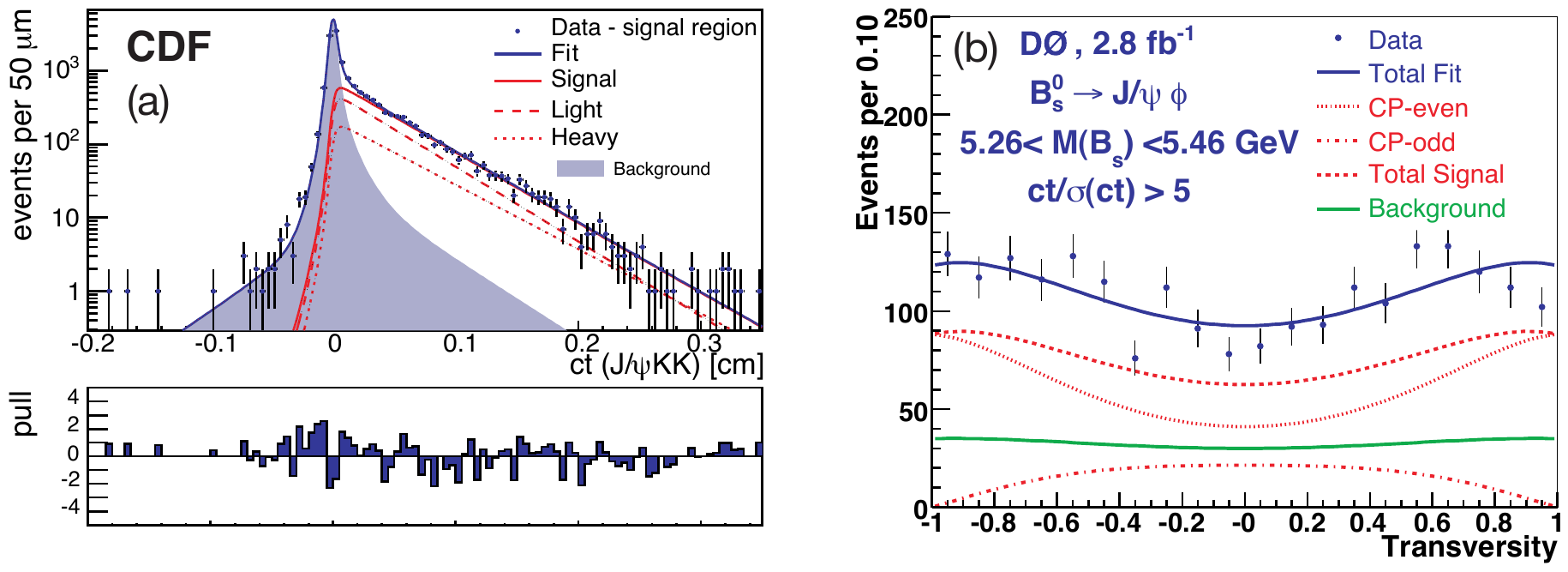}}
\hfill
\caption[]{(a) Example of a proper decay time fit projections for
light and heavy mass eigenstates from a CDF analysis\cite{CDF:2011af}
(6.5 fb$^{-1}$); and (b) example of fit to transversity angle of the
$\mu^+$ in the \Jpsi\ rest frame with respect to the $\phi$ decay
plane to \CP-even (``light") and \CP-odd (``heavy") components from a
\d0\ analysis\cite{Abazov:2008af} (2.8 fb$^{-1}$).}
\label{fig:jpsiphi}
\end{figure}

Determinations of \DGs\ and \Gs\ in specific \CP\ eigenstates were
discussed in Sect.~\ref{Bmesonlife}.  The best sensitivity to \DGs\
and \Gs\ (and hence the for the mean lifetime $\tau(B^0_s) = 1/\Gs$)
is achieved by time-dependent measurements of the $\Bs \ra \Jpsi\, \phi$
decay where the \CP-even and \CP-odd components are separated via a
full angular analysis of the $\Jpsi \ra \mu^+\mu^-$ and $\phi \ra
K^+K^-$ decay products.  Fit projections from two analyses are shown in
Fig.~\ref{fig:jpsiphi}. Both untagged and flavor-tagged analyses have
been pursued at the Tevatron, with earlier analyses assuming no CP
violation and then being optimized to measure the CP-violating phase
$\phi_s$. A complication is the possibility of a $S$-wave $K^+K^-$
amplitude in addition to the usual $P$-wave $\phi$ resonance in the
angular analysis, the fraction of which is fitted in later
analyses. The \d0\ Collaboration
finds\cite{Abazov:2011ry,Abazov:2008af,Abazov:2007zj,Abazov:2007tx,Abazov:2005sa}
$\DGs = 0.163^{+0.065}_{-0.064} \thinspace {\mathrm{ps}}^{-1}$
and $\tau(B^0_s) = 1/\Gs = 1.443 \pm 0.038 \thinspace {\mathrm{ps}}$,
while CDF measures\cite{Acosta:2004gt,Aaltonen:2007gf,Aaltonen:2007he,CDF:2011af,Aaltonen:2012ie}
$\DGs = 0.068 \pm 0.026 \, ({\mathrm{stat}}) \pm 0.007 \, ({\mathrm{syst}}) \thinspace {\mathrm{ps}^{-1}}$ and 
$\tau(B^0_s) = 1/\Gs = 1.528 \pm 0.019 \, ({\mathrm{stat}}) \pm 0.009 \, ({\mathrm{syst}}) \thinspace {\mathrm{ps}}$. These can be compared with the theory prediction\cite{Lenz:2011ti} of
$\DGs = 0.087 \pm 0.021 \thinspace {\mathrm{ps}}^{-1}$ for the most stringent
test of the validity of HQE.

%% file: CPViolation.tex
%
%
%
\section{CP Violation}
\label{CPViolating}

One of the most prominent questions in particle physics is the source
of the baryon-antibaryon asymmetry observed in the universe. One of
the requirements for this asymmetry is \CP\ violation. The SM
naturally includes \CP\ violation (CPV) in the quark sector through
the presence of a single complex phase in the CKM matrix, which in
turn determines the strength of flavor transitions through the weak
interaction.  However, the degree of CPV from this SM source is
insufficient to explain the cosmological matter
dominance.\cite{Huet:1994jb} Heavy flavor systems are ideal to search
for new phases and levels of CPV that depart from SM
predictions. There are three kinds of \CP\ violation, all explored at
the Tevatron: direct \CP\ violation where the probability for a particle
to decay to a given final state differs from the probability for the
antiparticle to decay to the charge conjugate state $|{\cal{A}}_f|^2 \neq
|{\cal{\bar{A}}}_{\bar{f}}|^2$, \CP\ violation in mixing where $|q/p|
\neq 1$ (see Eq.~\ref{eq:mixing}), and in the interference of decay
and mixing amplitudes.\footnote{test}

The advantage of carrying out CPV tests at the Tevatron is the
$p\bar{p}$ \CP-invariant initial state, in contrast to the $pp$ collisions at the LHC, 
where production asymmetries need to be taken into account.
In the case of the \d0\ detector\cite{Abazov:2005pn}, the polarities
of the toroidal and solenoidal magnetic fields were reversed on
average every two weeks so that the four solenoid-toroid polarity
combinations are exposed to approximately the same integrated
luminosity. This allows for a cancellation of first-order effects
related to instrumental charge asymmetries, particularly for tracking,
to achieve levels of precision that would be difficult to reach otherwise.

      \subsection{CP violation in charm}

\begin{figure}[tb]
\centerline{\includegraphics[width=0.95\linewidth]{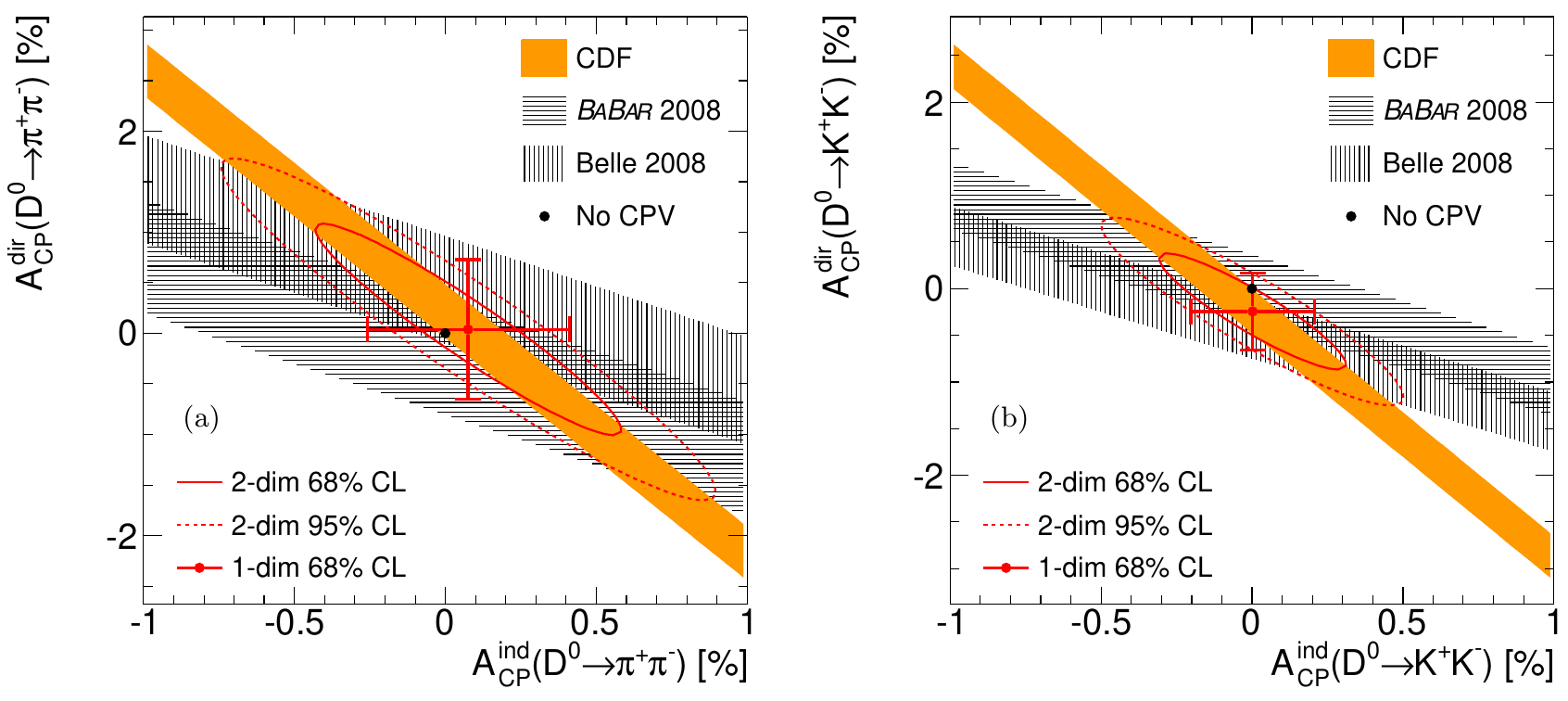}}
\hfill
\caption[]{(a) CDF results on \CP\ violation in $D^0 \rightarrow
  \pi^+\pi^-$  decays.  The asymmetries from direct and indirect \CP\
  violation.
  The allowed band is a function of the mean decay time of the sample.
   (b) Results for $D^0 \rightarrow K^+K^-$  decays.}
\label{fig:CPD0hh}
\end{figure}

CDF has studied \CP\ violation in $D^0$ decays using the soft pion
from $D^{*\pm}$ decays to tag the $D$-meson flavor at the time of
production as in the mixing measurement.  Also similar to the mixing
measurement, the component related to $D^0$ mesons produced in bottom
decays is subtracted using the $D^0$ impact parameter information.
Following initial studies\cite{Acosta:2004ts} with small data samples,
CDF searched\cite{Aaltonen:2011se} in a 5.9\,\ifb\ sample for \CP\ violation in the decays to
\CP\ eigenstates $D^0 \rightarrow K^+K^-$ and $D^0 \rightarrow
\pi^+\pi^-$ and found results that were consistent with both the
standard model and significant \CP\ violation.  The individual
asymmetries from direct and indirect \CP\ violation cannot be
determined from a time-integrated measurement. The figure shows that
band that is allowed based on the mean decay time of the sample.

Theoretical calculations\cite{Grossman:2006jg} indicated that CPV in
the charm system should be small, of order $10^{-3}$.  Thus a large
difference $\Delta A_{CP}$ between the asymmetries for $\pi\pi$ and $KK$ would
indicate a large asymmetry in one mode or both.  Taking the difference
also has the advantage of canceling production asymmetries as well as
production biases.  LHCb was able to to use this technique to find the
first evidence \cite{Aaij:2011in} of \CP\ violation in
charm decays.  CDF followed up\cite{Aaltonen:2012qw} using the full
dataset and including events with a looser selection than the original
analysis.  The three observables were the $\pi\pi$ asymmetry and $KK$
asymmetry using the original selections as well as the difference in
asymmetries from the subsample with the looser selection.  The new
result improved the individual asymmetry measurements and confirmed the
LHCb measurement of a difference from zero in $\Delta A_{CP}$ at nearly
the $3\sigma$ level.  The CDF measurements of the individual
asymmetries are the most precise in the world.

CDF has also searched\cite{Aaltonen:2012nd} for CPV in $D^0
\rightarrow K^0_S \pi^+ \pi^-$ decays.  Two complementary approaches
are used: a full Dalitz fit employing the isobar model for the
involved resonances and a model-independent bin-by-bin comparison of
the $D^0$ and $\bar{D}^0$ Dalitz plots.  No evidence of \CP\ violation
is found.

The \d0\ Collaboration has searched for direct CPV in charm decay.
Since all of the contributing processes to each of the  decays $D_s^\pm \ra \phi
\pi^\pm$ and $D^{\pm} \ra K^{\mp} \pi^{\pm}$ 
have the same weak phase in each case, there should be no direct CPV for these
channels, so that any non-zero value could point towards new
physics.  The \CP\ asymmetry of these channels is also assumed to be zero for
a number of other heavy flavor \CP\ measurements. \d0\ has made the
most precise measurements\cite{Abazov:2013woa,Abazov:2014wga} of these \CP\ asymmetries,
and they are indeed consistent with zero.

      \subsection{Direct CP violation in $B$ decays}

Two-body charmless hadronic decays of $b$ hadrons are an important
avenue for the study of direct \CP\ violation.  While \CP\ asymmetries
of \Bd and \Bu mesons can be studied at $e^+e^-$ colliders, a complete
understanding can be achieved only by comparing results to those
for \Bs meson which are the exclusive domain of hadron colliders.  The
study of charmless modes of \Lb also offers the opportunity to search
for physics beyond the standard model.  The CDF SVT was designed for
the purpose of identifying two-body charmless decays in the
trigger.\cite{CDF_TDR} 

With a sample of only 180\,\ipb, CDF was able to make the first
observation\cite{Abulencia:2006psa} of a charmless \Bs decay mode.
While the signature of a two-particle decay that is displaced from the
beamline gives a quite clean signal for decays of the form $B_{(s)}\ra
h^+h^{\prime-}$ where $h$ and $h^\prime$ are either pions or kaons,
the signals from the various modes are nearly degenerate in mass.
Therefore, it is necessary to use particle identification coupled with
kinematic information to disentangle the various modes.  While
measurements of $dE/dx$ in the CDF drift chamber cannot provide
event-by-event identification, when included in a likelihood fit along
with the mass and the asymmetry between the $h^+$ and $h^{\prime-}$
momenta, the yields in the various modes can be determined.  Thus CDF
was able to observe $\Bs\ra K^+ K^-$ and to make measurements of the
branching ratios of $\Bd\ra\pi^+K^-$ and $\Bd\ra\pi^+\pi^-$ that were
nearly as good as those from \B factories at that time.\cite{PDG2006}  

\begin{figure}[tb]
\centerline{\includegraphics[width=0.50\linewidth]{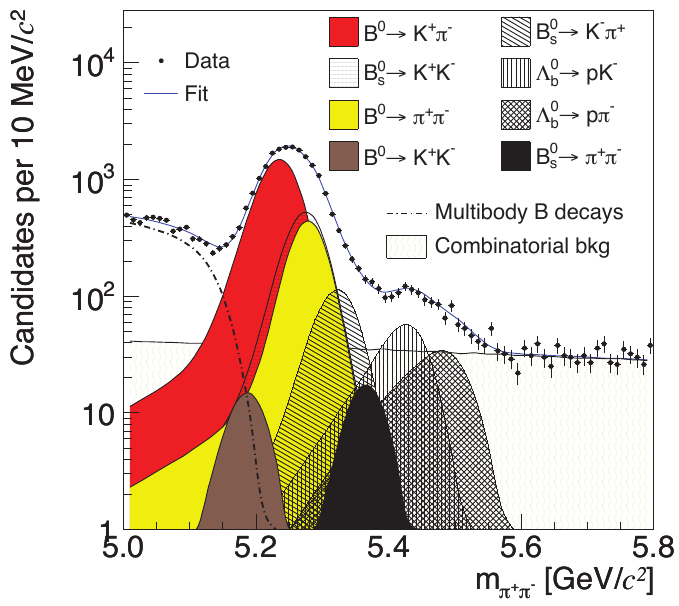}}
\hfill
\caption{ Mass distribution of reconstructed candidates in two-body
  charmless decays where the charged pion mass is assigned to both
  tracks in the full 9.6\,\ifb\ sample from CDF. The sum of the fitted
  distributions and the individual components (conjugate decay modes
  are also implied) of signal and background are overlaid on the data
  distribution. The degeneracy of the different components is broken
  using particle identification and kinematic information.}
\label{fig:BhhCDF}
\end{figure}

Using similar methods, in a sample of 1\,\ifb\ CDF was made the first
observation\cite{Aaltonen:2008hg} of the decay modes $\Bs \ra
K^-\pi^+$, $\Lb\ra p K^-$, and $\Lb\ra p \pi^-$, each with greater
than $6\sigma$ significance, and measured their branching ratios.
Differences between \CP\ asymmetries in $\Bd\ra K^+\pi^-$ and $\Bu\ra
K^+\pi^0$ observed at the $b$
factories\cite{Lin:2008zzaa,Aubert:2007mj,Aubert:2007hh} is
significantly larger than na\i ve expectations of the
SM.\cite{Keum:2002vi,Beneke:2003zv} Insight into theoretical
explanations of these results can be found from the asymmetry in
$\Bs\ra K^-\pi^+$ decays.  Using improved techniques in the 1\,\ifb
sample, CDF made the first measurement\cite{Aaltonen:2011qt} of this
quantity as well as improved measurements of the branching ratios of
$\Bd\ra\pi^+\pi^-$ and $\Bs\ra K^+K^-$ and the first measurements of
asymmetries in $\Lb\ra p K^-$, and $\Lb\ra p \pi^-$ decays.  Further
progress in \Bs\ decays came with 6.1\ifb\ and the first
evidence\cite{Aaltonen:2011jv} for the charmless annihilation decay
mode $B^0_s \to \pi^+\pi^-$.  Subsequently, LHCb made the first
observation \cite{Aaij:2013iua} of \CP\ violation in \Bs\ decays.
Using the full 9.6\,\ifb\ dataset, CDF
confirmed\cite{Aaltonen:2014vra} that observation.  The asymmetry
measurements in the charmless \Lb\ decay modes have better than 10\%
precision and show no significant asymmetry.  Those measurements
remain unique to CDF.

In the 180\,\ipb\ sample, CDF also searched\cite{Acosta:2005eu} for \CP\
violation in $B^+ \rightarrow  \phi K^+$ decays.  The asymmetry was
consistent with zero, but the measurement was competitive with those
from Belle\cite{Chen:2003jfa} and Babar.\cite{Aubert:2003hz}

The SM predicts that for $b \ra  c\bar{c}s$ decays, the tree and
penguin contributions have the same weak phase, and thus no direct \CP\
violation is expected in  $\Bu\ra\Jpsi\, K^{\pm}$ decays,
although there may be a tiny amount due to penguin loops. The \d0\
Collaboration has made the best precision
measurement\cite{Abazov:2013sqa,Abazov:2008gs} of the \CP\ asymmetry
between the width of $\Bu \ra \Jpsi\, K^+$ and its charge conjugate and
finds that is consistent with zero.

%
%

\begin{figure}[tb]
\centerline{\includegraphics[width=0.9\linewidth]{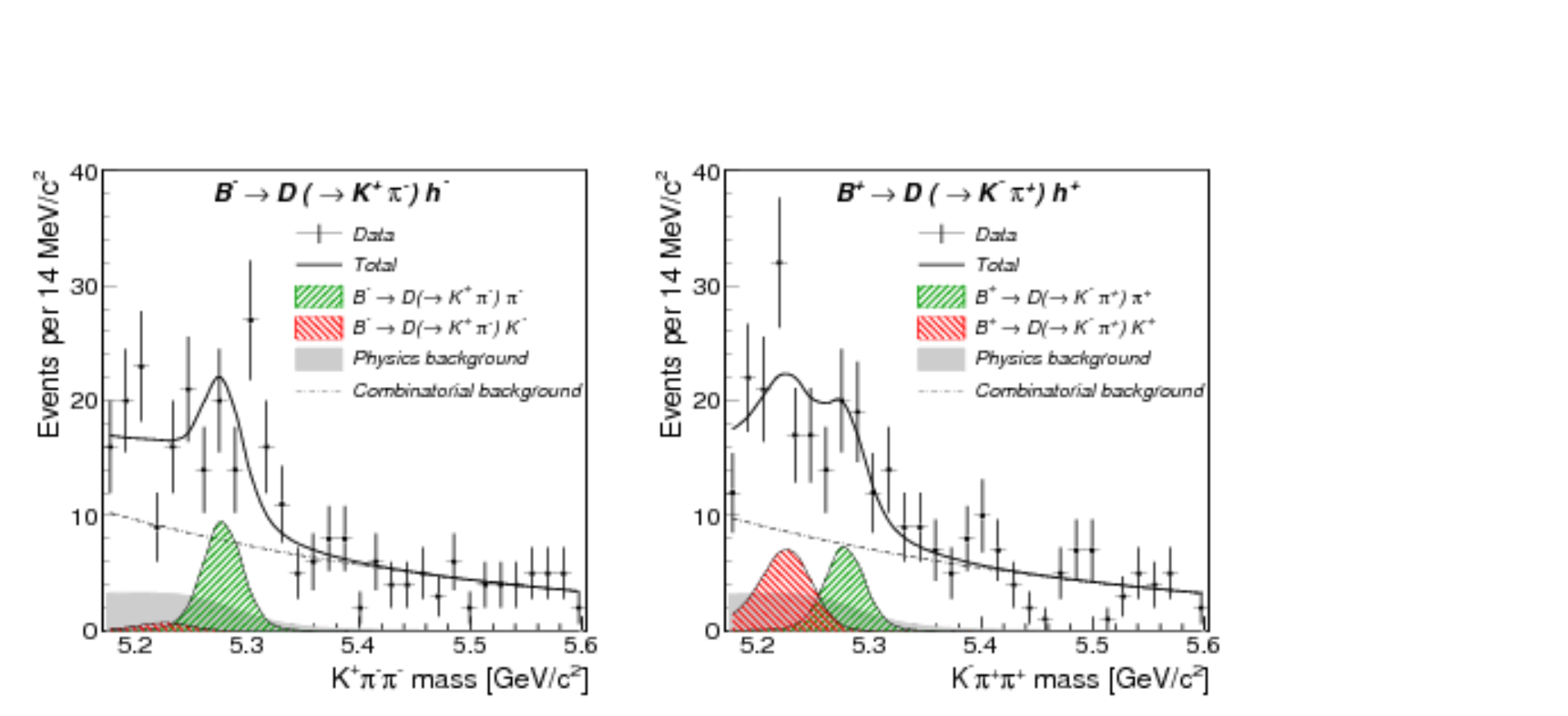}}
\hfill
\caption{Invariant mass distributions of the suppressed mode $B^\pm\ra D^0 h^\pm$.  The pion mass is assigned to the charged track from
the $B$ candidate decay vertex.}
\label{fig:ADS}
\end{figure}

Taking advantage of the displaced-track trigger, CDF has also studied
\CP\ violation in hadronic decay modes with $b\ra c$ transitions.  The
branching fractions and \CP\ asymmetries of $B^\pm\ra D^0 h^\pm$
modes, where $h$ is a pion or kaon, allow a theoretically-clean way of
measuring the CKM angle $\gamma$ which is the least well-known CKM
angle.\cite{PDG2012} The ADS method\cite{Atwood:1996ci,Atwood:2000ck}
takes advantage of the the large interference between the process in
which a $B^-\ra D^0 h^-$ decay through a color-allowed $b\ra c$
transition is followed by the doubly Cabibbo-suppressed $D^0\ra
K^+\pi^-$ decay and the process in which a $B^-\ra\bar{D}^0 h^-$ decay
through a color-suppressed $b\ra u$ transition followed by the
Cabibbo-favored decay $\bar{D}^0\ra K^+\pi^-$.  This interference can
lead to significant \CP\ asymmetries from which $\gamma$ can be
extracted.  CDF observed \cite{Aaltonen:2009hz,Aaltonen:2011uu} both
the $B^-\ra[K^+\pi^-]_D K^-$ and $B^-\ra[K^+\pi^-]_D \pi^-$ modes with
greater than $3\sigma$ significance and measured the rates and
asymmetries.

      \subsection{CP violation in $B$ mixing }
      \label{CPV_mixing}

\CP\ violation can develop in the mixing of the neutral $B$ meson
system if the $2 \times 2$ mass and decay matrix described in
Sect.~\ref{DmsDGs} has a non-zero phase $\phi =
\arg[-M_{12}/\Gamma_{12}]$ between the off-diagonal elements that are
responsible for the mixing.  The time-integrated flavor-specific
semileptonic charge asymmetry is defined as
\begin{equation}
\label{eq:semilep_ch_asymm}
a^{d(s)}_{\mathrm{sl}} = \frac{\Gamma(\bar{B}^0_{(s)} \ra {B}^0_{(s)} \ra \ell^+X) - \Gamma({B}^0_{(s)} \ra \bar{B}^0_{(s)} \ra \ell^-X)}{\Gamma(\bar{B}^0_{(s)} \ra {B}^0_{(s)} \ra \ell^+X) + \Gamma({B}^0_{(s)} \ra \bar{B}^0_{(s)} \ra \ell^-X)}
= \frac{\Delta \Gamma_q}{\Delta m_q} \tan \phi_q
\end{equation}
that is also equivalent to $(|p/q|^2- |q/p|^2)/(|p/q|^2+ |q/p|^2)$.

By measuring the asymmetry between the number of reconstructed $B^0
\ra D^{(*)-} \mu^+ X$ decays compared to the charge conjugate
$D^{(*)+} \mu^-$ in bins of visible proper decay length and then
correcting for detector asymmetries as shown in
Fig.~\ref{fig:CPBs}(a), the \d0\ Collaboration
measured\cite{Abazov:2012hha} $a^d_{\mathrm{sl}} = [0.68 \pm 0.45 \,
{\mathrm{(stat.)}} \pm 0.14 \, {\mathrm{(syst.)}}]\%$, which is the
single most precise measurement of this parameter, with uncertainties
smaller than the previous world average of $B$ factory measurements,
and consistent with the SM prediction\cite{Lenz:2011ti} of $<
10^{-3}$. The \d0\ Collaboration performed a similar analysis using
time-integrated $B_s^0 \ra D_s^- \mu^+ X$ fitting simultaneously to
the sum and the difference of the two charge-conjugate processes as
shown in Fig.~\ref{fig:CPBs}(b,c) to
measure\cite{Abazov:2012zz,Abazov:2009wg} $a^s_{\mathrm{sl}} =
[−1.12 \pm 0.74 \, {\mathrm{(stat)}} \pm 0.17 \,
{\mathrm{(syst)}}]\%$, the most precise measurement at the time,
consistent with the current LHCb measurement\cite{Aaij:2013gta} and small value of the SM
prediction.\cite{Lenz:2011ti}

\begin{figure}[tb]
\centerline{\includegraphics[width=0.90\linewidth]{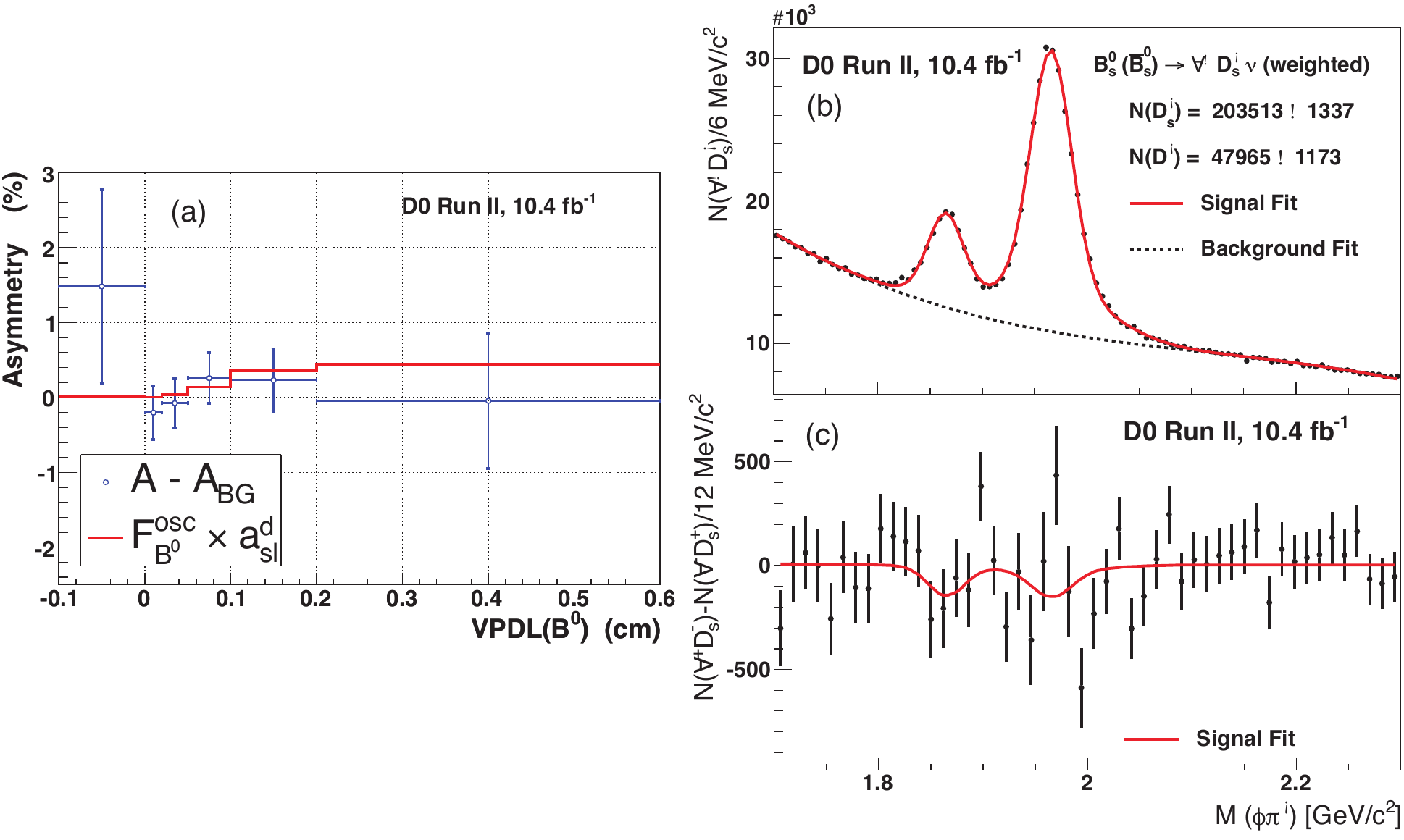}}
\hfill
\caption[]{(a) Background-corrected semileptonic charge asymmetry versus
visible proper decay length of $B^0$.  (b) Sum of $B^0_s \thinspace
(\bar{B}^0_s) \ra D_s^{\mp} \mu^{\pm}$; and (c) difference between the
two conjugate states.}
\label{fig:CPBs}
\end{figure}

One of the few ways for direct physics ({\it i.e.} excluding pion and
kaon decays in flight) to result in a pair of
same-sign muons is when a $b$ hadron
directly decays semileptonically, while a neutral $B$ meson from the
other produced $b$ quark oscillates before decaying semileptonically.
\CP\ violation in mixing can be expressed as $\Gamma(B^0_{(s)} \rightarrow
\bar{B}^0_{(s)} \rightarrow \mu^- X) \neq \Gamma(\bar{B}^0_{(s)}
\rightarrow {B}^0_{(s)} \rightarrow \mu^- X)$ and can be explored by
studying the semileptonic asymmetry ${\cal{A}}^b_{\mathrm{sl}} =
[N_b(\mu^+\mu^+) - N_b(\mu^-\mu^-)]/{\mathrm{sum}}$ by forming the raw
asymmetry, correcting for background asymmetries using independent
data samples, and determining the fraction of muons from $b$ quarks.
This asymmetry is a linear combination of the semileptonic charge
asymmetries of $B^0$ and $B^0_s$, {\it i.e.} ${\cal{A}}^b = C_d
a^d_{\mathrm{sl}} + C_s a^s_{\mathrm{sl}}$.

In previous
publications\cite{Abazov:2011yk,Abazov:2010hj,Abazov:2010hv,Abazov:2006qw},
the \d0\ Collaboration measured ${\cal{A}}^b$ with increasingly larger
datasets with values representing up to a $3.9\sigma$ deviation from
the SM prediction. The analysis~\cite{Abazov:2013uma} with the full
Run 2 dataset added a more detailed study of the asymmetry dependence
on the impact parameter (IP), $p_T$, and $|\eta|$ of each muon, as
well as including an additional \CP-violating process to interpret
results~\cite{Borissov:2013wwa}.  Measurements in the dimuon sample
shown in 
Fig.~\ref{fig:CPresults}(a) give a result of $A_{\mathrm{CP}}$ that
represents a $3.6\sigma$ deviation from the SM prediction which is the
largest observed deviation in the heavy flavor physics program at the
Tevatron.  Since the fractional mix of $B^0$ and $B^0_s$ is
different in each $({\mathrm{IP_1}},{\mathrm{IP_2}})$ bin, the
semileptonic charge asymmetries can be extracted as shown in
Fig.~\ref{fig:CPresults}(b).  The result deviates
from the SM by $3.0\sigma$. The results are consistent with the
independent \d0\ measurements\cite{Abazov:2012hha,Abazov:2012zz} of $a^d_{\mathrm{sl}}$ and
$a^s_{\mathrm{sl}}$  described above, and the
combination of all \d0\ results are also shown in
Fig.~\ref{fig:CPresults}(b).  These are the most precise determinations
of these quantities so far from a single measurement.
The observed dimuon charge asymmetry also has a contribution
from the \CP\ violation in the interference of decay amplitudes for the 
decay $B^0 (\bar{B}^0) \rightarrow c\bar{c}d\bar{d}$ with and 
without mixing.\cite{Borissov:2013wwa}
Since this contribution is proportional to the quantity ${\Delta\Gamma_d}/{\Gamma_d}$,
the measurement also allows for the extraction of
${\Delta\Gamma_d}/{\Gamma_d} = (0.50 \pm 1.38)\%$ (and $(0.79 \pm 1.15)\%$
when combining all the \d0\ measurements). This combination
is still consistent with all other measurements, and also stresses the
importance of having more independent measurements of
${\Delta\Gamma_d}/{\Gamma_d}$.

\begin{figure}[tb]
\centerline{\includegraphics[width=0.90\linewidth]{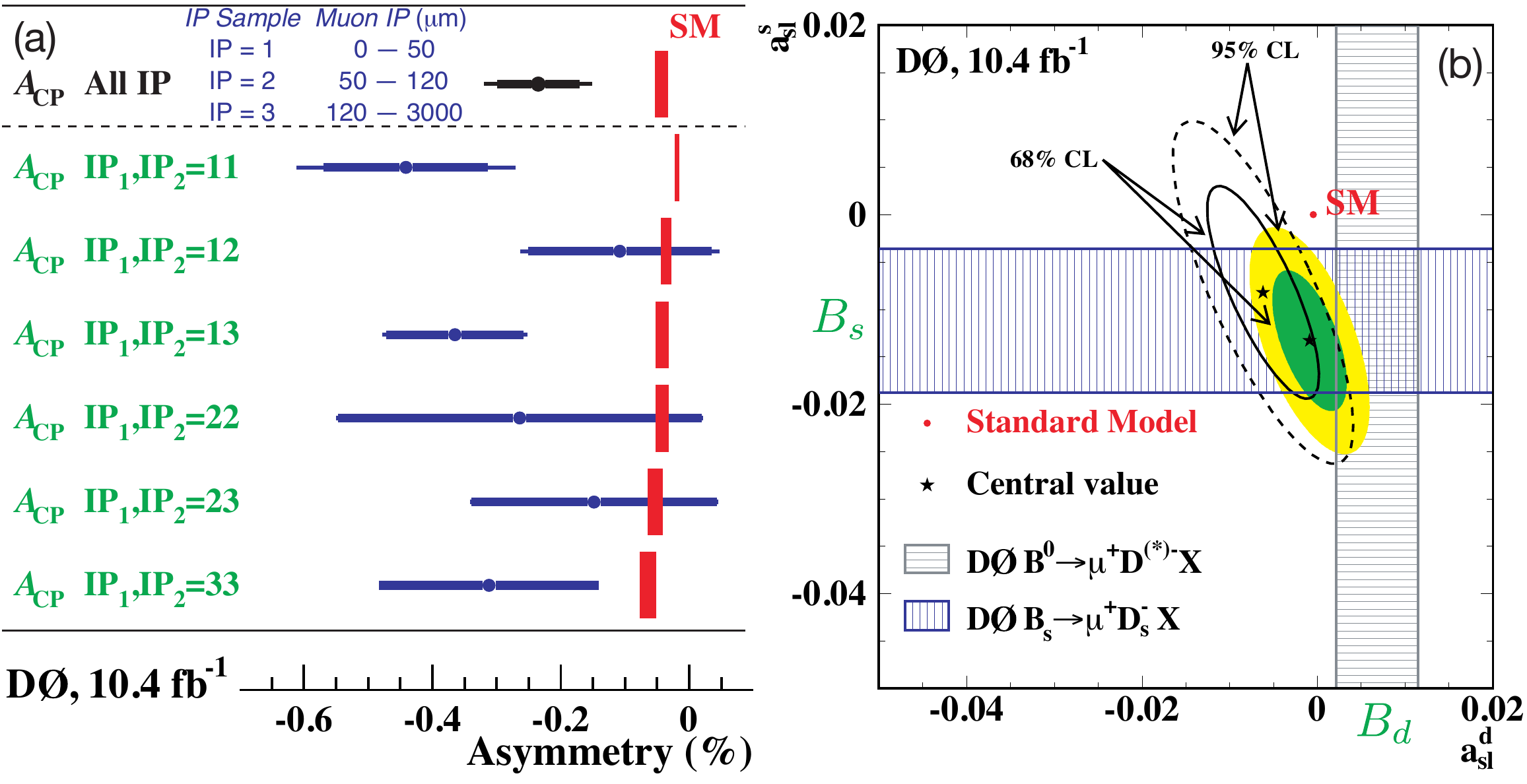}}
\hfill
\caption[]{(a) Dimuon charge \CP\ asymmetry ${\cal{A}}_{\mathrm{CP}}$
in bins of impact parameter (IP) of each of the two muons and over
entire sample. (b) Result (black dashed and solid line contours) of
dimuon charge asymmetry as a function of semileptonic \CP\ charge
asymmetry $a^s_{\mathrm{sl}}$ and $a^d_{\mathrm{sl}}$. Solid colored
contours after combination with independent direct measurements
(hashed bands) from \d0.}
\label{fig:CPresults}
\end{figure}

     \subsection{CP violation in interference between decay and $B$ mixing}

In the ``golden mode" of the $B$ factories, the final state of
$\Jpsi\,K^0$ can be reached directly from \Bd\ decay or after a flavor
oscillation, {\it i.e.} $\Bd \, (\ra \bar{B}^0) \ra \Jpsi\, K^0$. There is
interference since the same final state can be reached by two
different decay paths, and CPV can occur in
this interference via phases $\phi^{N\! P}$ arising from new
physics. This decay explores the unitarity triangle formed from the
first and third columns of the CKM matrix and can be used to measure
precisely the angle $\beta$. The corresponding golden modes at the
Tevatron are decays of the type $B^0_s \, (\ra \bar{B}^0_s) \ra \Jpsi\,
\phi$ that probe the ``squashed" unitarity triangle for the \Bs\
system formed from the second and third columns of the CKM matrix
characterized by the tiny angle\cite{Lenz:2011ti} $\beta_s^{S\!M} =
\arg[-V_{ts}V^*_{tb}/V_{cs}V^*_{cb}] \simeq 0.02$.
In the absence of new physics, the CPV phase measured in these
analyses will give $\phi_s = 2\beta_s^{S\!M}$, too small to resolve
with the sensitivities of the Tevatron experiments, while giving
$\phi_s = 2\beta_s^{S\!M} + \phi^{N\! P}$ in the presence of new
phases that potentially could be observed.

\CP\ violation can therefore be manifested in a difference between the
decays $\Bs \, (\ra \bar{B}^0_s) \ra \Jpsi\, \phi$ and $\bar{B}^0_s \,
( \ra \Bs) \ra J/\psi\, \phi$. The angular distributions of the decays
of the two vector particles in the final state $\Jpsi \ra \mu^+ \mu^-$
and $\phi \ra K^+ K^-$ can be used to disentangle the \CP-even and
\CP-odd components as a function of proper decay time. With a sizable
lifetime difference \DGs, there is sensitivity to $\phi_s$ through the
interference terms between the \CP-even and \CP-odd states even without
flavor-tagging the initial state.  In the approach pursued in early
analyses by \d0 \cite{Abazov:2007zj,Abazov:2007tx} and
CDF,\cite{Aaltonen:2007gf} the lifetime difference \DGs, average
lifetime $\bar{\tau}_s = 1/\Gs$, and $\phi_s$ were measured. These
early analyses generated considerable excitement as both Tevatron
experiments indicated modest deviations from the SM, both to negative
values as shown in Fig.~\ref{fig:combo_phis} that when
combined\cite{Amhis:older} gave a 2.2$\sigma$ deviation from the SM,
and 2.7$\sigma$ when combined with the value of $a^s_{\mathrm{sl}}$ at
that time, as shown in Fig.~\ref{fig:combo_phis}(a). Later
analyses\cite{CDF:2011af,Aaltonen:2012ie,Abazov:2011ry,Abazov:2008af}
included initial-state flavor tagging as well as taking into account
any ${\cal{S}}$-wave $K^+K^-$ component under the $\phi$ mass
peak. Final analyses give results consistent with both later
measurements at LHC and with the SM (see
Fig.~\ref{fig:combo_phis}(b)).

\begin{figure}[tb]
\centerline{\includegraphics[width=0.95\linewidth]{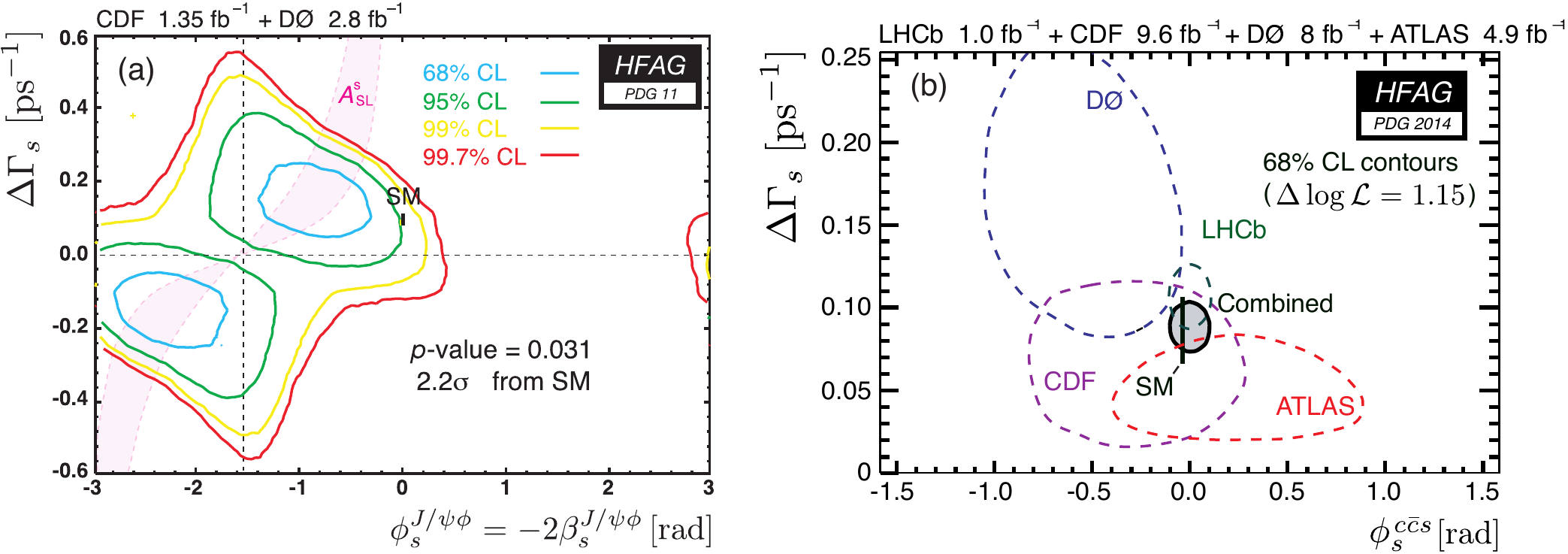}}
\hfill
\caption[]{(a) Combined CDF and \d0\ constraints on the \Bs\ lifetime difference \DGs\ and \CP-violating phase $\phi_s$ for PDG 2011; and (b) individual final constraints compared to and combined with results from LHC experiments.}
\label{fig:combo_phis}
\end{figure}

In the 180\,\ipb\ sample, CDF made the first
observation\cite{Acosta:2005eu} of a charmless \Bs\ decay, measuring
the branching ratio of $B^0_s \rightarrow \phi\phi$.  Because there
are two vector particles in the final state, the angular distributions
for this decay mode can be expressed in terms of three complex
amplitudes.  The full description includes the time evolution of the
heavy and light \CP\ eigenstates of the \Bs.  The general expression
for the angular distributions contains terms whose coefficients are
zero in the SM, thus providing another avenue to search for new
physics.  CDF has measured\cite{Aaltonen:2011rs} the amplitudes of the
angular distributions and searched for asymmetries in components
sensitive to new physics.  No significant asymmetries were found.  The
branching ratio is ${\rm BR}(B_s^0 \ra
\phi\phi)=[2.32\pm0.18 \, ({\mathrm{stat}})\pm0.82 \, ({\mathrm{syst}})]\times10^{-5}$.  The
longitudinal fraction of decays can be used to understand the details
of the decay process and is found to be $f_L = 0.348 \pm 0.041 ({\rm
stat}) \pm 0.021 ({\rm syst})$.

%% file: rare_decays.tex
%
%
%
\section{Rare Decays}
   
\begin{figure}[tb]
\centerline{\includegraphics[width=0.70\linewidth]{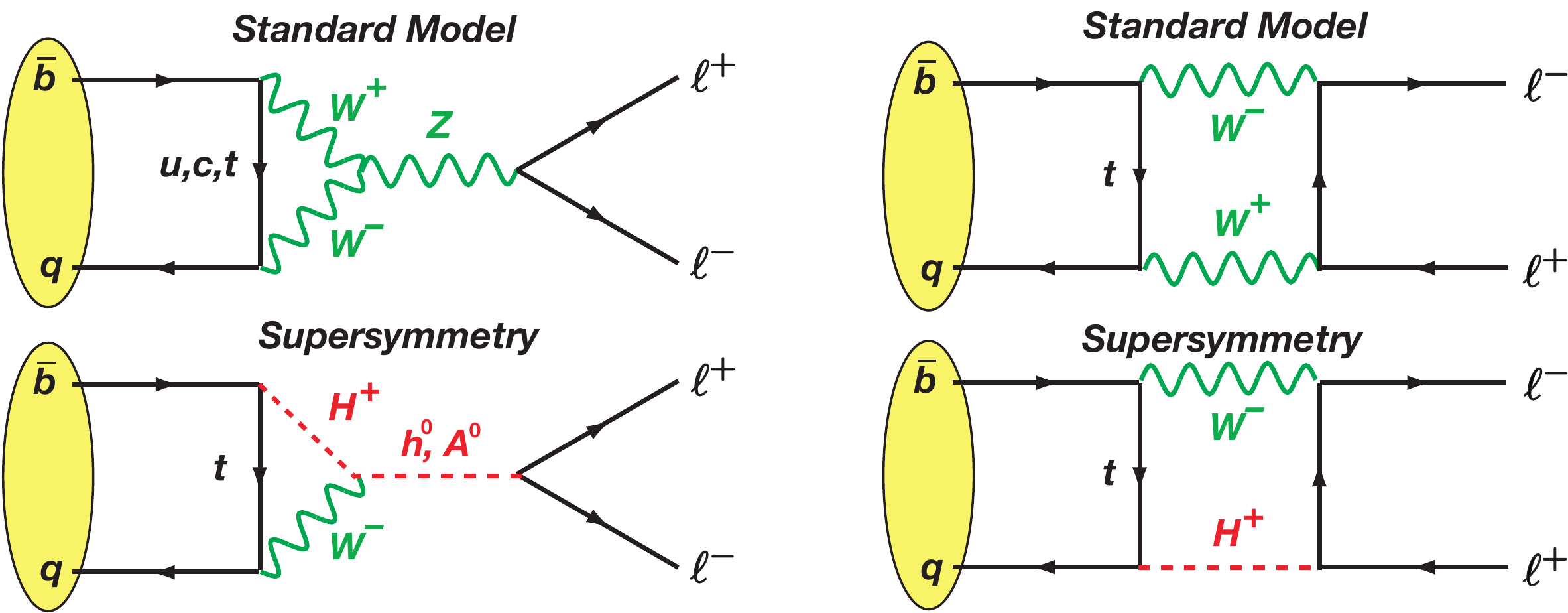}}
\hfill
\caption[]{In the SM, FCNC decays can proceed through box and loop $W$
and $Z$ diagrams, while in SM extensions new particles can contribute
to the processes and increase the branching fractions.}
\label{fig:FCNC_diagrams}
\end{figure}

Flavor-changing neutral current (FCNC) decay modes are forbidden at
the tree level in the SM by the GIM mechanism. They can proceed
through higher-order effective FCNC currents as shown in
Fig.~\ref{fig:FCNC_diagrams}. While highly suppressed in the SM, many
extensions to the SM allow for the branching fractions to be
increased, which makes these decays highly sensitive probes for
physics beyond the SM.

   \subsection{FCNC decays of charm}

The GIM suppression in $D$ meson decays is significantly stronger than
for $B$ mesons because of the smaller quark masses, and the SM
branching fractions are expected to be lower, leaving a large window
of opportunity still available to search for new physics in charm
decays. CDF searched\cite{Aaltonen:2010hz} for the decay
$D^0\ra\mu^+\mu^-$ and set an upper limit on the branching fraction
$3.0\times10^{-7}$ at the 95\% confidence level using the
kinematically similar $D^0\ra\pi^+\pi^-$ channel for normalization.
The \d0\ Collaboration made the first observation of the decay $D_s^+
\ra \phi \pi^+ \ra \mu^+\mu^- \pi^+$ and the first evidence for the
decay $D^+$ to the same final state.\cite{Abazov:2007aj}. The search
for the $c \ra u \mu^+\mu^-$ transition in the decay $D^+ \ra
\mu^+\mu^- \pi^+$ was performed in the continuum region outside of the
$\phi$ resonance to set a limit of ${\cal{B}}(D^+ \ra \pi^+ \mu^+
\mu^-) < 3.9 \times 10^{-6}$ at the 90\% C.L., the most stringent on
this transition at the time.
         
   \subsection{FCNC decays of \texorpdfstring{$b$}{b} hadrons}

In $B$ mesons,  internal quark annihilation decays are suppressed
 by $(f_B/m_B)^2 \simeq 2 \times 10^{-3}$ relative
to the electroweak penguin $b \ra s \gamma$ decay.  Helicity
suppression factor factors then push the SM branching fractions for
${\cal{B}}(\Bs \ra \mu^+\mu^-)$ and ${\cal{B}}(\Bd \ra \mu^+\mu^-)$
down to $(3.2 \pm 0.3) \times 10^{-9}$ and $(1.1 \pm 0.1) \times
10^{-10}$, respectively.\cite{Buras:2012ru}

Aside from a weak constraint\cite{Acciarri:1996us} from the L3
Collaboration at LEP, the Tevatron experiments provided the only
significant bounds on ${\cal{B}}(B^0_q \ra \mu^+\mu^-)$ for decades,
contributing strongly to placing powerful constraints on BSM
physics. As an example, the decay rate for $\Bs \ra \mu^+\mu^-$ is
proportional to $(\tan\beta)^6$ in the minimal supersymmetric standard
model and to $(\tan\beta)^4$  in more generic two-Higgs doublet
models, where $\tan\beta$ is the ratio of the vacuum expectation
values of the two Higgs fields, so that decay rate can be enhanced
relative to the SM by over two orders of magnitude at large
$\tan\beta$ values. An overview of these constraints on various models
as well as a model-independent treatment can be found in
Ref.~\citen{Altmannshofer:2012az}.

\begin{figure}[tb]
\centerline{\includegraphics[width=0.90\linewidth]{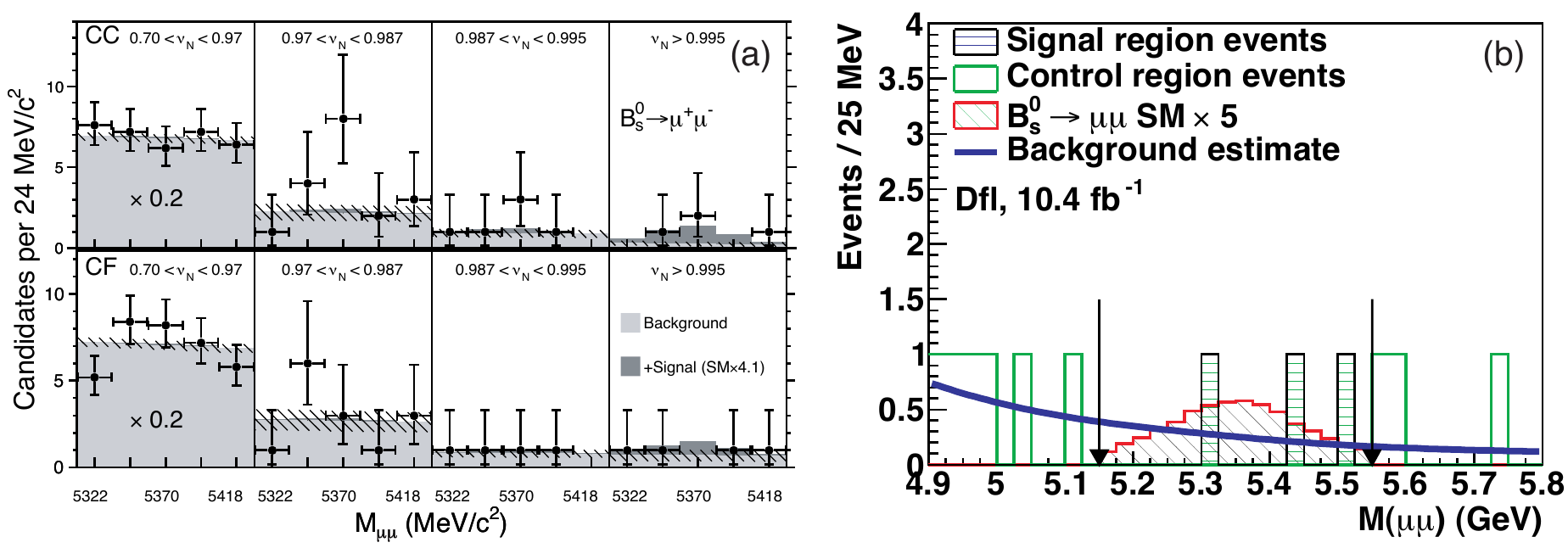}}
\hfill
\caption[]{(a) Results of CDF search for $\Bs \ra \mu^+\mu^-$ showing
  the dimuon mass distribution for four bins of neural net score and
  two sets of muon detectors. The light shaded region is the predicted
  background with the hatching showing its uncertainty, the points
  show the data, and the dark shading is the best fit for the
  signal. (b) \d0\ dimuon mass distribution in the blinded region for
  the full dataset after BDT selections are applied.}
\label{fig:Bsmumusigs}
\end{figure}

The CDF and \d0\ Collaborations carried out a series of increasingly
sophisticated searches for $\Bs \ra \mu^+\mu^-$ over time as more data
became available.  The analyses focused on reduction of combinatorial
and physics background through \Bs\ lifetime significance, muon
isolation, and other kinematic criteria.  CDF's first Run~2
result\cite{Acosta:2004xj} yielded a branching ratio 95\% CL upper
limit of $6\times10^{-7}$.  This was quickly surpassed by the first
Run~2 \d0\ analysis.\cite{Abazov:2004dj} Subsequent analyses by both
experiments used multivariate analyses and more data with the two
experiments achieving similar sensitivities as a result of their
relative strengths, CDF's tracking resolution versus \d0's greater
muon acceptance.  The later \d0\ analyses used a likelihood ratio
selection\cite{Abazov:2007iy}, Bayesian neural
nets\cite{Abazov:2010fs}, and finally, with the full data
set\cite{Abazov:2013wjb} two separate boosted decision trees treating
different categories of physics backgrounds, resulting in the limit
${\cal{B}}(\Bs \ra \mu^+\mu^-) < 1.5 \times 10^{-8}$ at the 95\%
C.L. as shown in Fig.~\ref{fig:Bsmumusigs}(b) and
Fig.~\ref{fig:Bslimits}.  Since the \d0\ detector did not have the
necessary mass resolution to separate \Bd\ from \Bs\ decays, it was
assumed that there are no contributions from $\Bd \ra \mu^+\mu^-$
decays, since this decay is suppressed by $|V_{td}/V_{ts}|^2 \simeq
0.04$.  Multivariate techniques substantially improved CDF's
sensitivity where using a likelihood technique improved the
limit\cite{Abulencia:2005pw} by a factor of 4 after only doubling the
integrated luminosity.  In the analysis\cite{Aaltonen:2007ad} of
2\,\ifb, CDF improved the muon selection and applied a neural net
classifier.  The limit was calculated using several bins of neural net
score with different background suppression factors in order to
improve sensitivity.  In a 7\,\ifb\ sample, CDF
found\cite{Aaltonen:2011fi} a small excess above expected backgrounds
that was more consistent with SM production than with the null
hypothesis, but not sufficiently significant to claim discovery.
Backgrounds in this study were verified in comparisons to data in
orthogonal samples.  This result was confirmed in the full dataset,
leading to the result\cite{Aaltonen:2013as} 
${\cal{B}}(\Bs\ra\mu^+\mu^-)=(1.3^{+0.9}_{-0.7})\times10^{-8}$.

\begin{figure}[tb]
\centerline{\includegraphics[width=0.60\linewidth]{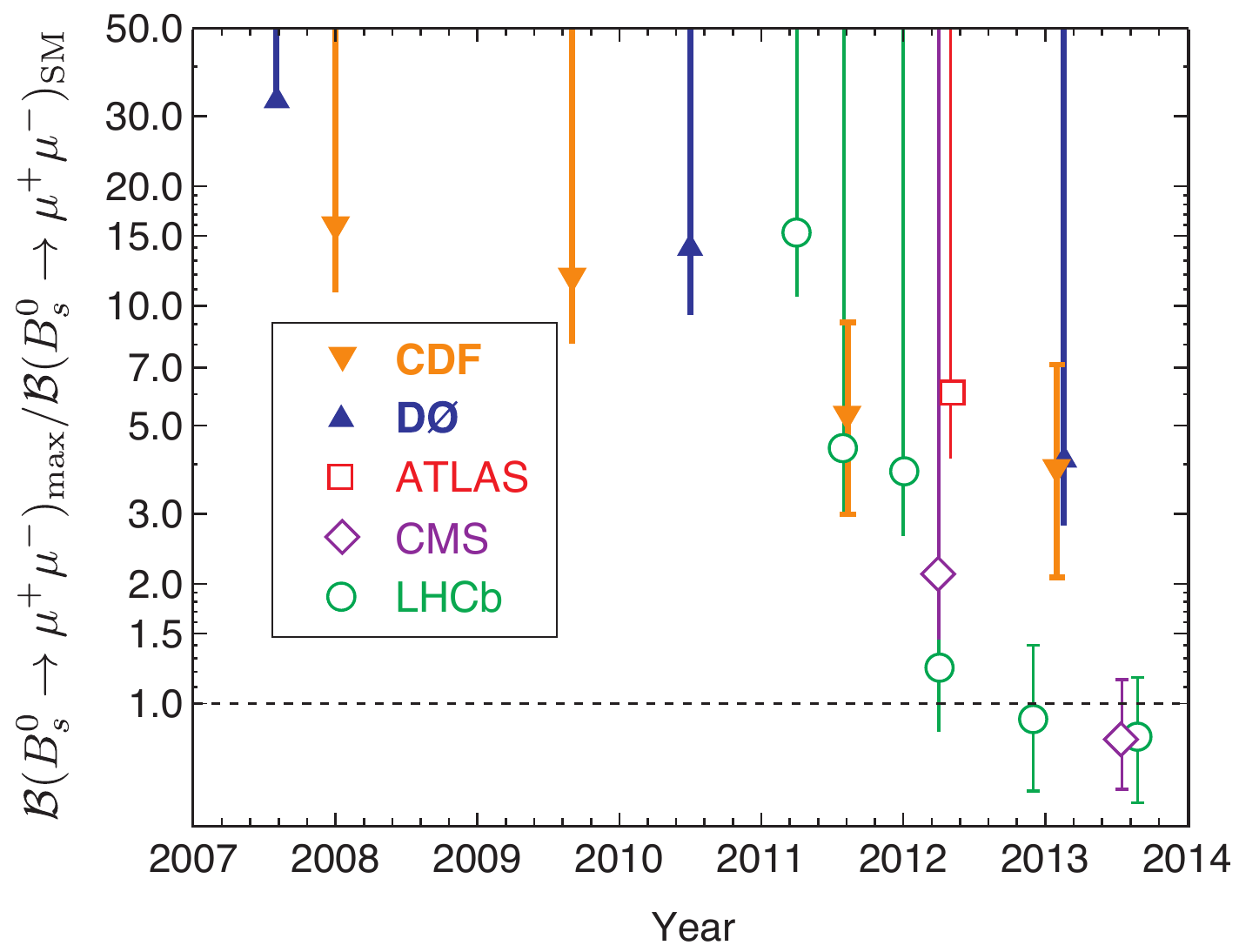}}
\hfill
\caption[]{History of ${\cal{B}}(\Bs\ra\mu^+\mu^-)$ results.  Symbols
  with one-sided bars represent 95\% C.L. limits, and symbols with
  two-sided error bars indicate measurements.}
\label{fig:Bslimits}
\end{figure}

To explore the quark-level transition $b \ra s \ell^+ \ell^-$,
inclusive FCNC decays like $\Bd \ra X_s \ell^+ \ell^-$ or $\Bd \ra X_s
\gamma$ are theoretically easier to calculate, but exclusive decays
such as $\Bd \ra K^* \ell^+ \ell^-$ with one hadron in the final state
are experimentally easier to study at the $B$ factories. The analogous
state $\Bs \ra \phi \ell^+ \ell^-$ tests the same transition and has
been explored at the Tevatron. The \d0\ Collaboration
searched\cite{Abazov:2006qm} for the decay $\Bs \ra \phi \mu^+ \mu^-$,
restricting the invariant mass of the dimuons to be outside the
charmonium resonances to avoid large interference effects with
dominant decay modes such as $\Bs \ra \Jpsi\, \phi$, and set a limit
that was the most stringent at the time.

CDF made the first observation\cite{Aaltonen:2008xf} of $b \rightarrow
s \mu^+\mu^-$ decays at a hadron collider in a sample of about
1\,\ifb\ and measured branching ratios for the $\Bu\ra K^+\mu^+\mu^-$,
$\Bd\ra K_S^0\mu^+\mu^-$, and $\Bd\ra K^{*0}\mu^+\mu^-$ decay modes.
With a factor of four increase in data, CDF made the first
observation\cite{Aaltonen:2011cn} of the decay
$B^0_s\rightarrow\phi\mu^+\mu^-$ and measured the differential decay
rate $d\Gamma/dq^2$ and the dimuon forward-backward asymmetry $A_{FB}$
as a function of $q^2\equiv M^2(\mu^+\mu^-)$ in $\Bu\ra
K^+\mu^+\mu^-$ and $\Bd\ra K^{*0}\mu^+\mu^-$ decays as well as the
the longitudinal polarization fraction $F_L$ as a function of $q^2$ in
the \Bd\ mode.  Because the presence of new particles mediating these
FCNC decays can change the internal dynamics of the decay,
distributions of quantities like $d\Gamma/dq^2$, $A_{FB}$, and $F_L$
provide important opportunities to search for the effects of new
physics.  The CDF measurements were quite competitive with those from
the $e^+e^-$ $B$ factories.\cite{Aubert:2008bi,Wei:2009zv}.

\begin{figure}[tb]
\centerline{
\includegraphics[width=0.47\linewidth]{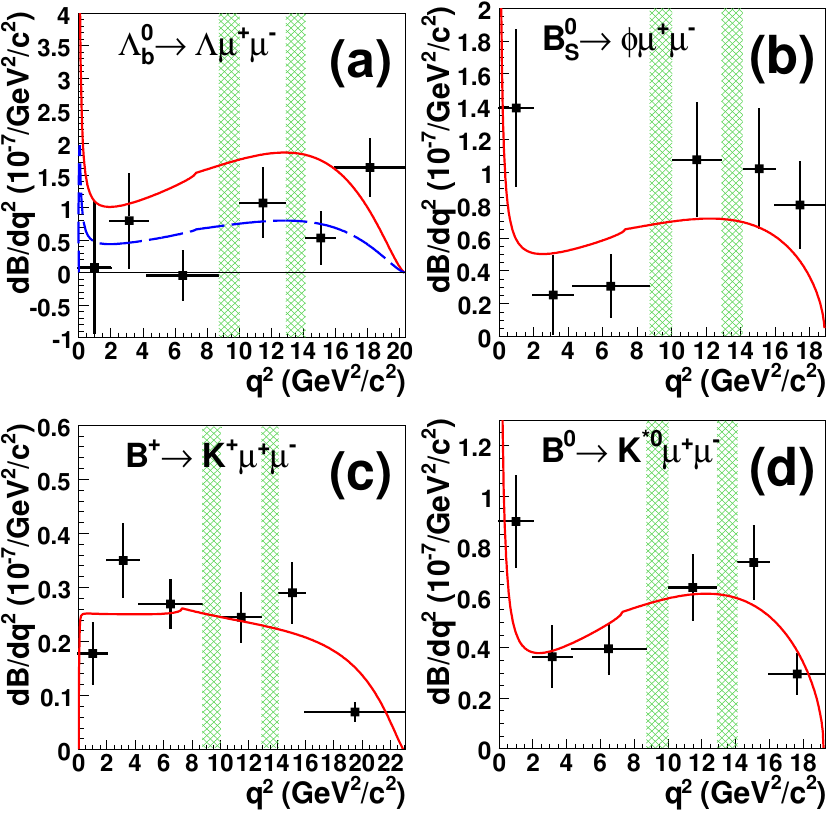}
\hfill
\includegraphics[width=0.49\linewidth]{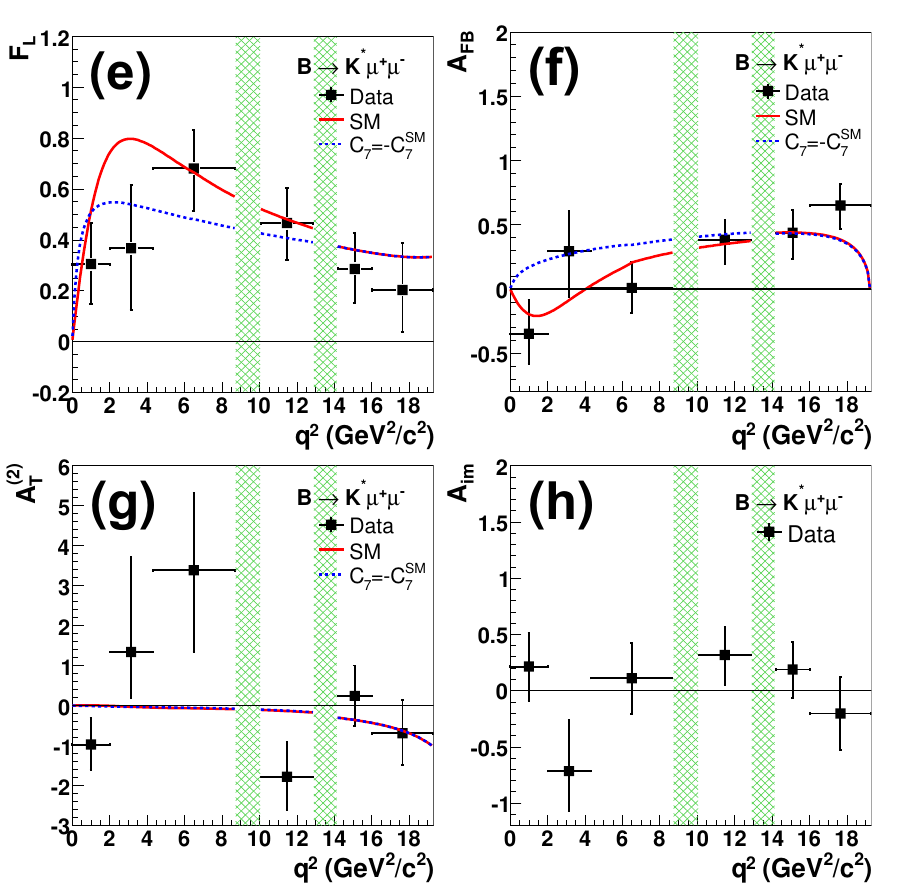}}
\caption[]{Distributions as a function of $q^2$ for various FCNC
decays measured by CDF. Differential branching ratios for 
{\it (a)} $\Lambda_b \rightarrow \Lambda \mu^+\mu^-$,
{\it (b)} $B^0_s\rightarrow\phi\mu^+\mu^-$
{\it (c)} $\Bu\ra K^+\mu^+\mu^-$,and
{\it (d)} $\Bd\ra K^{*0}\mu^+\mu^-$ decays show good agreement with the
standard model. 
Kinematic distributions in  $\Bd\ra K^{*0}\mu^+\mu^-$ decays
{\it (e)} $F_L$,
{\it (f)} $A_{FB}$, 
{\it (g)} $A_T^{(2)}$, and 
{\it (h)} $A_{im}$
are also consistent with the SM.
}
\label{fig:FCNC_CDF}
\end{figure}

With improved analysis methods, in a sample of 6.8\,\ifb, CDF made the
first observation\cite{Aaltonen:2011qs} of a FCNC decay of a baryon
for the decay $\Lambda_b \rightarrow \Lambda \mu^+\mu^-$ with a
significance corresponding to $5.8\sigma$.  CDF also improved on
branching ratios for the meson decay modes and made the first
measurement of $d\Gamma/dq^2$ in $B^0_s\rightarrow\phi\mu^+\mu^-$
decays.  Subsequently, CDF improved\cite{Aaltonen:2011ja} the
measurements of $A_{FB}$ and $F_L$ and made the first measurements of
transverse polarization asymmetry $A_T^{(2)}$ and the
time-reversal-odd charge-and-parity asymmetry $A_{im}$, including
$\Bu\ra K^{*+}\mu^+\mu^-$ decays as well.  Figure~\ref{fig:FCNC_CDF}
shows various distributions from this analysis.

\subsection{Lepton flavor violating decays}
      
While neutrino oscillations show that lepton number is violated, the
neutrino mass is too small for mixing to have an effect in hadron
decays.  Nevertheless, new particles in loops could lead to
substantial deviations from SM expectations.  For example, in the
Pati-Salam model,\cite{Pati:1974yy} leptoquarks carry quantum numbers
of both quarks and leptons.  CDF searched\cite{Aaltonen:2009vr} for
the lepton-flavor violating decay $B_{s,d}^0\rightarrow e^\pm\mu^\mp$
and set what was then the most stringent limit on the branching
ratios.  Those limits correspond to a 90\% CL lower limit on
leptoquark masses in the Pati-Salam model of 47.8\,\GeVcc
(59.3\,\GeVcc) for the \Bd\ (\Bs) decay mode.